\documentclass[12pt,a4paper]{article}
\usepackage{jcappub}

\usepackage{amsmath,amsfonts,amssymb}
\usepackage{epsfig,multicol,bbm,array}

\newcommand{\ud}{\mathrm{d}}

\newcommand{\om}{\omega_{\textrm{\tiny BD}}}

\title{Dynamical complexity of the Brans-Dicke cosmology}

\author[a]{Orest Hrycyna}
\author[b,c]{and Marek Szyd{\l}owski}
\affiliation[a]{Theoretical Physics Division, National Centre for Nuclear
Research,\\  Ho{\.z}a 69, 00-681 Warszawa, Poland}
\affiliation[b]{Astronomical Observatory, Jagiellonian University, \\Orla 171,
30-244 Krak{\'o}w, Poland}
\affiliation[c]{Mark Kac Complex Systems Research Centre, Jagiellonian
University, \\Reymonta 4, 30-059 Krak{\'o}w, Poland}

\emailAdd{orest.hrycyna@fuw.edu.pl}
\emailAdd{marek.szydlowski@uj.edu.pl}

\abstract{The dynamics of the Brans-Dicke theory with a quadratic scalar field potential function and barotropic matter is investigated. The dynamical system methods are used to reveal complexity of dynamical evolution in homogeneous and isotropic cosmological models. The structure of phase space crucially depends on the parameter of the theory $\om$ as well as barotropic matter index $w_{m}$. In our analysis these parameters are treated as bifurcation parameters. We found sets of values of these parameters which lead to generic evolutional scenarios. We show that in isotropic and homogeneous models in the Brans-Dicke theory with a quadratic potential function the de Sitter state appears naturally. Stability conditions of this state are fully investigated. It is shown that these models can explain accelerated expansion of the Universe without the assumption of the substantial form of dark matter and dark energy. The Poincare construction of compactified phase space with a circle at infinity is used to show that phase space trajectories in a physical region can be equipped with a structure of a vector field on nontrivial topological closed space. For $\om<-3/2$ we show new types of early and late time evolution leading from the anti-de Sitter to the de Sitter state through an asymmetric bounce. In the theory without a ghost we find bouncing solutions and the coexistence of the bounces and the singularity. Following the Peixoto theorem some conclusions about structural stability are drawn.
}

\keywords{modified gravity, dark energy theory, dark matter theory}
\arxivnumber{1310.1961}

\begin{document}
\maketitle

\section{Introduction}

In cosmology, which is understood as the physics of the Universe, at the very
beginning there appears the problem of indeterminacy of initial conditions for
the Universe. To tackle it, there are two complementary strategies employed in
modern cosmological investigations. Theoretical cosmology is dedicated to
the presentation of theoretical possibilities admissible for any initial
conditions, whereas the main aim of observational cosmology is to constrain
these theories as well as models derived from them. The latter usually
assumes the form of an effective theory of the Universe and the task left to
observational cosmology is to estimate the cosmological parameters from
astronomical data.

Theory of gravity governing cosmic expansion of the Universe constitutes a
very complicated system of nonlinear differential equations and only under some
simplified assumptions it can be solved. In cosmology, the cosmological principle
(i.e.~the assumption of the homogeneity and isotropy of the space) is just an
idealization of the Universe on the large scale. This principle determines the
universal cosmic time and hence one can consider the dynamics of expansion of
the Universe. From the mathematical point of view, the complicated dynamical
evolution is reduced to the dynamical system $\ud x/\ud t = f(x)$, where $x$ is a
state of the system vector and $f(x)$ is a smooth function. The solution of the
system is a function of both the cosmological time $t$ and initial conditions
$x_0$, i.e. $x(t,x_0)$. In this way, the dynamical system methods are available for
investigation of the degree of complexity of cosmic evolution. There are two main
advantages of using these methods in the context of cosmology. First, they
offer a mathematical language of studying all solutions of the system
for all admissible initial conditions. Second, all knowledge concerning the
dynamical behavior can be acquired without knowledge of its exact solution.

In cosmology, dynamical system methods have long tradition of successful
applications \cite{Wainwright:book, Belinskii:1985, Belinskii:1987}. A cosmic evolution determined by a model
is represented by trajectories in the phase space (a space of all states at any
instant of time $t$). The special role in organization of the phase space is
played by the
critical points which, from the physical point of view, correspond to stationary
states of the system. The trajectories representing system evolution as well as
critical points constitute the global phase portrait of the dynamics. This
portrait contains all information about dynamics modulo homeomorphism
preserving direction of time along the trajectories. Dynamical systems methods
give us mathematical tools of description of the system's trajectories in a
neighborhood of the non-degenerate critical points only in terms of its
linearization.

In the present paper we apply the dynamical system methods to investigation of
cosmological dynamics in the Brans-Dicke theory. This study extends previous
investigations of Brans-Dicke cosmological systems
\cite{Kolitch:1994kr,Kolitch:1994qa,Santos:1996jc,Holden:1998qg} to the case of models with a scalar field
with a potential. We additionally postulate the presence of a barotropic matter and due to a
suitable choice of state variables reduce it to the form of a three dimensional
dynamical system. In the special case of a quadratic potential the dynamical
system can be considered on the phase plane.

Our study of the dynamics in the phase plane is complete in this sense that
all representative phase diagrams are draw. It is important because the right hand sides of
the dynamical system under consideration depend on some parameters. We show how phase
diagrams qualitatively change with respect to the values of model's parameters like
the Brans-Dicke parameter $\om$ and the equation of state coefficient $w_{m}$ for the barotropic matter.
There are two cosmological reasons for studying global dynamics for all initial
conditions. First, we are looking for an alternative explanation of the
conundrum of acceleration of the Universe, alternative to the explanation in 
terms of the substantial dark energy of unknown form. We also explain dark matter in
terms of the Brans-Dicke theory and the parameter $\om$. Second, we reconstruct the
relation of variability of the effective gravitational constant.

The full analysis of dynamical behavior requires analysis of the system at infinity. If right hand sides of a dynamical system are given in a polynomial form then the projective variables can be used for such an analysis. For the system with a quadratic form of the potential function, the circle at infinity, in generic cases, consists of degenerate critical points. We show that this is the result of chosen phase space variables. This observation gives rise to a new phase space which possesses non-trivial topology. On such a manifold system is smooth and the phase space is finite. Then all the critical points are hyperbolic and the system is structurally stable. From the physical point of view this property can be interpreted as property of the model itself -- stability of the model under small perturbation.

We demonstrate how the standard cosmological model ($\Lambda$CDM
model) is derived from the Brans-Dicke cosmological models. On the level of a model
(rather then theory) the linearization of the dynamical system around a critical point representing the de Sitter state reproduces, but not exactly, the $\Lambda$CDM model. While
the equation of state coefficient for barotropic matter can assume an
arbitrarily small value, the case of the fine tuning of a dust matter is excluded. In other
words the correspondence of the Brans-Dicke cosmology with the $\Lambda$CDM model is
achieved for all matter besides dust $w_{m}=0$.

Different values of the bifurcation parameter corresponding to low energy string 
theory limit $\om=-1$ and the so-called ``O'Hanlon theory'' or ``massive
dilaton gravity'' \cite{ohanlon:1972,Capozziello:book} with $\om=0$ are studied. The conformal coupling value $\om=-3/2$ is excluded form our investigations since this value can not be crossed by continuous change of the parameter of the theory and this value leads to pathologies 
\cite{Dabrowski:2005yn}. The case $\om=0$ is much better known as $f(R)$ gravity. The quadratic potential function considered in this paper leads to $f(R)\propto R^{2}$ gravity for which de Sitter solutions exist for all values of the scalar curvature $R$. The stability of these solutions for the most general inhomogeneous and anisotropic space-times was proven in \cite{Starobinsky:1987zz}.

In general case the phase space is $3$-dimensional, for a very popular quadratic potential it 
reduces to a $2$-dimensional case which enables us to represent the global dynamics on
the phase plane. We study this dynamics in details and observe that inclusion
the cosmological constant in the model leads to structurally unstable system
\cite{Andronov:1937,Perko:2001,Wiggins:2003}
because of the presence of a degenerated non-hyperbolic critical point. 
 
The paper is organized as follows: in section \ref{sec:intro} we formulate the Brans-Dicke 
cosmological models on the flat Friedmann-Robertson-Walker background. 
Introducing new phase space 
variables we reduce dynamics of the model to a $3$-dimensional dynamical system. In section 
\ref{sec:quadratic} we present detailed qualitative dynamical analysis of the system with a 
quadratic potential function. This form of the potential constitutes an invariant sub-manifold 
of the original system. In subsection \ref{subsec:phase_diagrams} we present phase space 
diagrams for all regions of the parameter space $(w_{m},\om)$  while in subsection 
\ref{subsec:singularities} we present the analysis of dynamics of the Brans-Dicke scalar field.
The results are summarized in section \ref{sec:conclusions}.

\section{The Brans-Dicke cosmology}
\label{sec:intro}

The action for the Brans-Dicke theory \cite{Brans:1961sx} in the so-called Jordan
frame is of the following form \cite{Faraoni:book}
\begin{equation}
\label{eq:action}
S = \int\ud^{4}x\sqrt{-g}\left\{\phi R
-\frac{\om}{\phi}\nabla^{\alpha}\phi\nabla_{\alpha}\phi - 
2\,V(\phi)\right\} + 16\pi S_m
\end{equation}
where the barotropic matter is described by
\begin{equation}
S_{m} = \int\ud^{4}x\sqrt{-g}\mathcal{L}_{m}\,,
\end{equation}
and $\om$ is a dimensionless parameter of the theory.

Varying the total action (\ref{eq:action}) with respect to the metric
tensor we obtain the field equations for the theory
\begin{equation}
\label{eq:feq}
\begin{split}
& \phi\left(R_{\mu\nu}-\frac{1}{2}g_{\mu\nu}R\right) =\\
& =\frac{\om}{\phi}\left(\nabla_{\mu}\phi\,\nabla_{\nu}\phi -
\frac{1}{2}\,g_{\mu\nu}\,\nabla^{\alpha}\phi\,\nabla_{\alpha}\phi\right)
- g_{\mu\nu}V(\phi)
-\left(g_{\mu\nu}\Box\phi-\nabla_{\mu}\nabla_{\nu}\phi\right) + 8\pi\,
T_{\mu\nu}^{(m)}\,,
\end{split}
\end{equation}
where the energy momentum tensor for the matter content is 
\begin{equation}
T_{\mu\nu}^{(m)} = -\frac{2}{\sqrt{-g}}\,\frac{\delta}{\delta
g^{\mu\nu}}\left(\sqrt{-g}\mathcal{L}_{m}\right)\,.
\end{equation}
Taking the trace of \eqref{eq:feq} one obtains
\begin{equation}
\label{eq:trace}
R = \frac{\om}{\phi^{2}}\,\nabla^{\alpha}\phi\,\nabla_{\alpha}\phi +
4\frac{V(\phi)}{\phi} + 3 \frac{\Box\phi}{\phi} - \frac{8\pi}{\phi} T^{m}.
\end{equation}
Variation of the action with respect to $\phi$ gives the dynamical equation for the scalar field
\begin{equation}
\Box\phi = \frac{1}{2\phi}\nabla^{\alpha}\phi\,\nabla_{\alpha}\phi -
\frac{\phi}{2\om}\left(R - 2 V'(\phi)\right)\,,
\end{equation}
and using \eqref{eq:trace} to eliminate the Ricci scalar $R$, one obtains
\begin{equation}
\Box\phi = -\frac{2}{3+2\om}\left(2 V(\phi) - \phi V'(\phi)\right) +
\frac{8\pi}{3+2\om} T^{m}\,.
\end{equation}

In the model under considerations we use the spatially flat Friedmann-Robertson-Walker metric
\begin{equation}
\ud s^{2} = -\ud t^{2} + a^{2}(t)\big(\ud x^{2}+\ud y^{2} + \ud z^{2}\big),
\end{equation}
and the matter content is described by the barotropic matter with equation of state $p_{m}= w_{m }\rho_{m}$, where $p_{m}$ and $\rho_{m}$ are the pressure and the energy density of the matter, and $w_{m}$ is the equation of state parameter.

Then, the energy conservation condition is
\begin{equation}
\label{eq:encon:1}
3 H^{2} = \frac{\om}{2}\frac{\dot{\phi}^{2}}{\phi^{2}} + \frac{V(\phi)}{\phi} -
3 H \frac{\dot{\phi}}{\phi} + \frac{8\pi}{\phi}\rho_{m}\,
\end{equation}
the acceleration equation has the following form
\begin{equation}
\label{eq:acc:1}
\dot{H} =
-\frac{\om}{2}\frac{\dot{\phi}^{2}}{\phi^{2}}-\frac{1}{3+2\om}\frac{2\,V(\phi)-\phi
V'(\phi)}{\phi} + 2 H \frac{\dot{\phi}}{\phi} -
\frac{8\pi}{\phi}\rho_{m}\frac{2+\om(1+w_{m})}{3+2\om}\,,
\end{equation}
and the dynamical equation for the Brans-Dicke scalar reduces to 
\begin{equation}
\ddot{\phi}+3H\dot{\phi}=2\frac{2\,V(\phi)-\phi\,V'(\phi)}{3+2\om} +
8\pi\rho_{m}\frac{1-3w_{m}}{3+2\om}\,.
\end{equation}

In what follows we introduce following energy phase space variables
\begin{equation}
x\equiv \frac{\dot{\phi}}{H \phi}\,,\qquad y\equiv
\sqrt{\frac{V(\phi)}{3\phi}}\frac{1}{H}\,,\qquad
\lambda\equiv-\phi\frac{V'(\phi)}{V(\phi)} \,,
\end{equation}
the energy conservation condition (\ref{eq:encon:1}) can be presented as
\begin{equation}
\Omega_{m}=\frac{8\pi \rho_{m}}{3 \phi H^{2}} = 1+x-\frac{\om}{6}x^{2} - y^{2}\,,
\label{eq:concond}
\end{equation}
and the acceleration equation (\ref{eq:acc:1}) as
\begin{equation}
\label{eq:acc:2}
\begin{split}
	\frac{\dot{H}}{H^{2}} = \hspace{1mm} & 2x-\frac{\om}{2}x^{2}
	-\frac{3}{3+2\om}y^{2}(2+\lambda) \\ &
-
3\left(1+x-\frac{\om}{6}x^{2}-y^{2}\right)\frac{2+\om(1+w_{m})}{3+2\om}\,.
\end{split}
\end{equation}

The dynamical system describing the evolution of the Brans-Dicke theory of
gravity with a scalar field potential and the barotropic matter has the
following form
\begin{align}
\label{eq:dynsys}
\frac{\ud x}{\ud \tau} & =  -3x - x^{2} - x \frac{\dot{H}}{H^{2}} +
\frac{6}{3+2\om}y^{2}(2+\lambda) +
3\left(1 + x -\frac{\om}{6}x^{2} - y^{2}\right)\frac{1-3w_{m}}{3+2\om}\,,
\nonumber \\
\frac{\ud y}{\ud \tau} & =  -y\left(\frac{1}{2}x(1+\lambda)+\frac{\dot{H}}{H^{2}}\right)\,, \\
\frac{\ud \lambda}{\ud \tau} & =  x\lambda\Big(1-\lambda(\Gamma-1)\Big)\,,\nonumber
\end{align}
where $\frac{\ud}{\ud \tau}=\frac{\ud}{\ud \ln{a}}$ and
\begin{equation}
\Gamma=\frac{V''(\phi)V(\phi)}{V'(\phi)^{2}}\,,
\end{equation}
where $()'=\frac{\ud}{\ud\phi}.$

To solve this system of equations we need to assume some form of the $\Gamma=\Gamma(\lambda)$ function. Then the critical points of
the system \eqref{eq:dynsys}, which represent asymptotic states of the system, depend on this $
\Gamma(\lambda)$ function. In our previous paper \cite{Hrycyna:2013hla} we have investigated dynamics of the system \eqref{eq:dynsys} in the vicinity of critical point corresponding to the de Sitter expansion. We were able to find stability conditions of this state whilst the dynamics 
of the Universe, for suitable initial conditions, mimics the standard cosmological model $\Lambda$CDM.

While the system \eqref{eq:dynsys} is the $3$--dimensional dynamical system we can postulate some general form of the $\Gamma(\lambda)$ function \cite{Hrycyna:2010yv} and search for critical points and their stability, one can notice that for arbitrary power-law potential functions of the type
$V(\phi)=V_{0}\,\phi^{n}$ system reduces to a $2$-dimensional one, i.e.~the power-law potentials are invariant submanifolds of the system \eqref{eq:dynsys}.

Simple inspection of the acceleration equation \eqref{eq:acc:2} and the system \eqref{eq:dynsys}
for the power-law potential $\lambda=-n$ confirms that the de Sitter expansion is possible only for 
the quadratic $n=2$ or the linear $n=1$ potential function. On the other hand, one can 
calculate the mass of the Brans-Dicke scalar field in both the Einstein and the Jordan frames \cite{Faraoni:2009km}. In the Einstein frame one has
\begin{equation}
\widetilde{m}^{2} = \frac{32\pi}{(3+2\om)G\,\phi}\Big(\frac{4}{\phi}V(\phi)-3 V'(\phi)+\phi V''(\phi)\Big) = \frac{32\pi}{(3+2\om)G}V_{0}(n-2)^{2}\,\phi^{n-2}\,,
\end{equation}
while in the Jordan frame
\begin{equation}
m^{2} = \frac{2}{3+2\om}\Big(\phi V''(\phi)-V'(\phi)\Big) = \frac{2}{3+2\om}V_{0}\,n(n-2)\,\phi^{n-1}\,.
\end{equation}
For the linear potential function the scalar field $\phi$ has a finite range in the Einstein frame and is tachyonic in the Jordan frame. Only the quadratic potential function with $n=2$ leads to the BD field $\phi$ which has an infinite range in both frames \cite{Faraoni:2009km}.

\section{A quadratic potential function}
\label{sec:quadratic}

In this section we present a detailed dynamical analysis of the system
(\ref{eq:dynsys}) with a quadratic potential function $V(\phi)= V_{0}\,\phi^{2}$.
Within this assumption we have $\lambda=-2$ and dynamics reduces the $2D$
dynamical system. 

Before we proceed to investigation of the system with the barotropic matter
content, let us find solutions for the vacuum case. These solutions play the
crucial role in our system, namely, they divide the phase space into two
distinct regions. The physical trajectories are located in one region while the
non-physical in the other.

In vacuum, the energy conservation condition \eqref{eq:concond} is
\begin{equation}
y^{2} = 1 + x -\frac{\om}{6}x^{2}\,,
\end{equation}
and the acceleration equation \eqref{eq:acc:2} simplifies to
\begin{equation}
\frac{\dot{H}}{H^{2}} = 2 x -\frac{\om}{2}x^{2}\,.
\end{equation}
Then the dynamics of the vacuum model with a quadratic potential function is
completely determined by the single equation
\begin{equation}
\frac{\ud x}{\ud \ln{a}} = -3 x\left(1+x-\frac{\om}{6}x^{2}\right)\,,
\end{equation}
which can easily be solved in the following parametric form
\begin{equation}
\begin{split}
\bigg(\frac{a}{a^{(i)}}\bigg)^{-3} & = \frac{x}{x^{(i)}} \sqrt{\frac{1+x^{(i)}-\frac{\om}{6}(x^{(i)})^{2}}{1+x-\frac{\om}{6}x^{2}}}\times\\ &\times\left(\frac{1+x^{(i)}-\frac{\om}{6}(x^{(i)})^{2}}{1+x-\frac{\om}{6}x^{2}}
\left(\frac{\sqrt{3(3+2\om)}-(\om x-3)}{\sqrt{3(3+2\om)}-(\om x^{(i)}-3)}\right)^{2}\right)^{\frac{1}{2}\sqrt{\frac{3}{3+2\om}}}\,,
\end{split}
\end{equation}
where $a^{(i)}$ and $x^{(i)}$ are the initial condition for the scale factor and the variable 
$x$.

In the case of so-called ``O'Hanlon theory'' with $\om=0$ this equation simplifies to
\begin{equation}
\bigg(\frac{a}{a^{(i)}}\bigg)^{-3} = \frac{x(1+x^{(i)})}{x^{(i)}(1+x)}\,.
\end{equation}
Now, using this solution one can find an exact form of the Hubble function
\begin{equation}
\left(\frac{H(a)}{H(a^{(i)})}\right)^{2} = \left(1+x^{(i)}-x^{(i)}\bigg(\frac{a}{a^{(i)}}\bigg)^{-3}\right)^{\frac{4}{3}}\,.
\end{equation}
Taking the initial conditions at present epoch close to de Sitter state, $x^{(i)}=x_{0} \ll1$, one can expand this function to
\begin{equation}
\left(\frac{H(a)}{H(a_{0})}\right)^{2} \approx 1 + \frac{4}{3}x_{0} - \frac{4}{3}x_{0}\bigg(\frac{a}{a_{0}}\bigg)^{-3}\,,
\end{equation}
which indicates that close to de Sitter state and for $x_{0}<0$ expansion of universe mimics expansion of the standard cosmological model --- the $\Lambda$CDM model. We can proceed now to investigation of the non-vacuum model. 

For the model with barotropic matter content the acceleration equation is
\begin{equation}
\label{eq:h2hquad}
\frac{\dot{H}}{H^{2}} = 2x-\frac{\om}{2}x^{2} - 3\left(1+x-\frac{\om}{6}x^{2}
- y^{2}\right)\frac{2+\om(1+w_{m})}{3+2\om},
\end{equation}
and the effective equation of state parameter is
\begin{equation}
\label{eq:weff}
w_{\textrm{eff}} = -1 - \frac{2}{3} \frac{\dot{H}}{H^{2}}\,.
\end{equation}

The dynamical system describing dynamics of the model with a quadratic potential function is in
the following form
\begin{eqnarray}
\label{eq:sysquad}
x' & = & -3x\left(
1+x-\frac{\om}{6}x^{2}-\Bigg(1+x-\frac{\om}{6}x^{2}-y^{2}\Bigg)
\frac{2+\om(1+w_{m})}{3+2\om}\right) \nonumber \\
 & & \hspace{1mm}+
3\Bigg(1+x-\frac{\om}{6}x^{2}-y^{2}\Bigg)\frac{1-3w_{m}}{3+2\om},\\
y' & = & 3y\left(-\frac{1}{2}x+\frac{\om}{6}x^{2} +
\Bigg(1+x-\frac{\om}{6}x^{2}-y^{2}\Bigg)
\frac{2+\om(1+w_{m})}{3+2\om}\right). \nonumber
\end{eqnarray}
The right hand sides of this system are polynomial in the phase space variables and
we are able to use standard dynamical system methods to investigate the phase space structure 
as well as linearized solutions in the vicinity of
the stationary states \cite{Perko:2001,Wiggins:2003}. In our analysis we treat
the model parameters, i.e.~the Brans-Dicke parameter $\om$ and the equation of
state parameter $w_{m}$ of barotropic matter, as bifurcation parameters and
search for values for which the dynamics changes dramatically.

In table \ref{tab:1} we have gathered the critical points of the system
(\ref{eq:sysquad}) with corresponding eigenvalues of the linearization matrix.

\begin{table}
\begin{center}
\begin{tabular}{|>{\scriptsize}c|>{\scriptsize}c|>{\scriptsize}c|}
\hline
& location & eigenvalues \\
\hline
A & $x^{*}_{1}=\frac{3\pm\sqrt{3(3+2\om)}}{\om}$, $y^{*}_{1}=0$ & $3(1-w_{m})+x^{*}_{1}$,
$3+\frac{3}{2}x^{*}_{1}$\\
B & $x^{*}_{2}=\frac{1-3w_{m}}{1+\om(1-w_{m})}$, $y^{*}_{2}=0$ &
$-\frac{3}{2}(1-w_{m})-\frac{1}{2}x^{*}_{2}$,
$\frac{3}{2}(1+w_{m})+x^{*}_{2}$\\
C & $x^{*}_{3}=0$ , $y^{*}_{3}=\pm1$ & $-3$, $-3(1+w_{m})$\\
D & $x^{*}_{4}=-\frac{3}{2}(1+w_{m})$,
$y^{*}_{4}=\pm\sqrt{\frac{1}{8}(5-3w_{m}+3\om(1-w_{m}^{2}))}$ &
$-\frac{3}{8}(1-3w_{m})\pm\frac{1}{2}\sqrt{\Delta}$\\
& & $\Delta = \frac{9}{16}(1-3w_{m})^{2}+36(y^{*}_{4})^{2}(1+w_{m})$\\
\hline
\end{tabular}
\end{center}
\caption{The critical points of the system under considerations (\ref{eq:sysquad}) with a 
quadratic potential function.}
\label{tab:1}
\end{table}

The critical points: $x^{*}_{1}=\frac{3\pm\sqrt{3(3+2\om)}}{\om}$, $y^{*}_{1}=0$
linearized solutions are
\begin{subequations}
\label{eq:lin_1}
\begin{eqnarray}
\label{eq:lin_1a}
x_{1}(a) & = & x^{*}_{1} + \Delta x
\left(\frac{a}{a^{(i)}}\right)^{\lambda_{1}}\,,\\
y_{1}(a) & = & \Delta y \left(\frac{a}{a^{(i)}}\right)^{\lambda_{2}}\,,
\end{eqnarray}
\end{subequations}
where the eigenvalues of the linearization matrix are 
\begin{equation}
\lambda_{1}= 3(1-w_{m}) + x^{*}_{1}\,, \qquad \lambda_{2}= 3+
\frac{3}{2}x^{*}_{1}\,,
\end{equation}
and $\Delta x = x^{(i)}_{1}-x^{*}_{1}$, $\Delta y = y^{(i)}_{1}-y^{*}_{1}$ are
the initial conditions, and $a^{(i)}$ is the initial value of the scale factor.
Using relations (\ref{eq:h2hquad}) and (\ref{eq:weff}) one can calculate the
effective equation of state coefficient at this critical point
\begin{equation}
w_{\textrm{eff}}\big|^{*}_{1} = 1 + \frac{2}{3}x^{*}_{1}.
\end{equation}
Now using (\ref{eq:lin_1}) we are able to calculate corresponding formula for
the Hubble function. The equation (\ref{eq:h2hquad}) up to linear terms in
initial conditions reduces to
\begin{equation}
\frac{\ud \ln{H^{2}}}{\ud \tau} \approx -2(3+x^{*}_{1}) +
\frac{2\om}{3+2\om}\big(1-3w_{m}-(1+\om(1-w_{m}))x^{*}_{1}\big)\Delta x
\exp{\lambda_{1}\tau},
\end{equation}
and after integration we obtain
\begin{equation}
\ln{\left(\frac{H(\tau)}{H(0)}\right)^{2}} \approx -2(3+x^{*}_{1})\tau +
\frac{2\om}{3-\om x^{*}_{1}}\Delta x (\exp{\lambda_{1}\tau} - 1)\,.
\end{equation}
Up to linear terms in $\Delta x$ this equation reduces to
\begin{equation}
\left(\frac{H(a)}{H(a^{(i)}_{1})}\right)^{2} \approx
\left(1-\frac{2\om}{3-\om x^{*}_{1}}\Delta x\right)\left(\frac{a}{a^{(i)}_{1}}\right)^{-6-2x^{*}_{1}} +
\frac{2\om}{3-\om x^{*}_{1}}\Delta x\left(\frac{a}{a^{(i)}_{1}}\right)^{-3(1+w_{m})-x^{*}_{1}}\,,
\end{equation}
where we reintroduced scale scale factor $\tau=\ln{\Big(\frac{a}{a^{(i)}_{1}}\Big)}$,
$a^{(i)}_{1}$ is the initial value of the scale factor in the vicinity of the
critical point. Now using the energy conservation condition \eqref{eq:concond} and the linearized solutions \eqref{eq:lin_1} we obtain the linearized solution for the barotropic matter density parameter
\begin{equation}
\Omega_{m} \approx (1-\frac{\om}{3}x^{*}_{1})\Delta x \left(\frac{a}{a^{(i)}_{1}}\right)^{\lambda_{1}} = \Omega_{m,i}\left(\frac{a}{a^{(i)}_{1}}\right)^{\lambda_{1}}\,.
\end{equation}
Finally we obtain
\begin{equation}
\left(\frac{H(a)}{H(a^{(i)}_{1})}\right)^{2} \approx
\left(1-\Omega_{M,i}\right)\left(\frac{a}{a^{(i)}_{1}}\right)^{-6-2x^{*}_{1}} +
\Omega_{M,i}\left(\frac{a}{a^{(i)}_{1}}\right)^{-3(1+w_{m})-x^{*}_{1}}\,,
\label{eq:H2_1}
\end{equation}
where
\begin{equation}
\Omega_{M,i} = \frac{2\om}{3-\om x^{*}_{1}}\Delta x = \frac{2\om}{3+2\om}\Omega_{m,i}\,.
\end{equation}
The Hubble function \eqref{eq:H2_1} resembles Hubble's formula for two component universe filled with stiff matter and barotropic matter, with small distinction due to the Brans-Dicke parameter $\om$ encoded in $x^{*}_{1}$.

Note that in the vicinity of this critical point the density parameter of a barotropic matter 
$\Omega_{m}$ behaves differently than in the general relativity and scales with the scale factor 
as $\propto a^{3(1-w_{m}) + x^{*}_{1}}$ instead of $\propto a^{-3(1+w_{m})}$ which indicates departure form standard general relativity. Additionally, density parameter of barotropic matter observed in the Hubble function can be larger or smaller than the matter content included in the model by hand depending on the value and sign of the Brans-Dicke parameter $\om$.

The critical point $x^{*}_{2}=\frac{1-3w_{m}}{1+\om(1-w_{m})}$, $y^{*}_{2}=0$ with
the effective equation of state parameter 
\begin{equation}
w_{\text{eff}}\big|^{*}_{2} =
w_{m}+\frac{1}{3}\frac{1-3w_{m}}{1+\om(1-w_{m})}\,.
\end{equation}
Using the linearized solutions in the vicinity of this critical point
\begin{subequations}
\begin{align}
x_{2}(a) & = x^{*}_{2} + \Delta x
\left(\frac{a}{a^{(i)}}\right)^{\lambda_{1}}\,,\\
y_{2}(a) & = \Delta y \left(\frac{a}{a^{(i)}}\right)^{\lambda_{2}}\,,
\end{align}
\end{subequations}
where the eigenvalues of the linearization matrix are
\begin{equation}
\lambda_{1} = -\frac{3}{2}(1-w_{m})-\frac{1}{2}x^{*}_{2}\,, \qquad \lambda_{2} =
\frac{3}{2}(1+w_{m}) + x^{*}_{2}
\end{equation}
and $\Delta x = x^{(i)}_{2}-x^{*}_{2}$, $\Delta y = y^{(i)}_{2}-y^{*}_{2}$ are
the initial conditions, and $a^{(i)}_{2}$ is the initial value of the scale
factor near the critical point, one can easy obtain corresponding formula for
the Hubble function. The acceleration equation \eqref{eq:h2hquad} up to linear terms in initial conditions is
\begin{equation}
\frac{\ud\ln{H^{2}}}{\ud \tau} \approx -3(1+w_{m}) - x^{*}_{2}\,,
\end{equation}
and after integration we obtain the following formula
\begin{equation}
\left(\frac{H(a)}{H(a^{(i)}_{2})}\right)^{2} \approx
\left(\frac{a}{a^{(i)}_{2}}\right)^{-3(1+w_{m})}\left(\frac{a}{a^{(i)}_{2}}\right)^{-x^{*}_{2}}
\,,
\end{equation}
where $x^{*}_{2}=\frac{1-3w_{m}}{1+\om(1-w_{m})}$ is the coordinate of the
critical point.

This Hubble's function describes evolution of the barotropic matter dominated
universe with contribution from Brans-Dicke parameter included in the exponent
$x^{*}_{2}$. For universe filled with radiation $w_{m}=\frac{1}{3}$ we obtain
pure radiation dominated evolution and the contribution form Brans-Dicke
parameter vanishes, while for different matter content effective evolution of
universe in the vicinity of this critical point can be quite different. In
the case of so-called ``O'Hanlon gravity'' with $\om=0$ and arbitrary matter
content universe effectively behaves also like the radiation dominated
universe. 

The critical points: $x^{*}_{3}=0$, $y^{*}_{3}=\pm1$ with the effective equation of state parameter 
\begin{equation}
w_{\text{eff}}\big|^{*}_{3} = -1\,,
\end{equation}
describe the de Sitter expansion and contraction, respectively.

The linearized solutions in the vicinity of these points are
\begin{subequations}
\begin{align}
\label{eq:dS_a}
x_{3}(a) & = \frac{1}{w_{m}}\,\frac{1+2\om w_{m}}{3+2\om}\bigg(\Delta x -2
y^{*}_{3}\frac{1-3w_{m}}{1+2\om
w_{m}}\Delta y\bigg)\bigg(\frac{a}{a^{(i)}_{3}}\bigg)^{\lambda_{1}}
- \nonumber \\
& -
\frac{1}{w_{m}}\,\frac{1-3w_{m}}{3+2\om}\bigg(\Delta x-2y^{*}_{3}\Delta y\bigg)
\bigg(\frac{a}{a^{(i)}_{3}}\bigg)^{\lambda_{2}}, \\
y_{3}(a) & = y^{*}_{3} + \frac{1}{2 y^{*}_{3}w_{m}}\frac{1+2\om w_{m}}{3+2\om}\Bigg\{
\bigg(\Delta x- 2 y^{*}_{3}\frac{1-3w_{m}}{1+2\om
w_{m}}\Delta y\bigg)
\bigg(\frac{a}{a^{(i)}_{3}}\bigg)^{\lambda_{1}} - \nonumber \\
& - \bigg(\Delta x-2 y^{*}_{3}\Delta y\bigg)
\bigg(\frac{a}{a^{(i)}_{3}}\bigg)^{\lambda_{2}}\Bigg\}\,,
\end{align}
\end{subequations}
where $\lambda_{1}=-3$ and $\lambda_{2}=-3(1+w_{m})$ are the eigenvalues of the
linearization matrix and $\Delta x = x^{(i)}_{3}-x^{*}_{3}$, $\Delta y =
y^{(i)}_{3}-y^{*}_{3}$ are the initial conditions. The critical point under considerations is stable during expansion of universe for all matter beyond phantom matter. For matter in form of cosmological constant $w_{m}=-1$ one of the eigenvalues vanishes leading to structurally unstable system.

Again, using these linearized solutions we are able to obtain the Hubble function in the vicinity of the critical point corresponding to de Sitter expansion or contraction in the following form
\begin{equation}
\label{eq:desitlcdm}
\left(\frac{H(a)}{H(a_{0})}\right)^{2} \approx 1 - \Omega_{DM,0} - \Omega_{M,0}
+ \Omega_{DM,0}\left(\frac{a}{a_{0}}\right)^{-3} +
\Omega_{M,0}\left(\frac{a}{a_{0}}\right)^{-3(1+w_{m})}\,,
\end{equation}
where
\begin{align}
\Omega_{DM,0} = & -\frac{4}{3}\bigg(\Delta x + \frac{1}{3+2\om}\frac{1-3w_{m}}{w_{m}}\,
\Omega_{m,i}\bigg)\bigg(\frac{a_{0}}{a^{(i)}_{3}}\bigg)^{-3}\,,\nonumber \\
\Omega_{M,0} = & \bigg(1+\frac{1}{3+2\om}\frac{(1-3w_{m})(4+3w_{m})}{3w_{m}(1+w_{m})}\bigg)\,\Omega_{m,0}\,,\nonumber
\end{align}
and in the linear approximation we can express the energy conservation condition \eqref{eq:concond} as
\begin{equation}
\Omega_{m,0} 
\approx (\Delta x - 2 y^{*}_{3}\Delta y)\left(\frac{a_{0}}{a^{(i)}}\right)^{-3(1+w_{m})} = \Omega_{m,i}\left(\frac{a_{0}}{a^{(i)}}\right)^{-3(1+w_{m})}\,.
\end{equation}

The Hubble function \eqref{eq:desitlcdm} in the vicinity of de Sitter state resembles the Hubble function for the $\Lambda$CDM model with the $\Omega_{DM,0}$ parameter playing the role of the dark matter density parameter. Note that inclusion of the dust matter $w_{m}=0$ leads to degenerated critical point (both eigenvalues assume the same value) and then terms proportional to the natural logarithm of the scale factor appear in the Hubble function.

The critical points: $x^{*}_{4}=-\frac{3}{2}(1+w_{m})$, $y^{*}_{4}=\pm\sqrt{\frac{1}{8}(5-3w_{m}+ 3\om(1-w_{m}^{2}))}$  with the effective equation of state parameter
\begin{equation}
w_{\text{eff}}\big|^{*}_{4} = -\frac{1}{2}(1-w_{m})\,,
\end{equation}
and the energy conservation condition \eqref{eq:concond} calculated at these points gives
\begin{equation}
\Omega_{m}\big|^{*}_{4} = -\frac{3}{8}(3+2\om)(1+w_{m})\,.
\end{equation}
The condition for the critical points to describe physical state is $\Omega_{m}\big|^{*}_{4}>0$.

We begin with solutions at the critical points. Using effective equation of state parameter we 
have
\begin{equation}
\left(\frac{H(a)}{H(a^{*}_{4})}\right)^{2} = \left(\frac{a}{a^{*}_{4}}\right)^{-\frac{3}{2}(1+w_{m})}\,,
\end{equation}
where $a^{*}_{4}$ is the initial value of the scale factor at the critical point. From the definition of the variable $x$ we obtain
\begin{equation}
\frac{\ud\ln{\phi}}{\ud\ln{a}} = x^{*}_{4}\,,
\end{equation}
and
\begin{equation}
\frac{\phi(a)}{\phi(a^{*}_{4})} = \left(\frac{a}{a^{*}_{4}}\right)^{-\frac{3}{2}(1+w_{m})}\,.
\end{equation}

We can also find the variability of the field $\phi$ with respect to the cosmological time $t$. The resulting differential equation is
\begin{equation}
\frac{\dot{\phi}}{\phi\sqrt{\phi}} = \sqrt{\frac{V_{0}}{3}}\frac{x^{*}_{4}}{y^{*}_{4}}\,,
\end{equation}
and the solution is 
\begin{equation}
\phi(t) = \frac{\phi^{*}_{4}}{\big(1-\frac{1}{2}\sqrt{\frac{V_{0}}{3}\phi^{*}_{4}}\frac{x^{*}_{4}}{y^{*}_{4}}\,(t-t_{0})\big)^{2}} = \frac{\phi^{*}_{4}}{\big(1+\frac{3}{4}(1+w_{m}) H(a^{*}_{4})\,(t-t_{0})\big)^{2}}\,.
\end{equation}
The solution for the scale factor can be presented in the following form
\begin{equation}
a(t) = a^{*}_{4}\Big(1+\frac{3}{4}(1+w_{m})H(a^{*}_{4})\,(t-t_{0})\Big)^{\frac{4}{3(1+w_{m})}}
\end{equation}
where $\phi^{*}_{4}$ is value of the field $\phi$ and $a^{*}_{4}$ is the value of the scale factor at time $t=t_{0}$.

Note that for the matter content in the form of cosmological constant $w_{m}=-1$ these solutions represent pure de Sitter state with $a(t)=a^{*}_{4}\,e^{H(a^{*}_{4})(t-t_{0})}$ and $\phi(t)=\phi^{*}_{4}=\text{const}$.

The linearized solutions in the vicinity of this critical point can be presented in the 
following form
\begin{subequations}
\begin{eqnarray}
\label{eq:lin_4a}
x_{4}(a) & = & x^{*}_{4} + \frac{1}{\det{P}}\Bigg(A_{1}(\Delta x- A_{2}\Delta y)\bigg(\frac{a}{a^{(i)}_{4}}\bigg)^{\lambda_{1}} 
- A_{2} (\Delta x - A_{1}\Delta y)\bigg(\frac{a}{a^{(i)}_{4}}
\bigg)^{\lambda_{2}}\Bigg)\,,\\
y_{4}(a) & = & y^{*}_{4} + \frac{1}{\det{P}}\Bigg((\Delta x- A_{2}\Delta y)\bigg(\frac{a}{a^{(i)}_{4}}\bigg)^{\lambda_{1}} 
-  (\Delta x - A_{1}\Delta y)\bigg(\frac{a}{a^{(i)}_{4}}
\bigg)^{\lambda_{2}}\Bigg)\,,
\end{eqnarray}
\end{subequations}
where the eigenvalues of the linearization matrix are
\begin{equation}
\begin{split}
\lambda_{1} = & -\frac{1}{2}\left(\frac{3}{4}(1-3w_{m}) + \sqrt{\frac{9}{16}(1-3w_{m})^{2} + 36 (y^{*}_{4})^{2}(1+w_{m})}\right)\, \\
\lambda_{2} = &-\frac{1}{2}\left(\frac{3}{4}(1-3w_{m}) - \sqrt{\frac{9}{16}(1-3w_{m})^{2} + 36 (y^{*}_{4})^{2}(1+w_{m})}\right)\,,
\end{split}
\end{equation}
and the matrix $P$ is given by
\begin{equation}
P=
\left(\begin{array}{cc}
A_{1} & A_{2} \\
1 & 1 
\end{array}\right)\,,
\end{equation}
and 
\begin{equation}
\begin{split}
A_{1}=&\frac{2}{3}\frac{6(y^{*}_{4})^{2}(2+\om(1+w_{m})) + \lambda_{1}(3+2\om)}{y^{*}_{4}(1-\om(1-w_{m})(1+\om(1+w_{m})))}\,,\\
A_{2}=&\frac{2}{3}\frac{6(y^{*}_{4})^{2}(2+\om(1+w_{m})) + \lambda_{2}(3+2\om)}{y^{*}_{4}(1-\om(1-w_{m})(1+\om(1+w_{m})))}\,.
\end{split}
\end{equation}
Using the linearized solutions and the acceleration equation \eqref{eq:h2hquad} we obtain the following form of the Hubble function
\begin{equation}
\label{eq:hubble_4}
\left(\frac{H(a)}{H(a^{(i)}_{4})}\right)^{2} \approx 
\left(\frac{a}{a^{(i)}_{4}}\right)^{-\frac{3}{2}(1+w_{m})}\Bigg(1-\Omega_{1,i} - \Omega_{2,i} + \Omega_{1,i}\left(\frac{a}{a^{(i)}_{4}}\right)^{\lambda_{1}} + 
\Omega_{2,i}\left(\frac{a}{a^{(i)}_{4}}\right)^{\lambda_{2}}\Bigg)
\end{equation}
where
\begin{equation}
\begin{split}
\Omega_{1,i} =& 2(y^{*}_{4})^{2}\frac{6+\om\big(3(1+w_{m})+4\lambda_{1}\big)}{(3+2\om)(\lambda_{1}-\lambda_{2})\lambda_{1}}\big(\Delta x - A_{2}\Delta y\big)\,,\\
\Omega_{2,i} =&-2(y^{*}_{4})^{2}\frac{6+\om\big(3(1+w_{m})+4\lambda_{2}\big)}{(3+2\om)(\lambda_{1}-\lambda_{2})\lambda_{2}}\big(\Delta x - A_{1}\Delta y\big)\,.
\end{split}
\end{equation}

One can easily check that the conditions for the critical point under considerations be stable (a stable node or a stable focus) give that the matter content in the model must be in the form of the phantom matter $w_{m}<-1$. Then we obtain that as universe expands the Hubble function tends to infinity.

\subsection{Phase space diagrams}
\label{subsec:phase_diagrams}

We now proceed to the detailed discussion of the phase space diagrams presented in this subsection.

Dynamics of the model described by system \eqref{eq:sysquad} depends on the two parameters: 
the equation of state parameter $w_{m}$ and the parameter of the theory $\om$. The character of the critical 
points of system \eqref{eq:sysquad} crucially depends on the values of these parameters. We treat both those 
parameters as bifurcation parameters. In order to obtain 
maximal information about dynamics for various values of the model parameter we divided space of 
the parameters $(w_{m},\om)$ into a few regions which enables us to present the specific dynamics 
with respect to values of this parameters.

\begin{figure}
\begin{center}
\epsfig{file=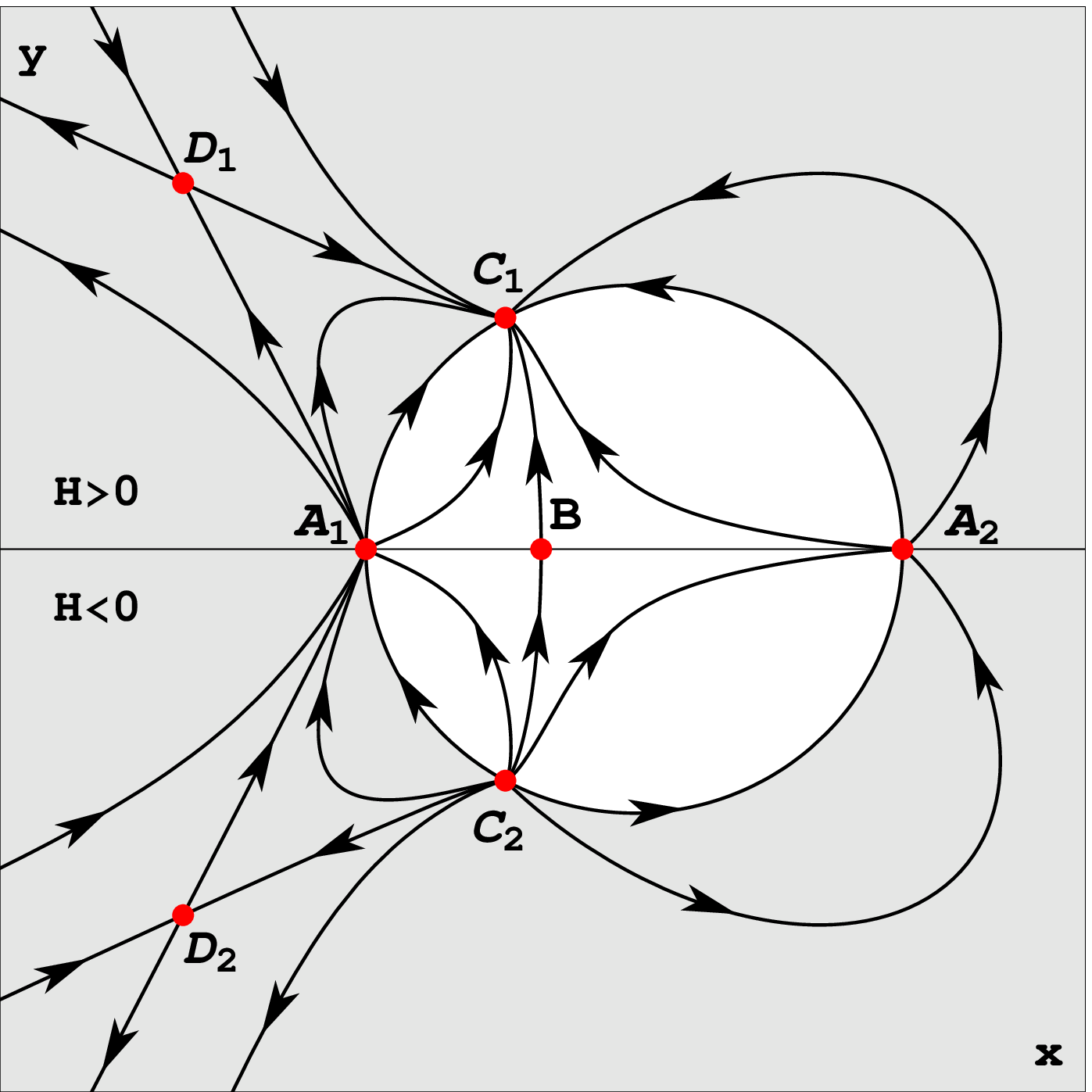,scale=0.5}
\epsfig{file=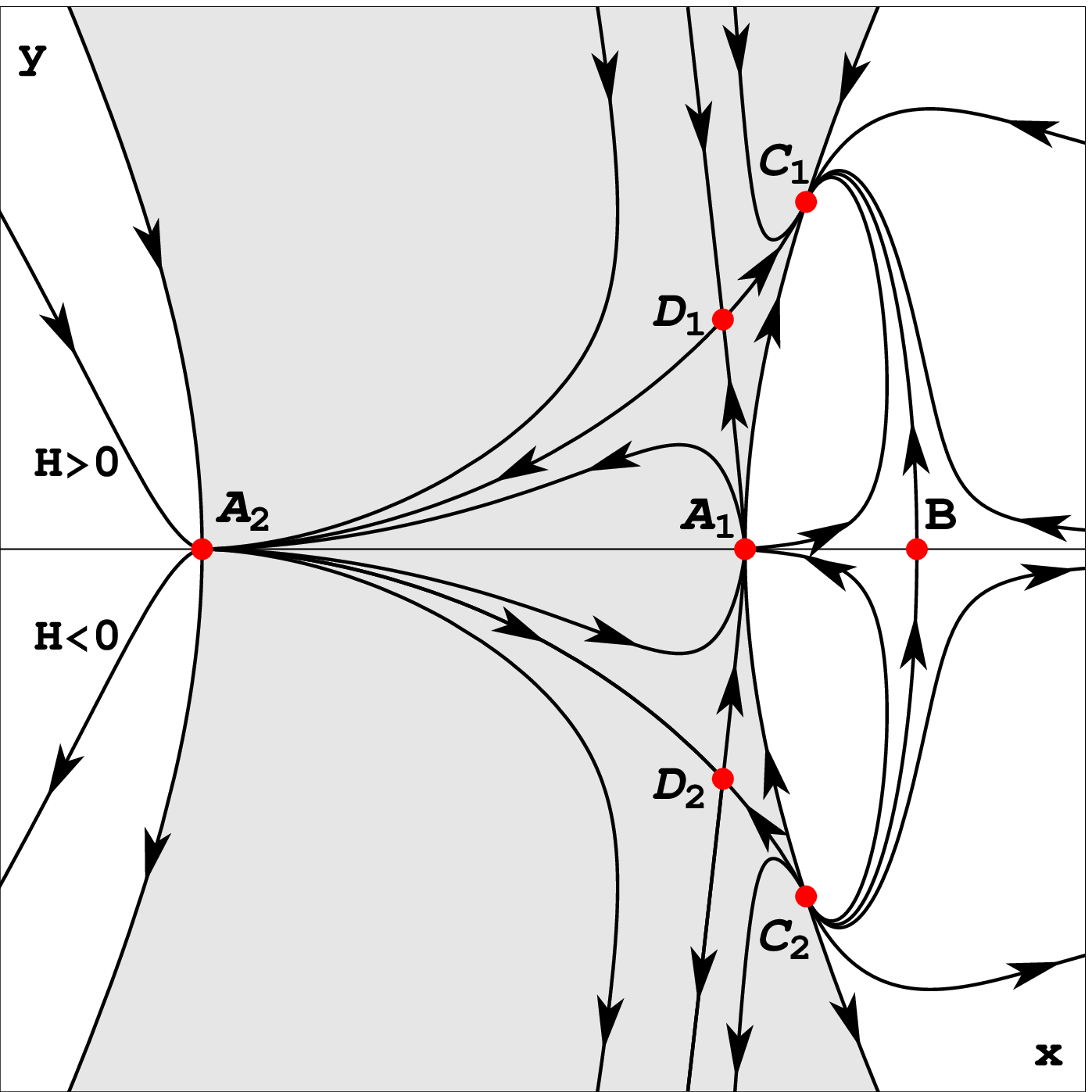,scale=0.5}
\epsfig{file=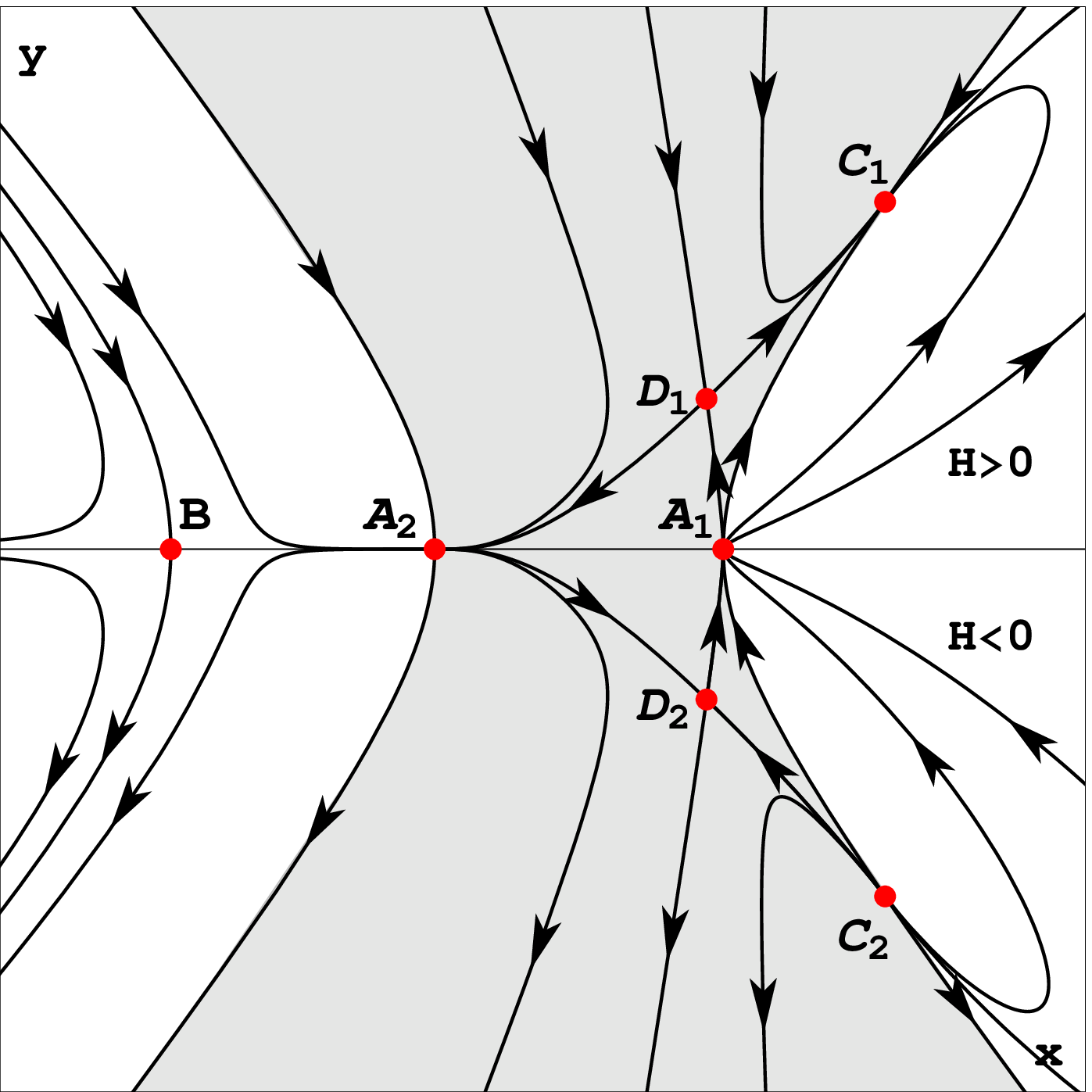,scale=0.5}
\epsfig{file=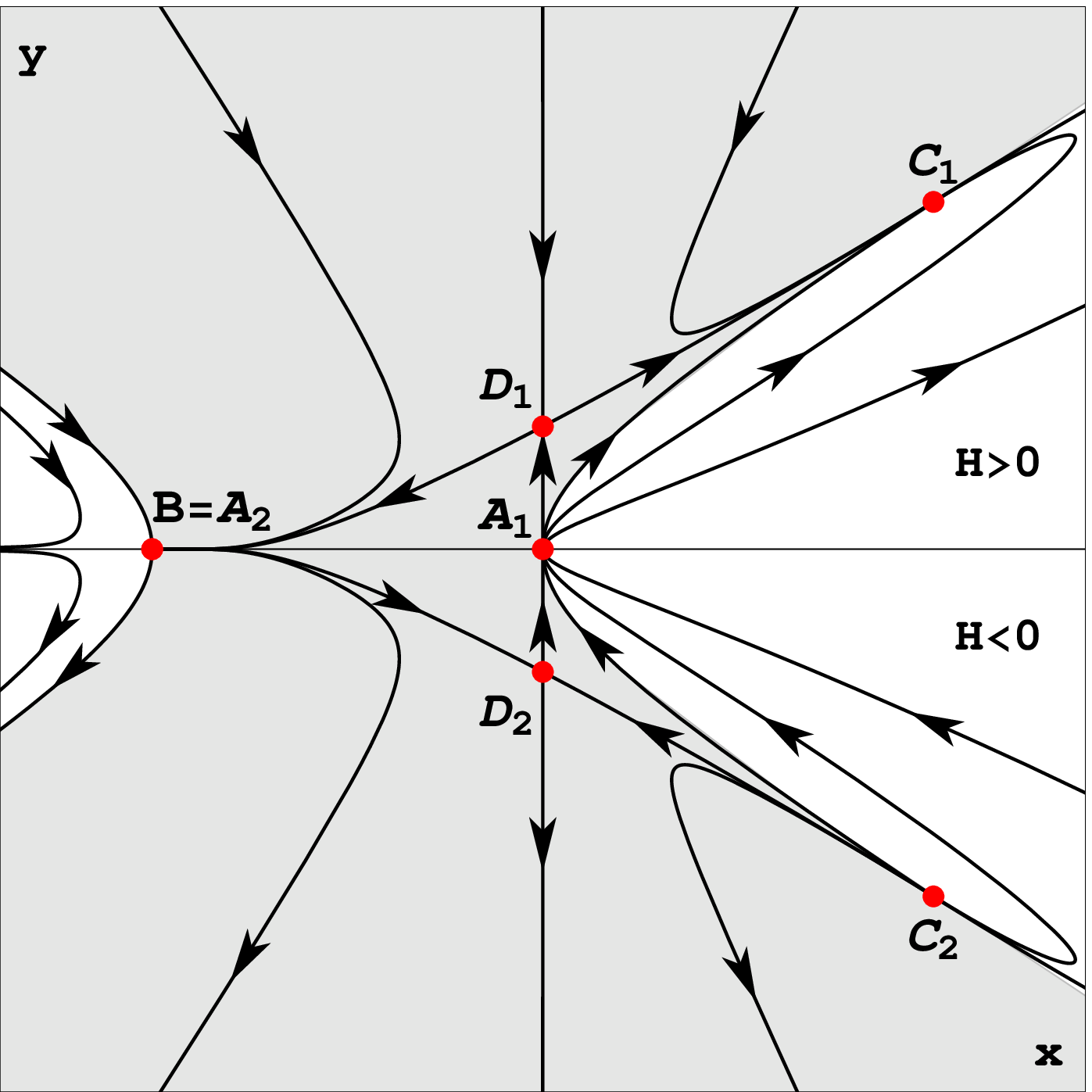,scale=0.5}
\caption{The phase plane diagrams for system \eqref{eq:sysquad} filled with the dust matter 
$w_{m}=0$ and $\om>0$ (top left), $-1<\om<0$ (top right), $-4/3<\om<-1$ (bottom left), 
$\om=-4/3$ (bottom right).}
\label{fig:1}
\end{center}
\end{figure}

\begin{figure}
\begin{center}
\epsfig{file=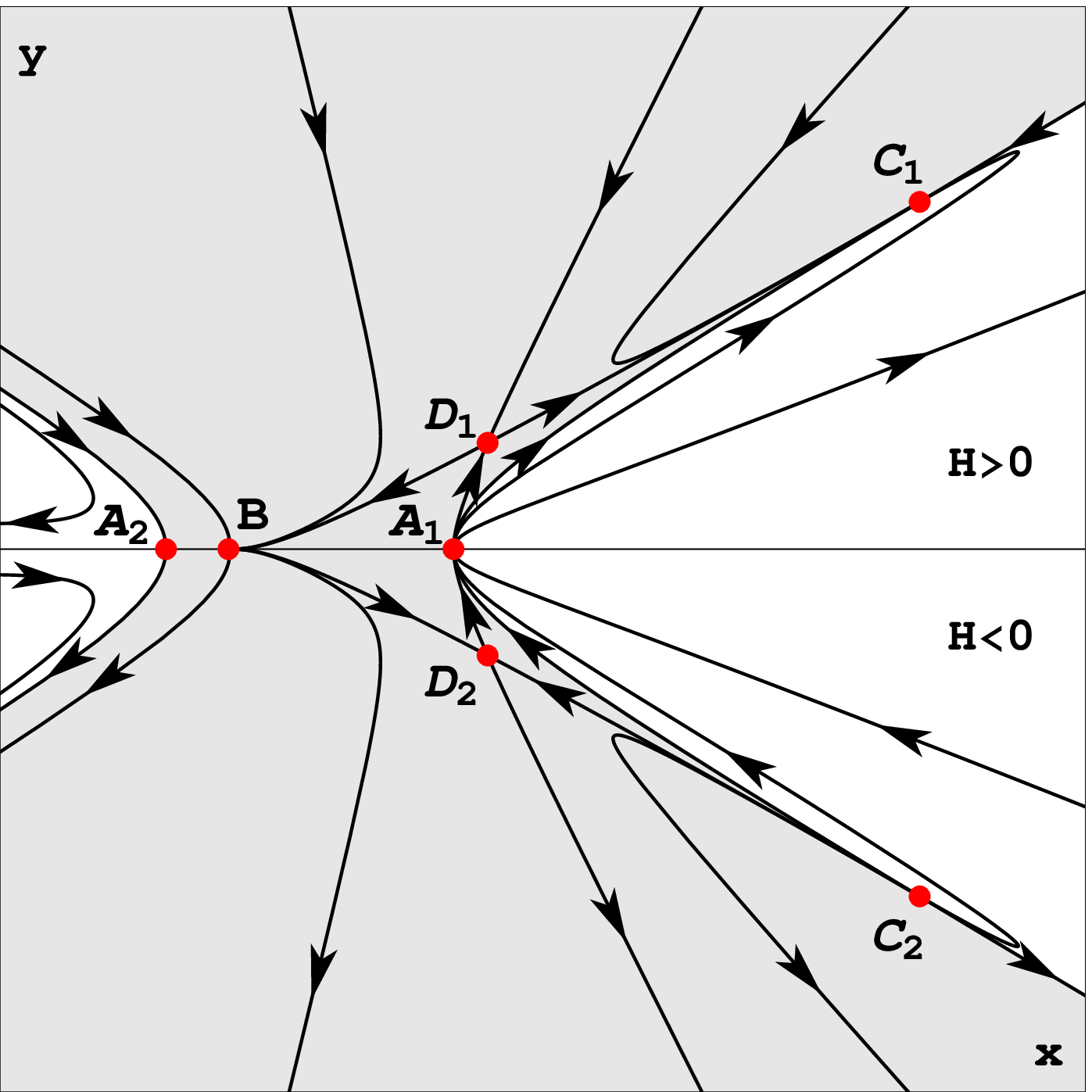,scale=0.5}
\epsfig{file=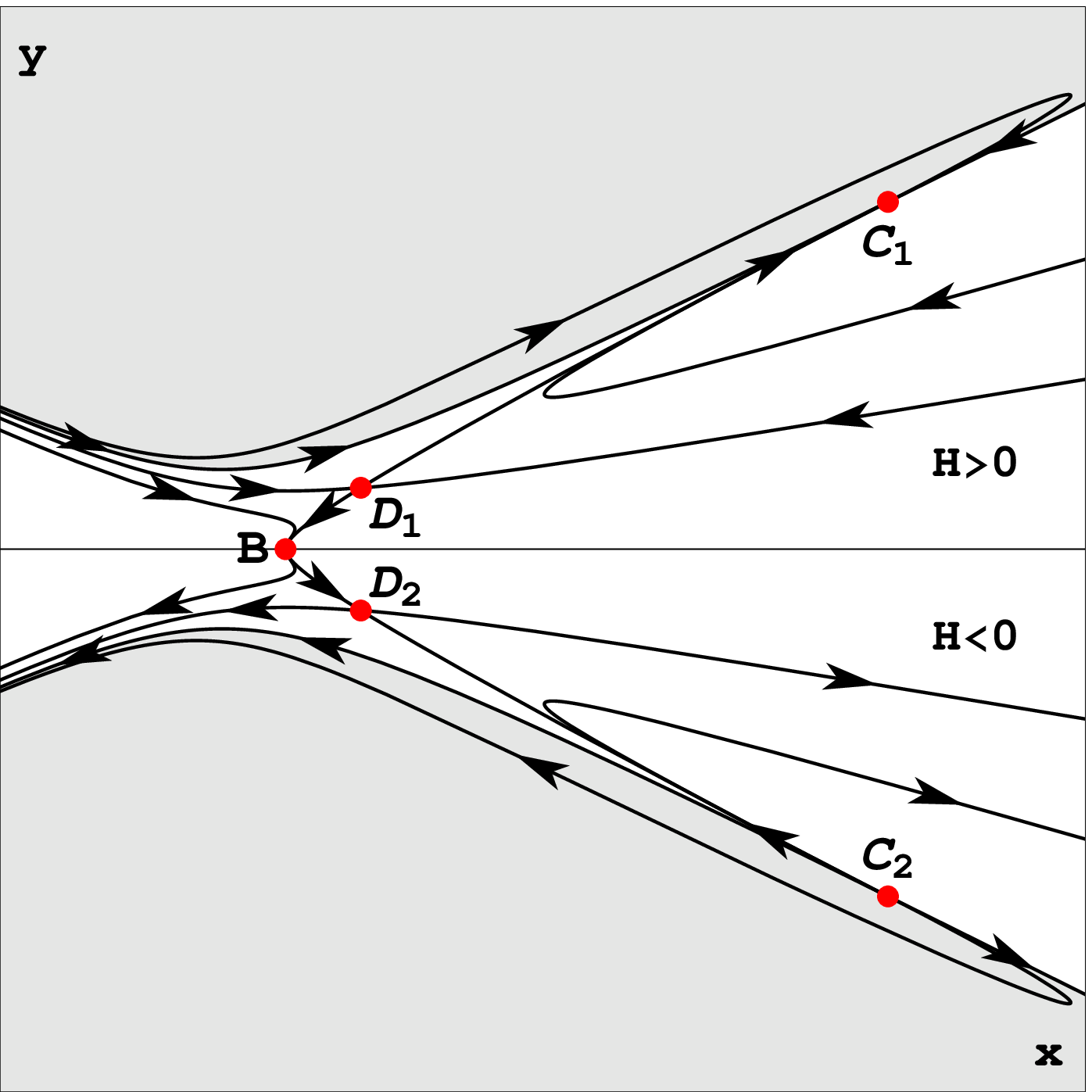,scale=0.5}
\epsfig{file=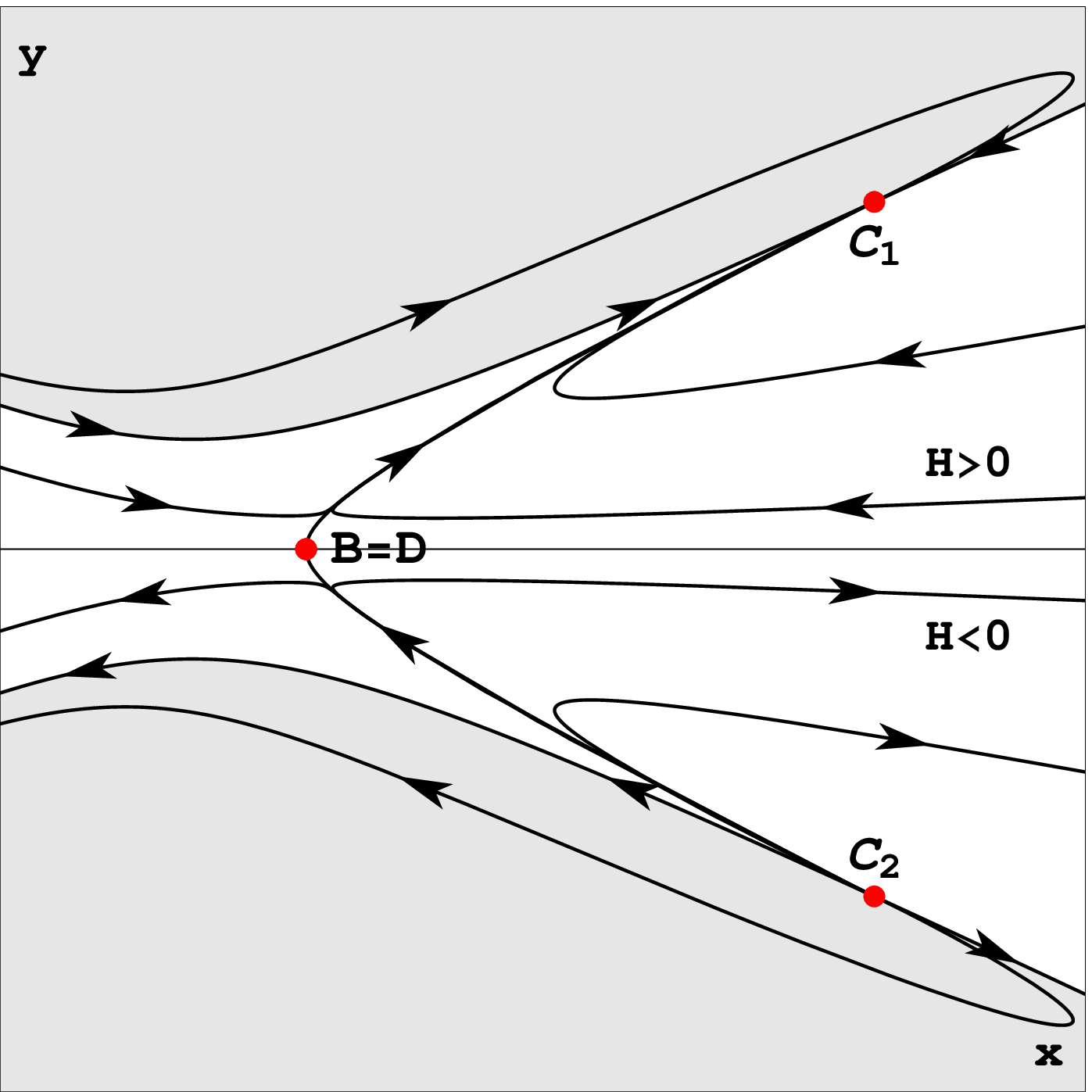,scale=0.5}
\epsfig{file=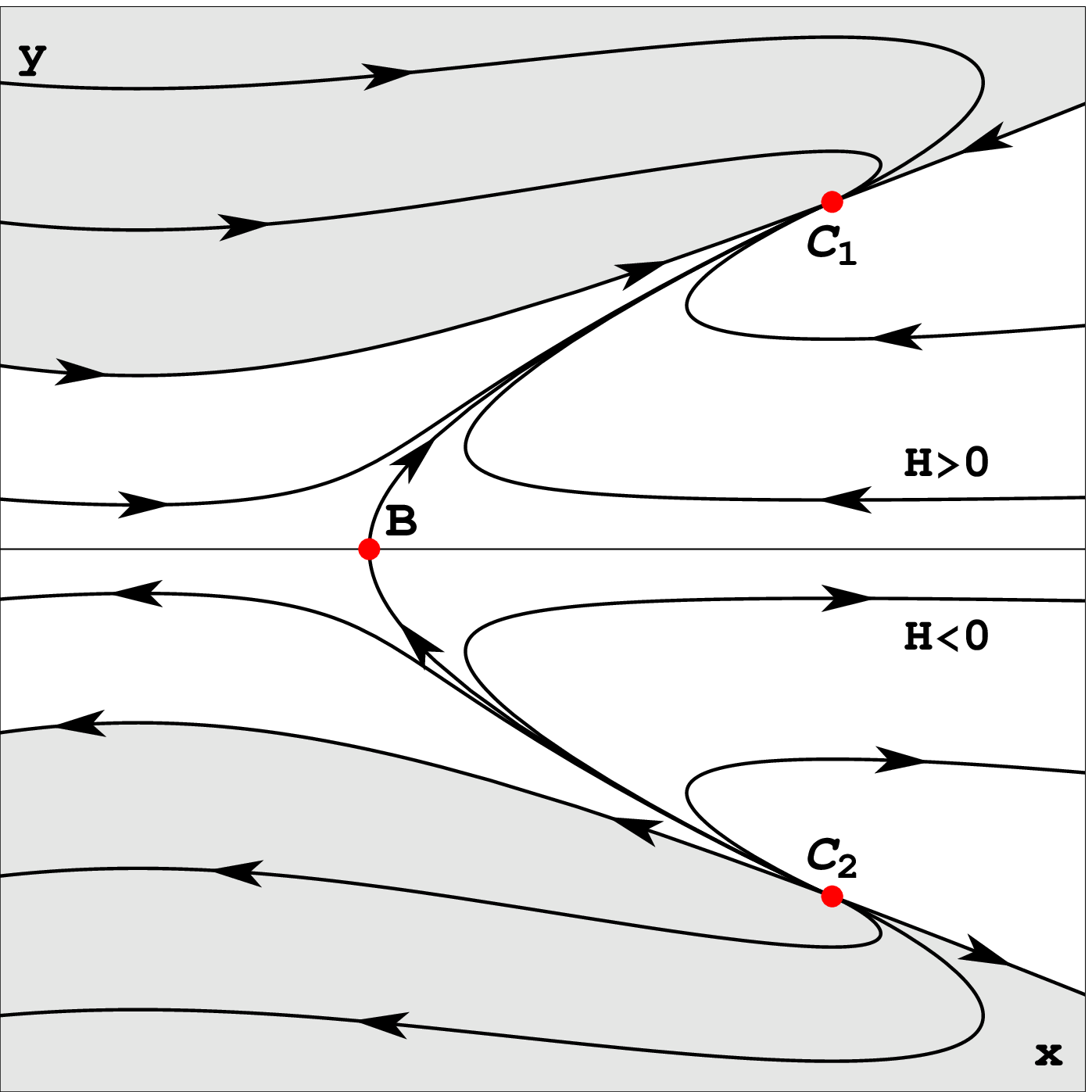,scale=0.5}
\caption{The phase space diagrams for system \eqref{eq:sysquad} filled with the dust matter
$w_{m}=0$ and for the values of the Brans-Dicke parameter: $-3/2<\om<-4/3$ (top left), 
$-5/3<\om<-3/2$ (top right), $\om=-5/3$ (bottom left), $\om<-5/3$ (bottom right).}
\label{fig:2}
\end{center}
\end{figure}

\begin{figure}
\begin{center}
\epsfig{file=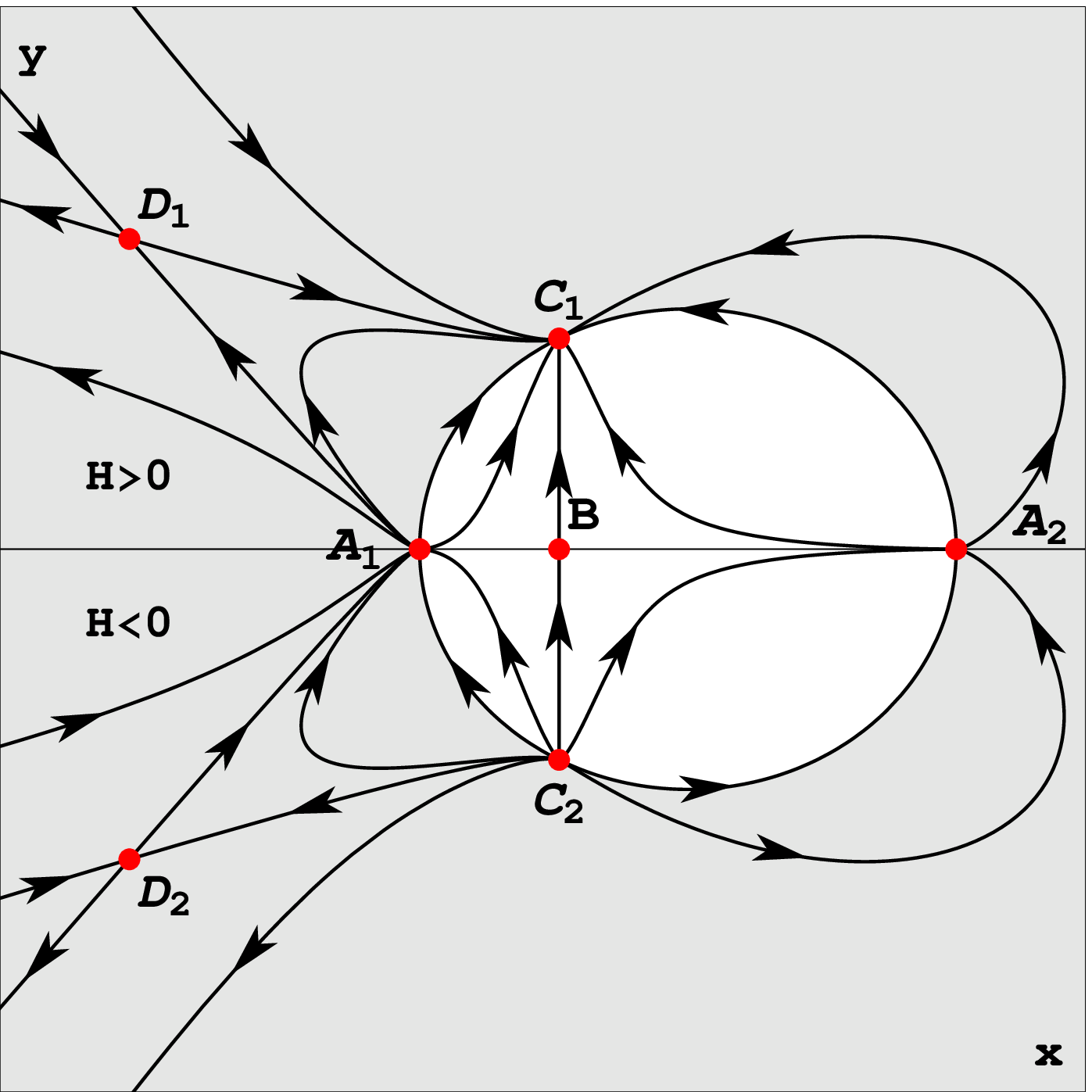,scale=0.5}
\epsfig{file=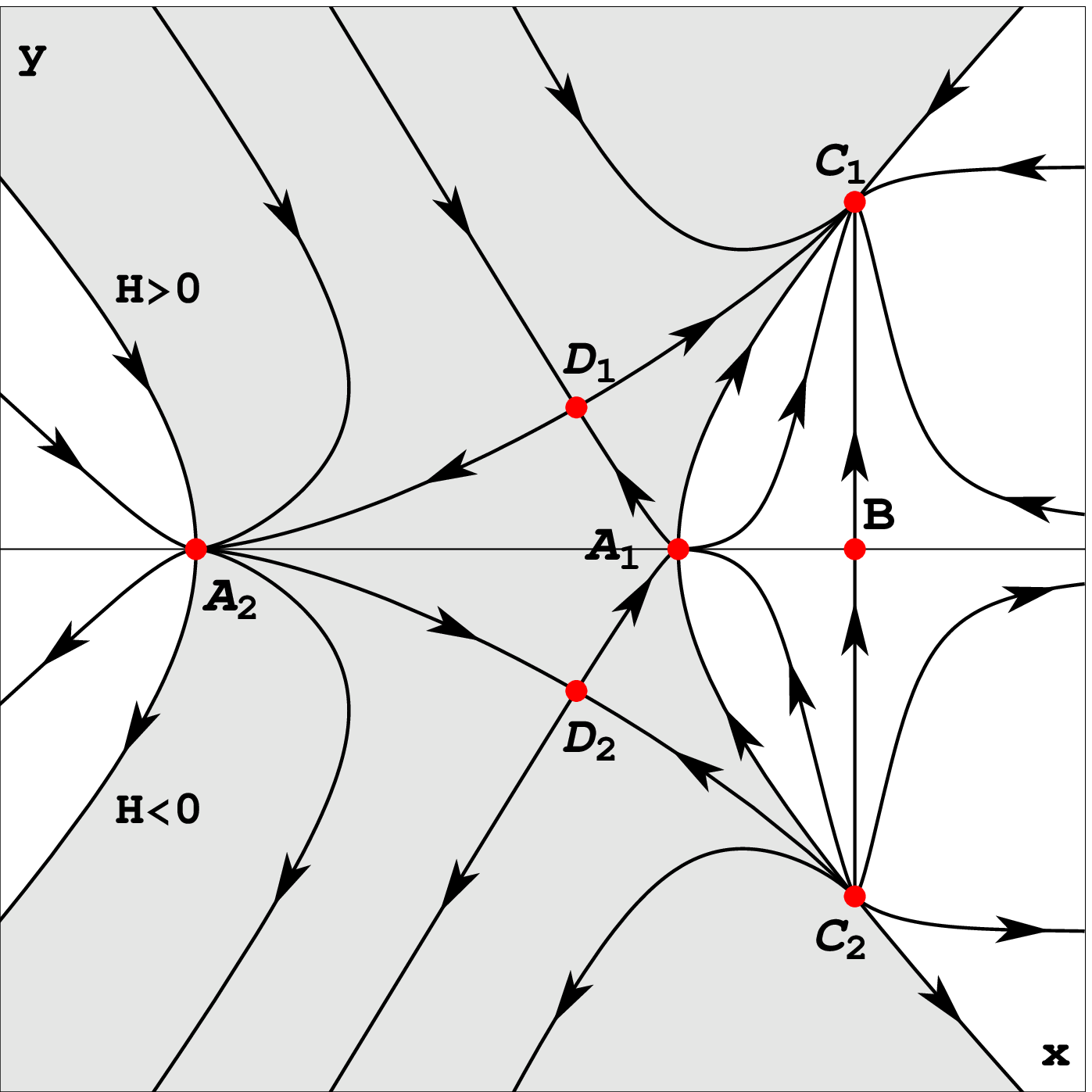,scale=0.5}
\epsfig{file=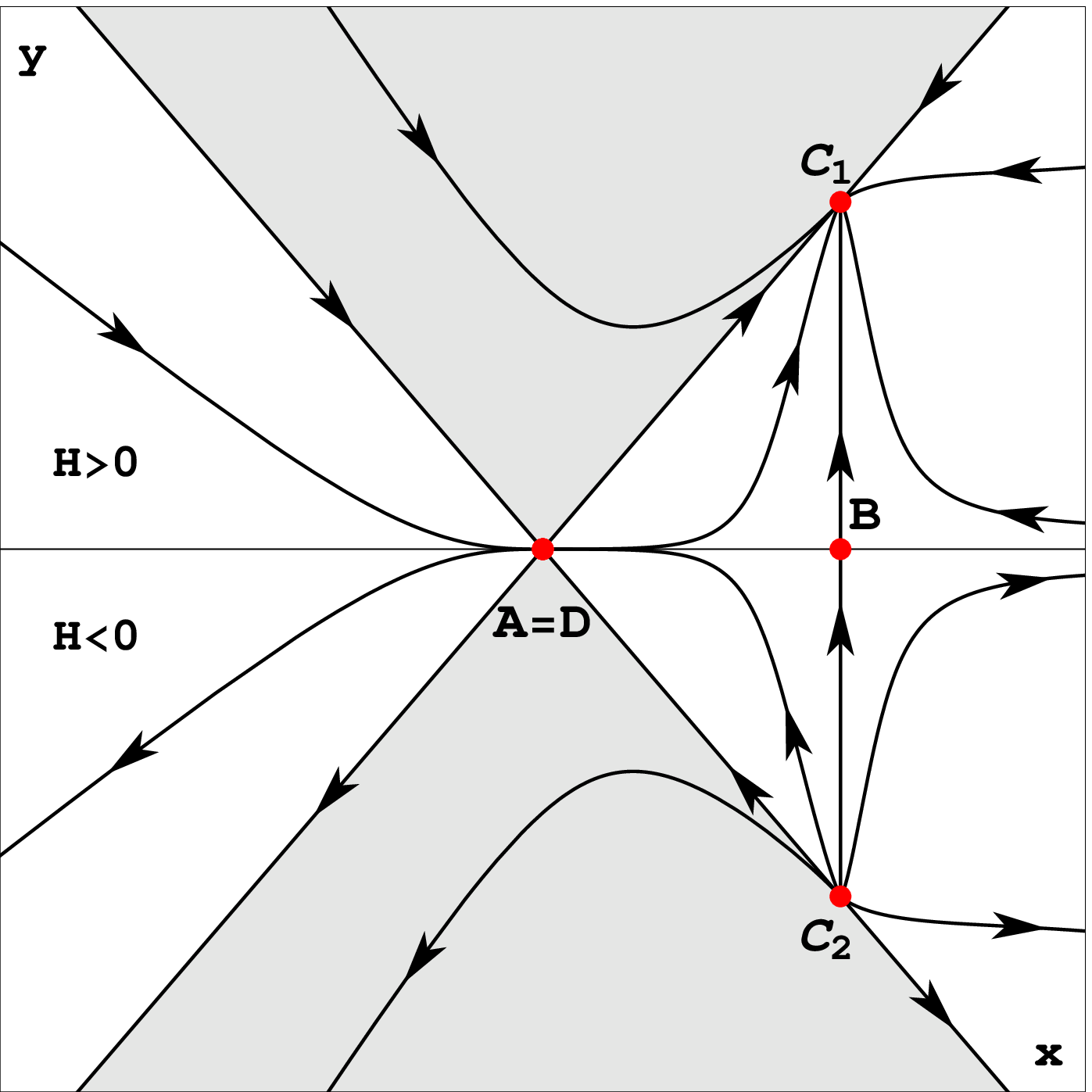,scale=0.5}
\epsfig{file=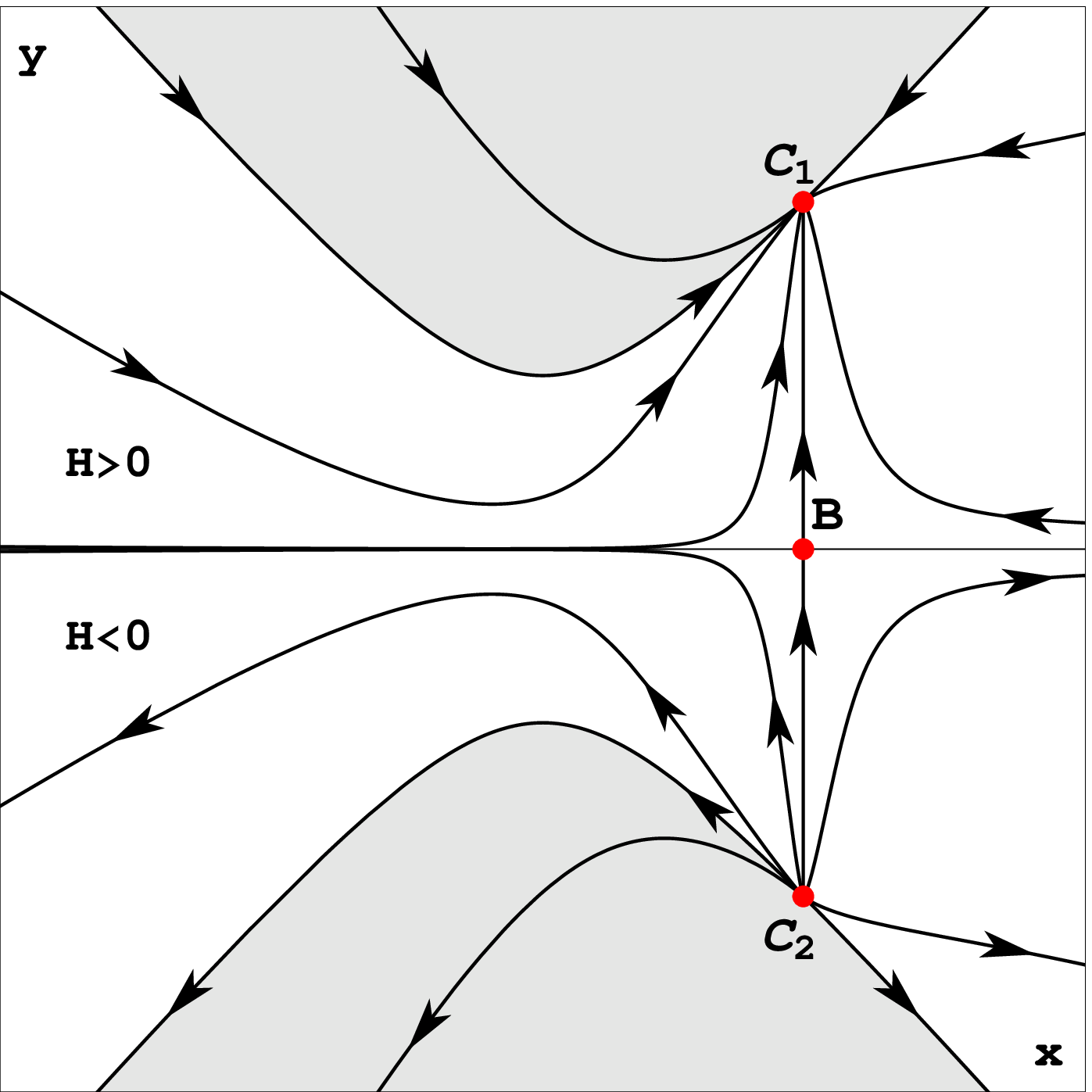,scale=0.5}
\caption{The phase plane diagrams for system \eqref{eq:sysquad} filled with the 
relativistic matter $w_{m}=1/3$ and the Brans-Dicke parameter: $\om>0$ (top left), $-3/2<\om<0$ 
(top right), $\om=-3/2$ (bottom left), $\om<-3/2$ (bottom right).}
\label{fig:3}
\end{center}
\end{figure}

\begin{figure}
\begin{center}
\epsfig{file=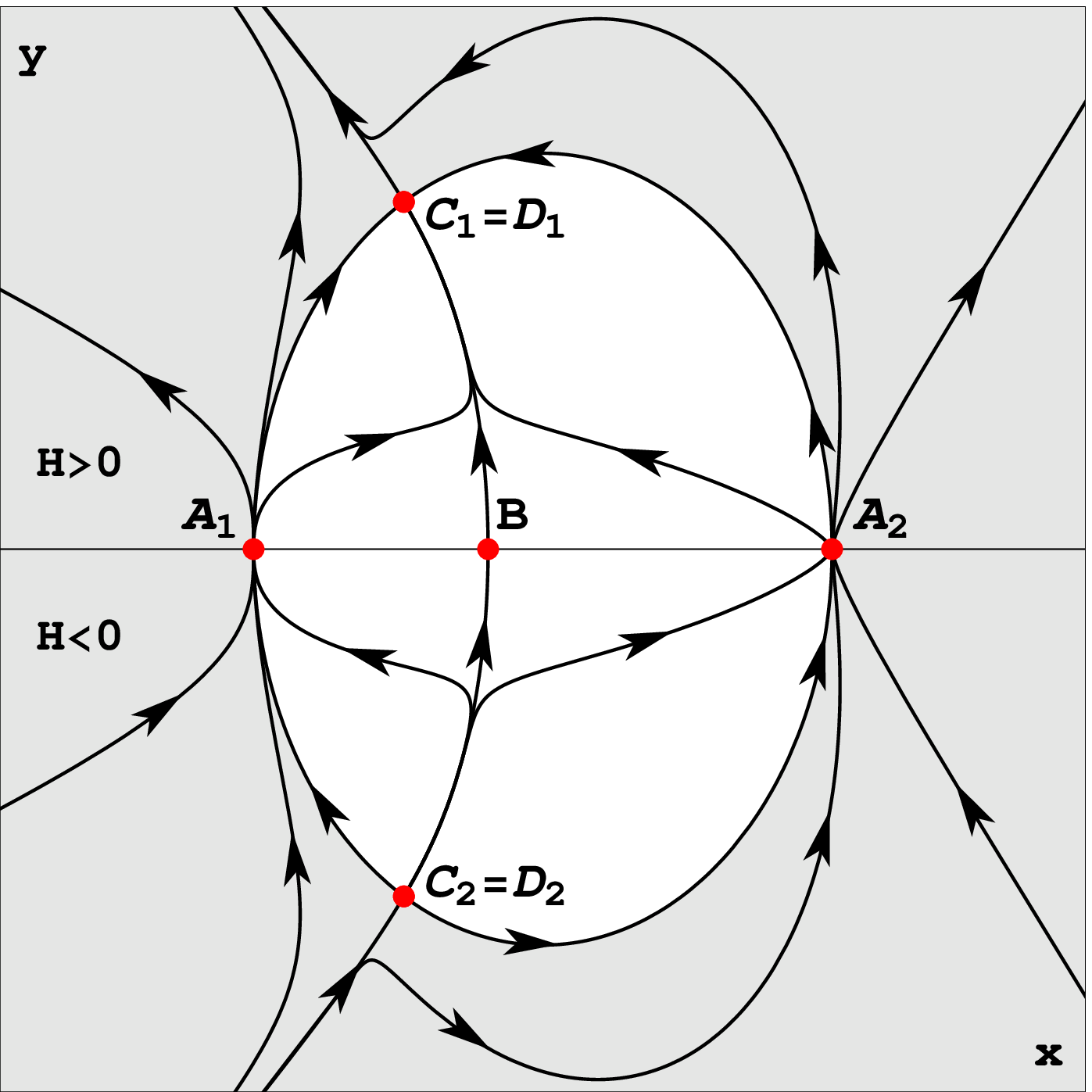,scale=0.5}
\epsfig{file=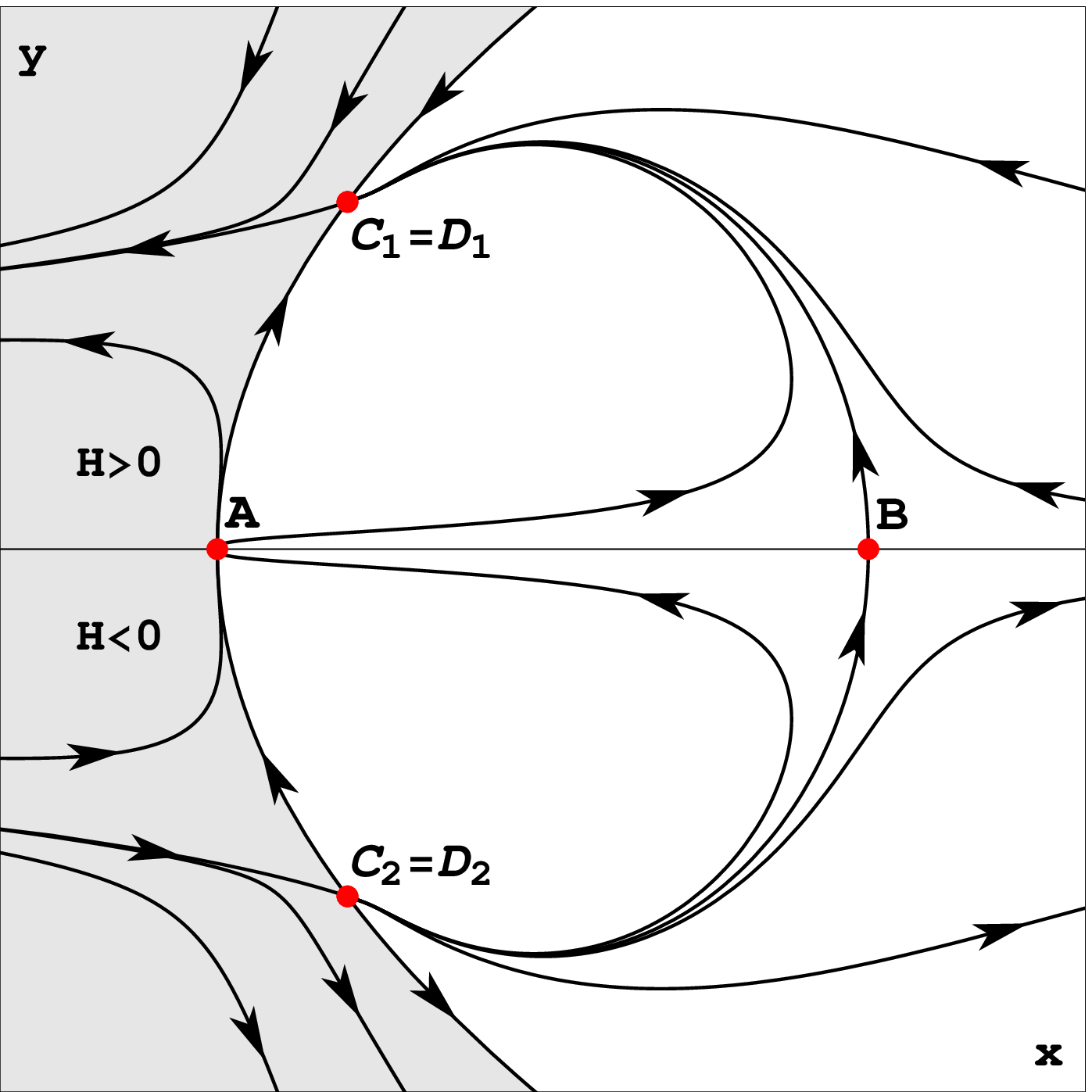,scale=0.5}
\epsfig{file=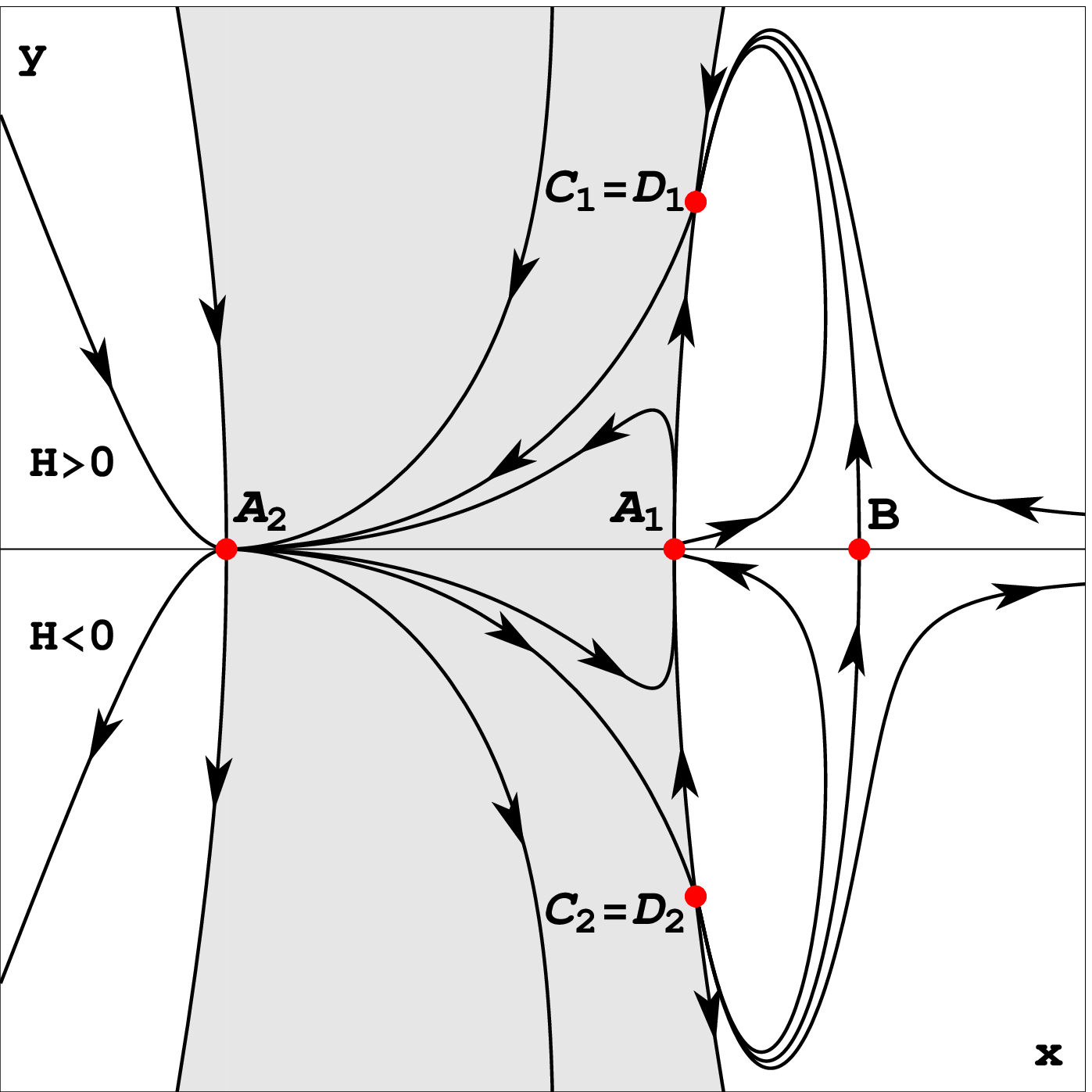,scale=0.5}
\epsfig{file=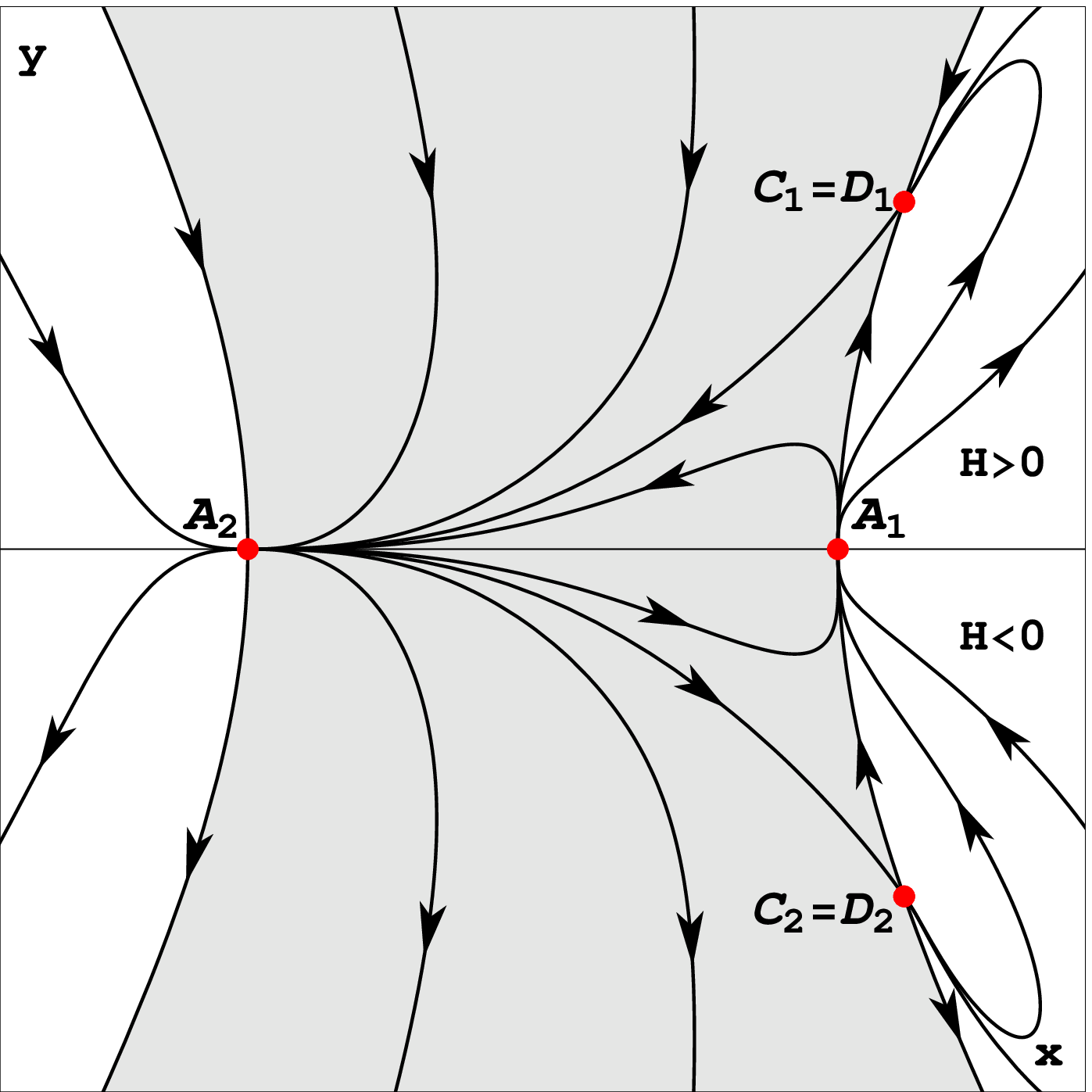,scale=0.5}
\caption{The phase plane diagrams for system \eqref{eq:sysquad} filled with the 
cosmological constant $w_{m}=-1$ and $\om>0$ (top left), $\om=0$ (top right), $0>\om>-1/2$
(bottom left), $\om=-1/2$ (bottom right).}
\label{fig:4}
\end{center}
\end{figure}

\begin{figure}
\begin{center}
\epsfig{file=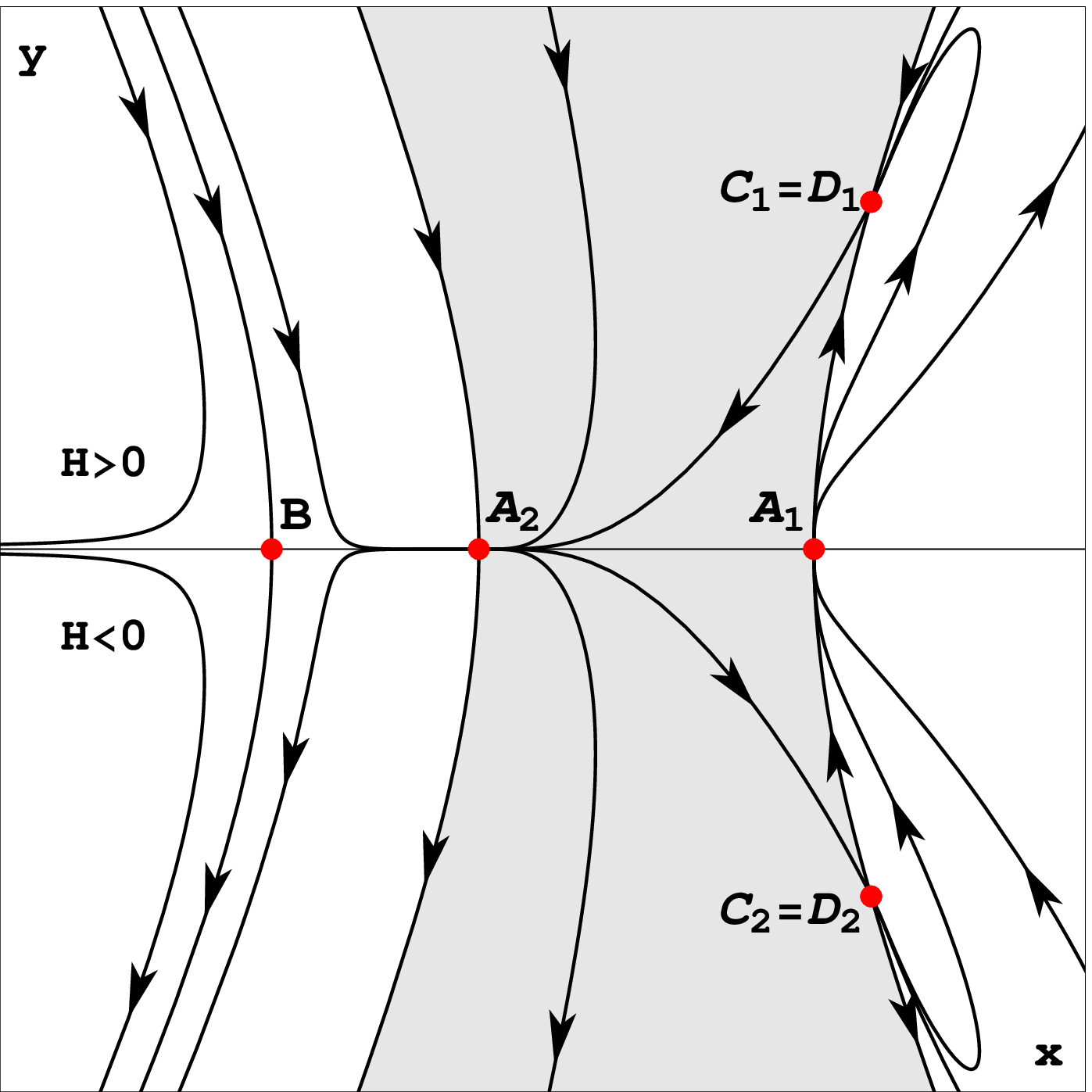,scale=0.5}
\epsfig{file=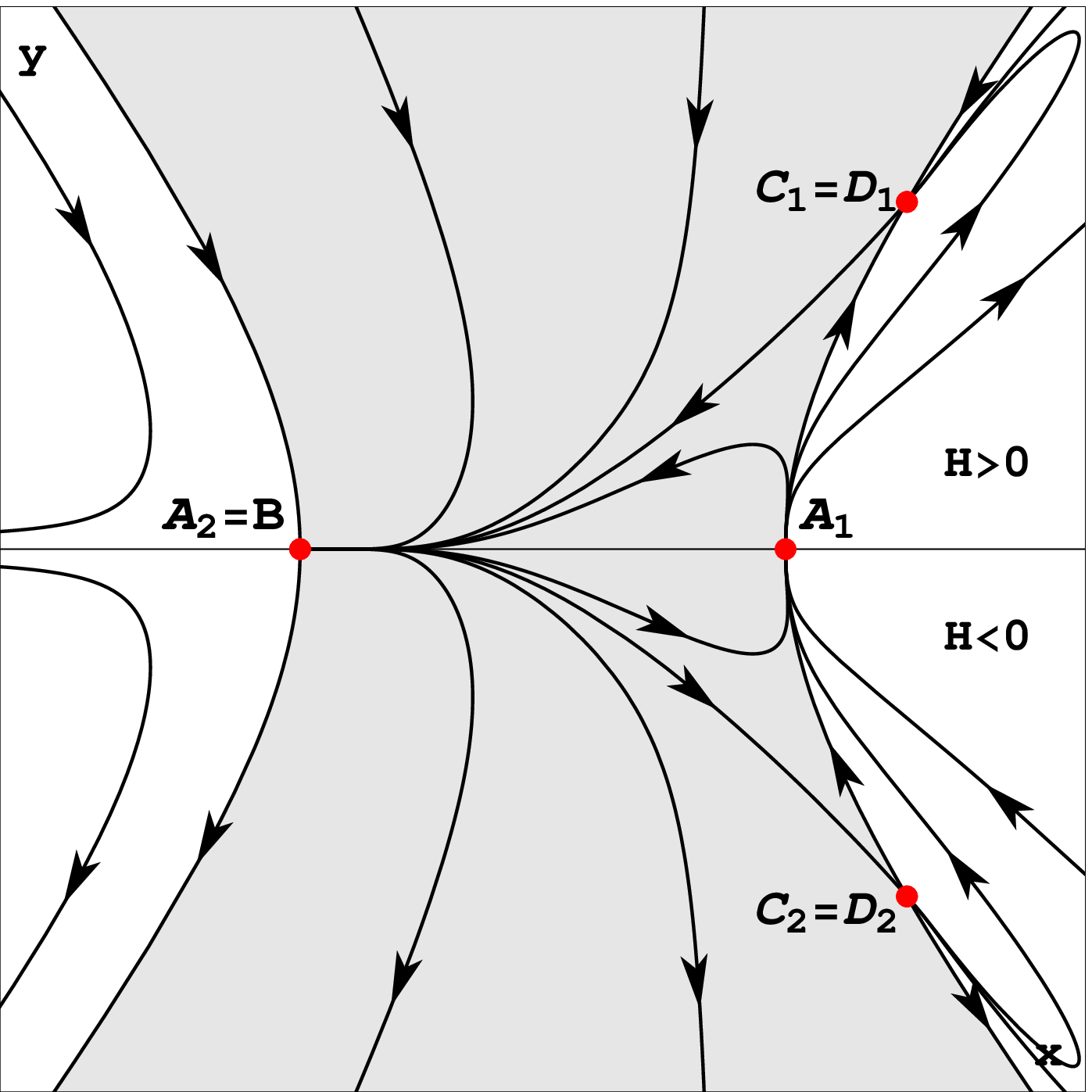,scale=0.5}
\epsfig{file=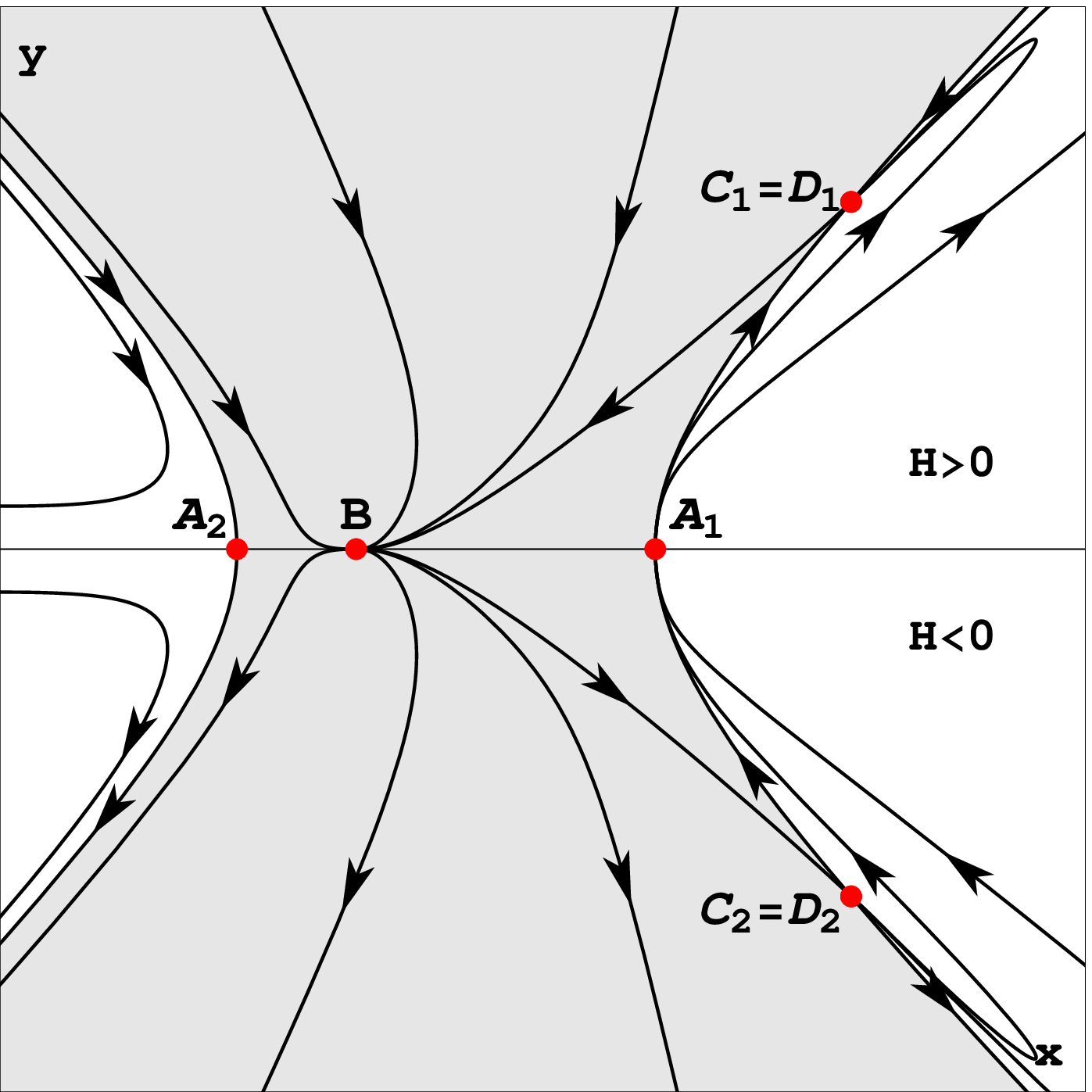,scale=0.5}
\epsfig{file=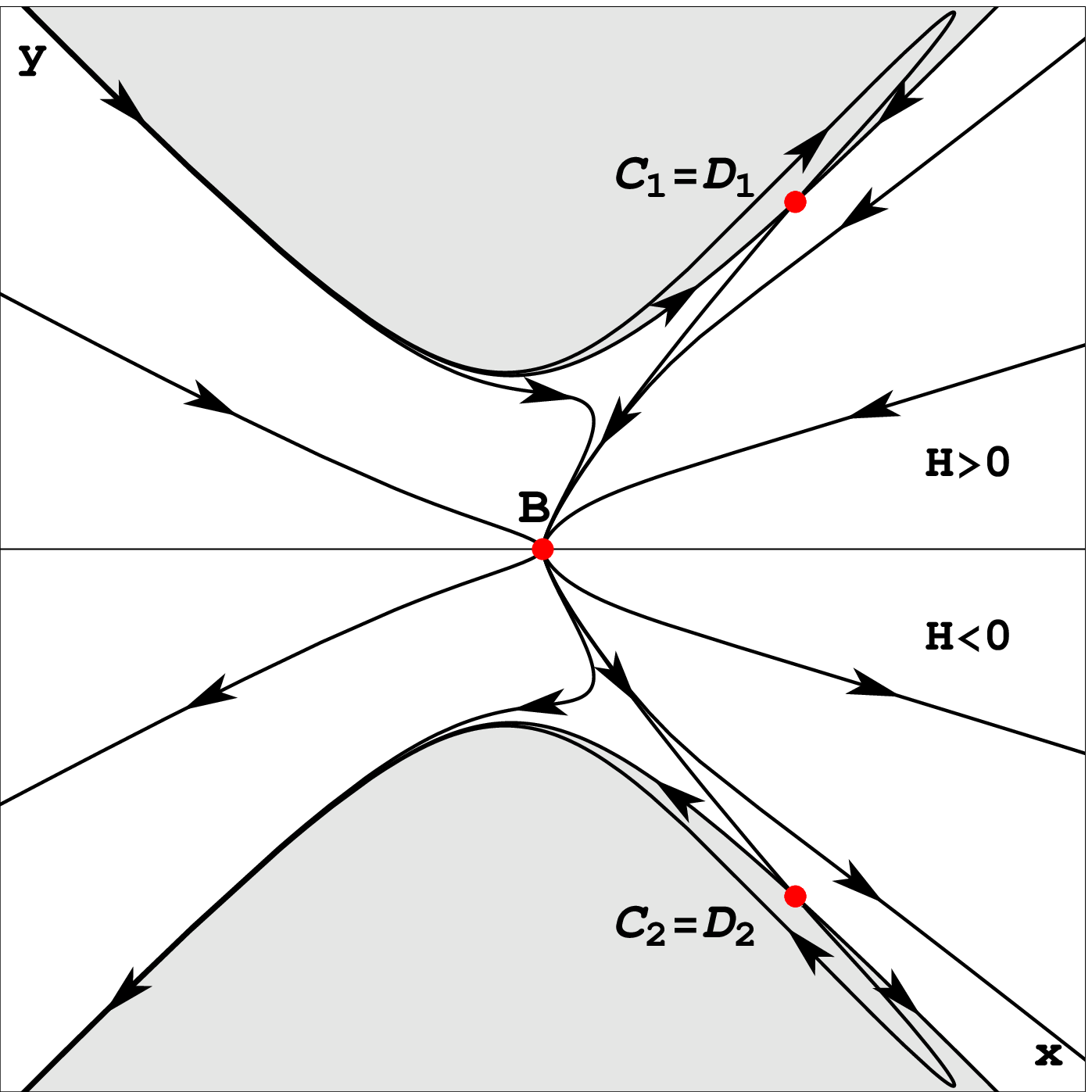,scale=0.5}
\caption{The phase plane diagrams for system \eqref{eq:sysquad} filled with the 
cosmological constant $w_{m}=-1$ and $-1/2>\om>-5/6$ (top left), $\om=-5/6$ (top right), $-5/6>\om>-3/2$ (bottom left), $\om<-3/2$ (bottom right).}
\label{fig:5}
\end{center}
\end{figure}

\begin{figure}
\begin{center}
\epsfig{file=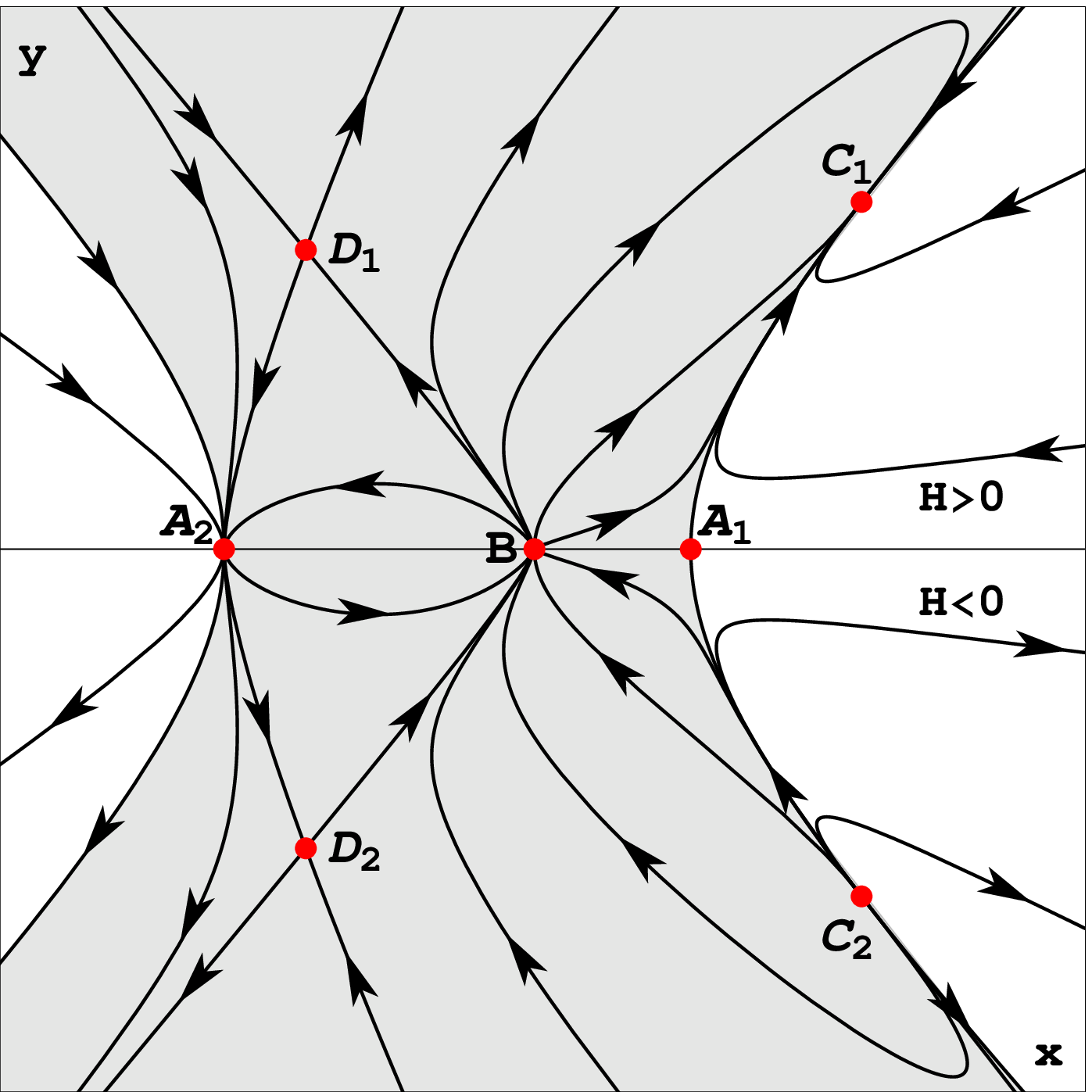,scale=0.5}
\epsfig{file=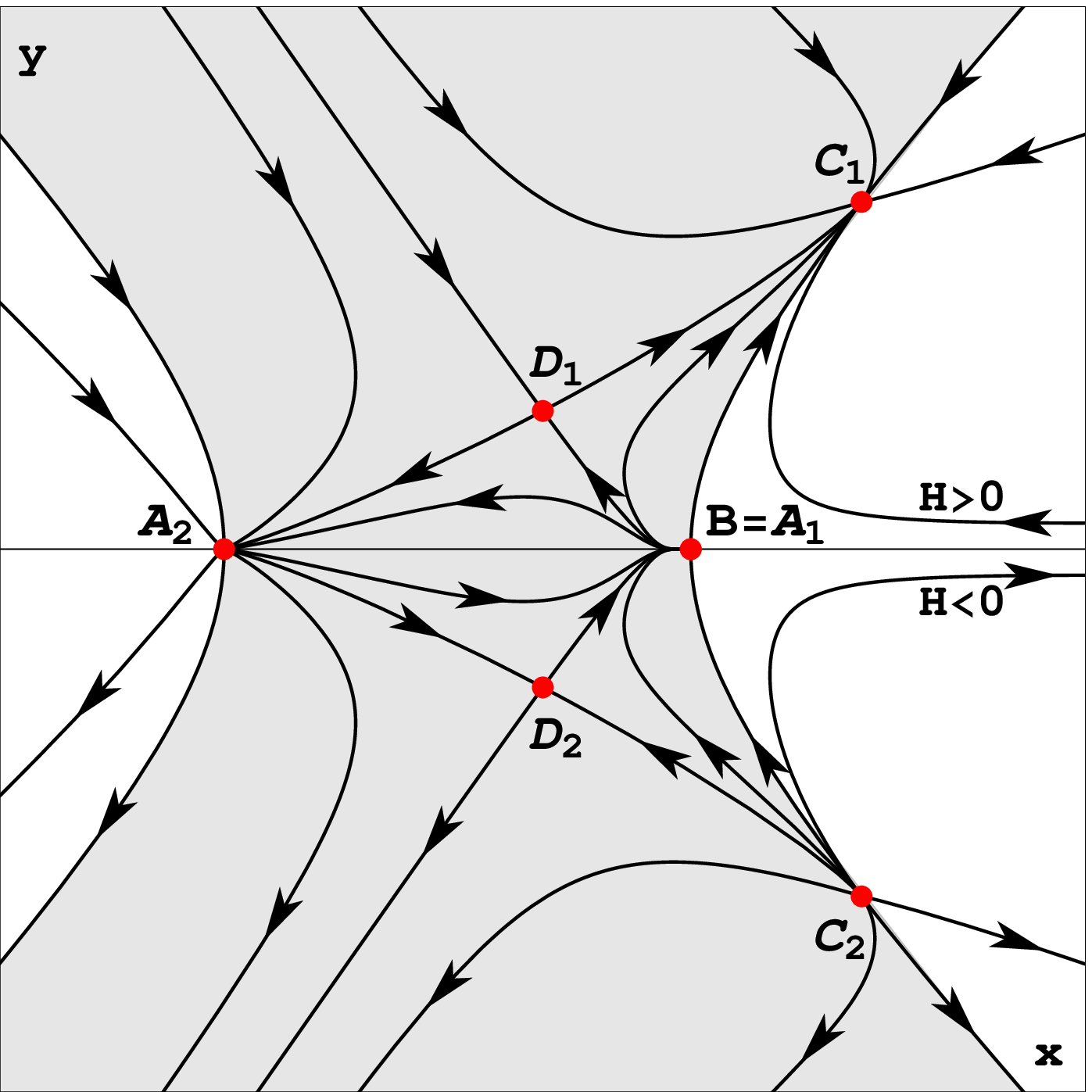,scale=0.5}
\epsfig{file=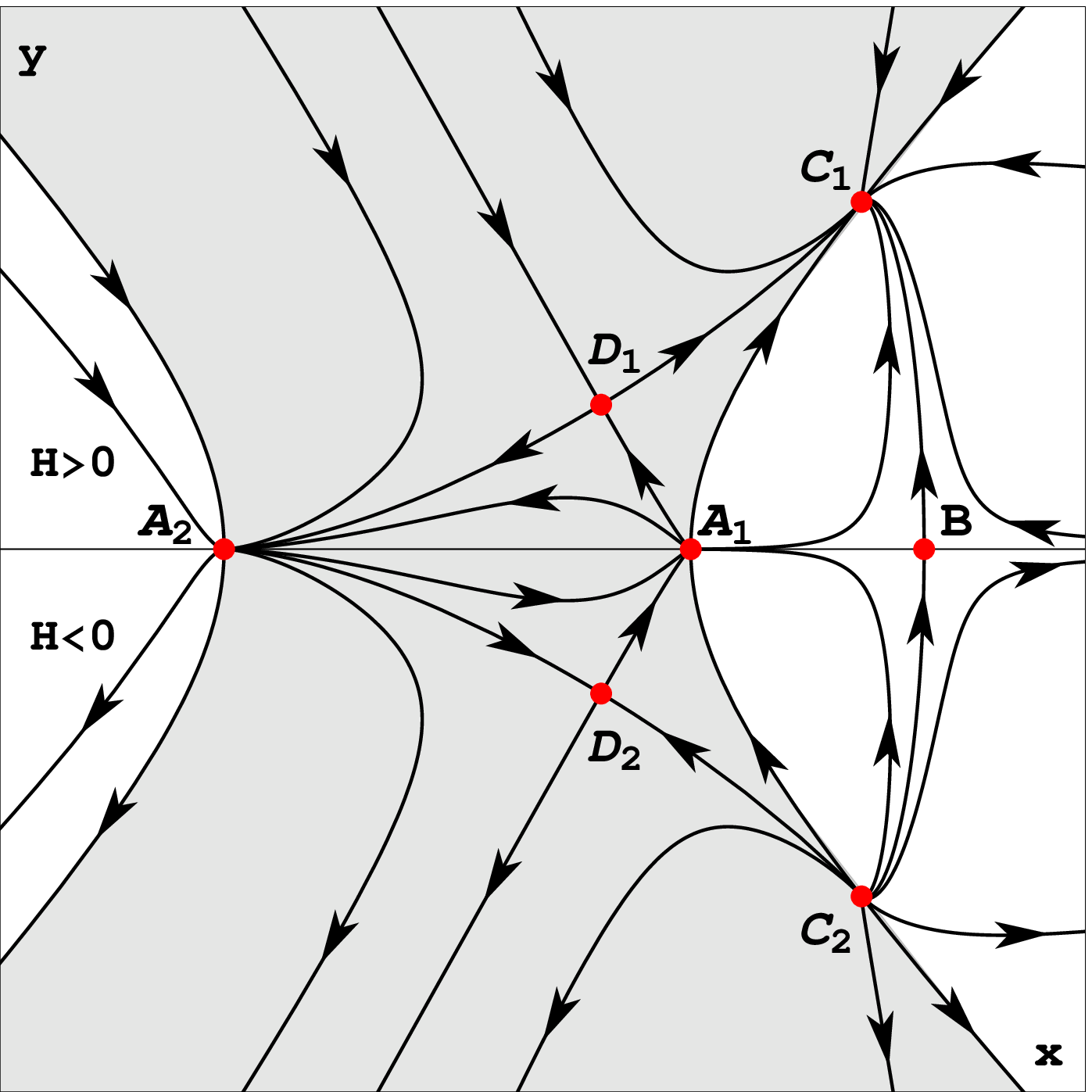,scale=0.5}
\epsfig{file=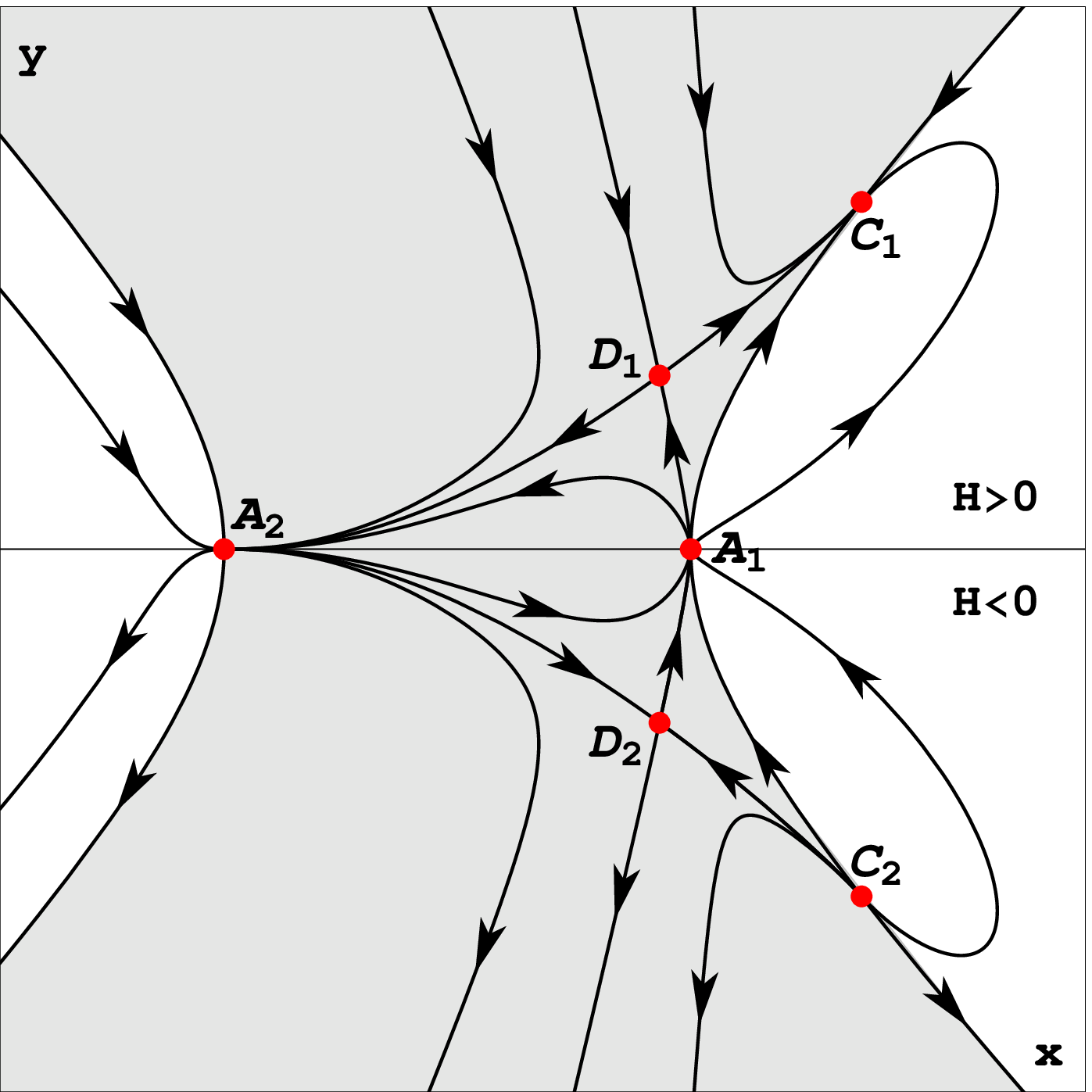,scale=0.5}
\caption{The phase plane diagrams for system \eqref{eq:sysquad} in low energy string theory limit $\om=-1$ for various matter content: $w_{m}>\frac{\sqrt{3}}{3}$ (top left), 
$w_{m}=\frac{\sqrt{3}}{3}$ (top right), $0<w_{m}<\frac{\sqrt{3}}{3}$ (bottom left), 
$w_{m}=0$ (bottom right).}
\label{fig:6}
\end{center}
\end{figure}

\begin{figure}
\begin{center}
\epsfig{file=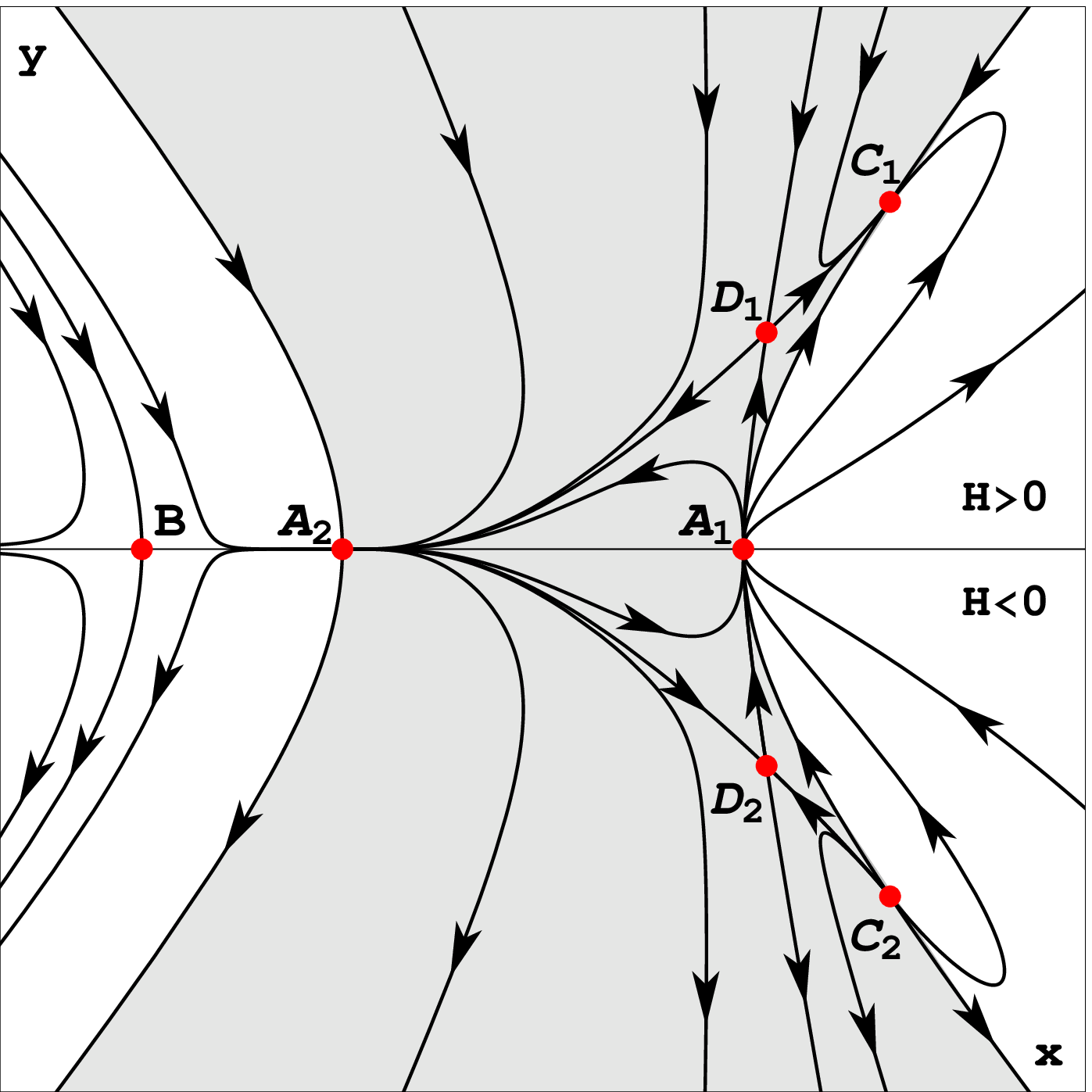,scale=0.5}
\epsfig{file=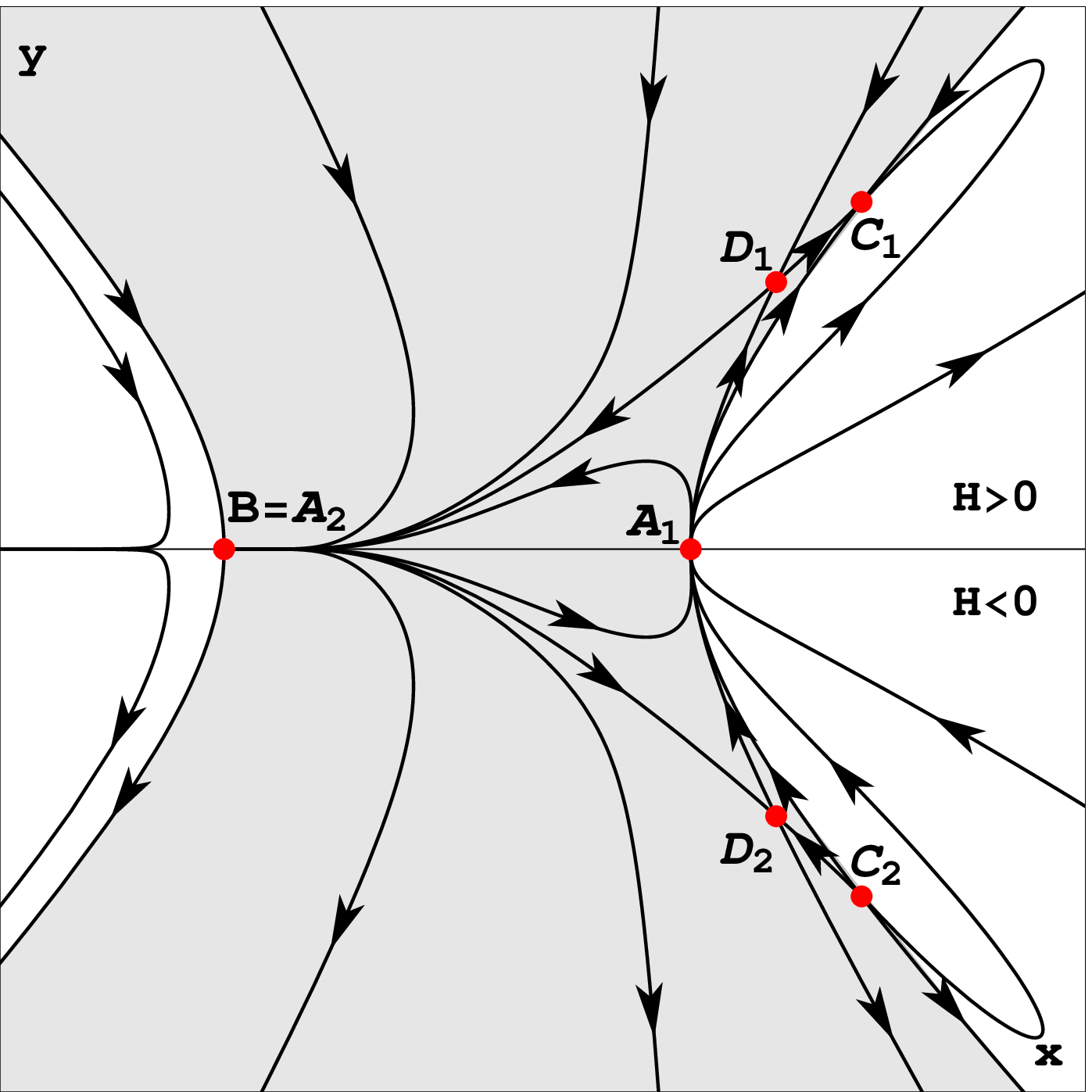,scale=0.5}
\epsfig{file=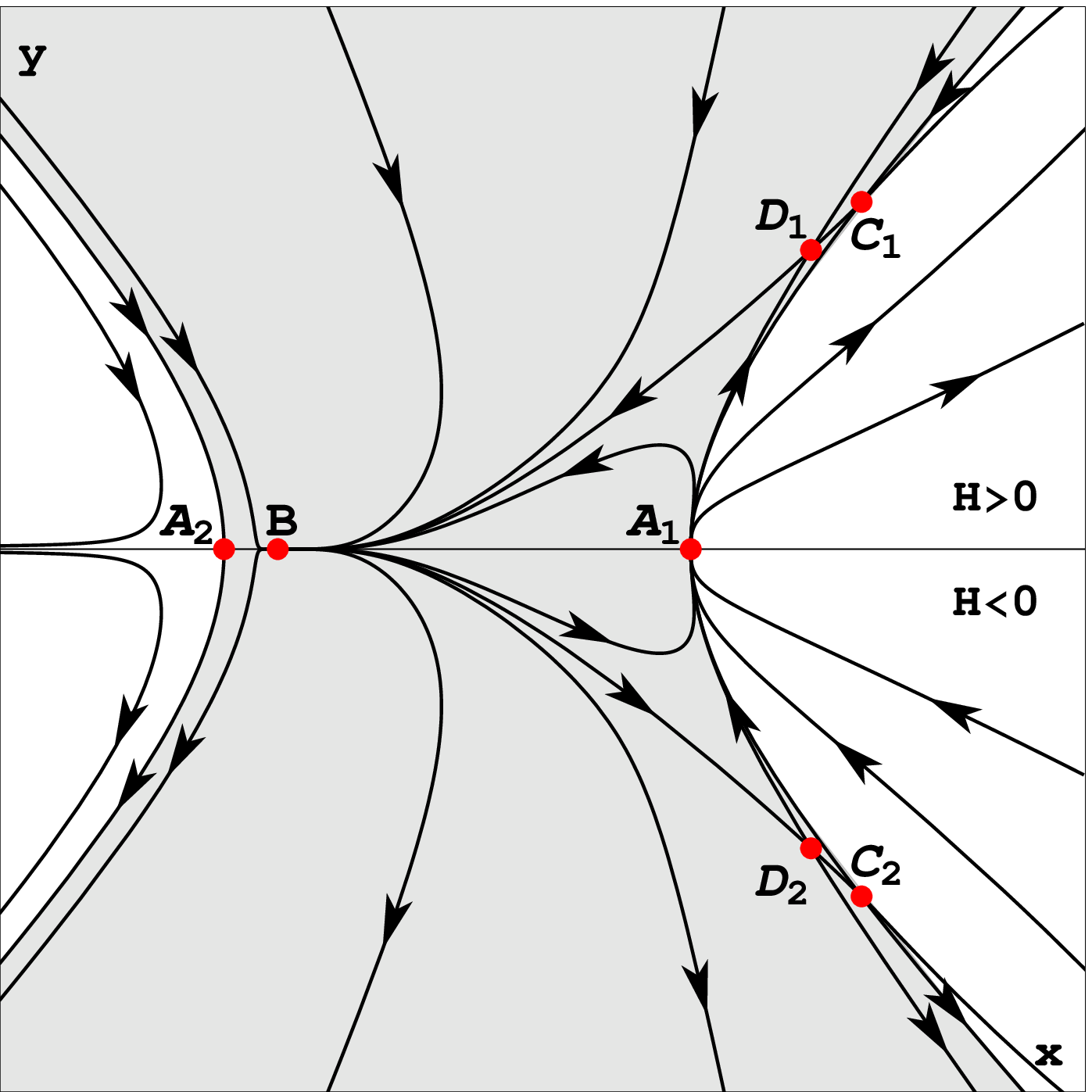,scale=0.5}
\epsfig{file=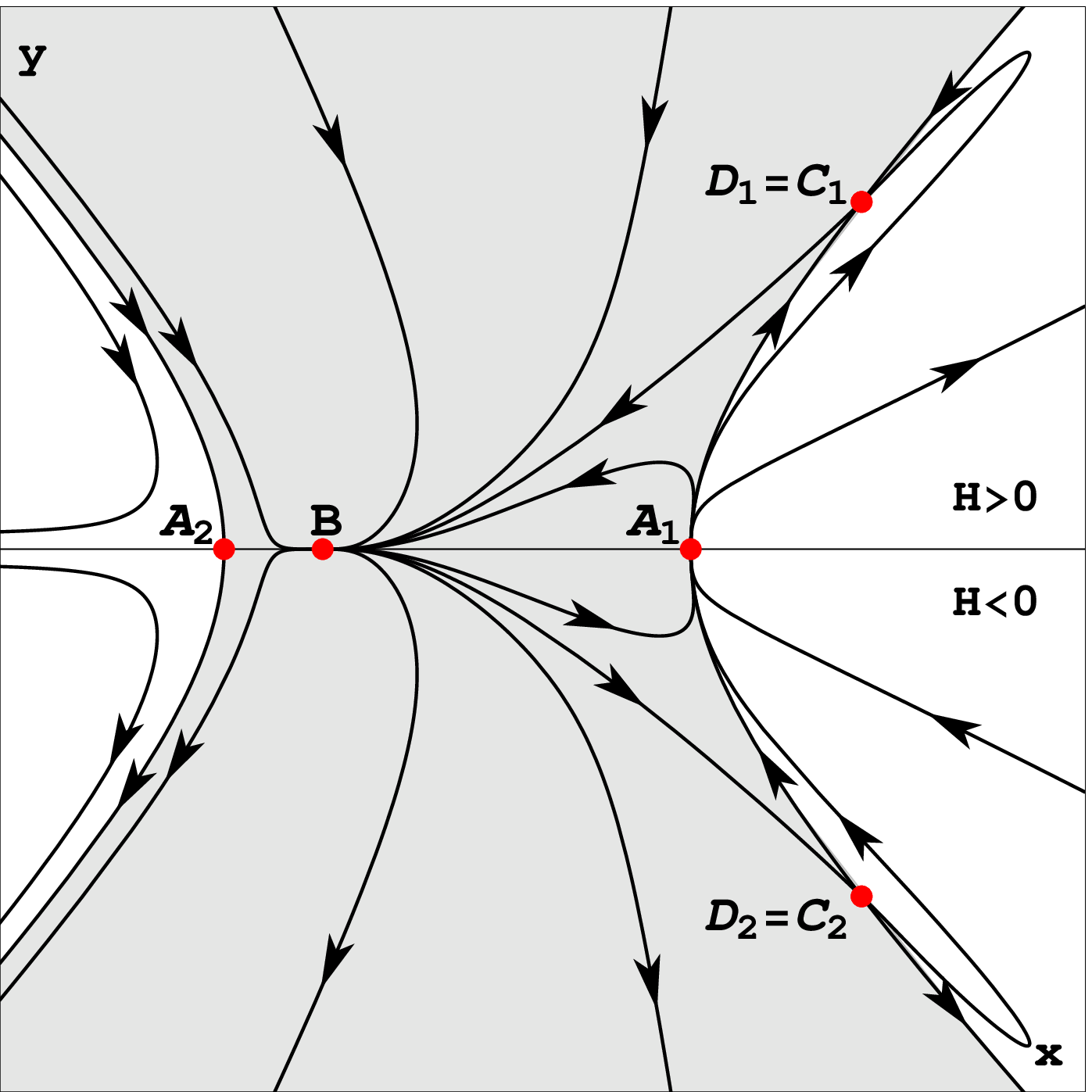,scale=0.5}
\caption{The phase plane diagrams for system \eqref{eq:sysquad} in low energy string theory limit $\om=-1$ for various matter content: $-\frac{\sqrt{3}}{3}<w_{m}<0$ (top left), 
$w_{m}=-\frac{\sqrt{3}}{3}$ (top right), $-1<w_{m}<-\frac{\sqrt{3}}{3}$ (bottom left), 
$w_{m}=-1$ (bottom right).}
\label{fig:7}
\end{center}
\end{figure}

\begin{figure}
\begin{center}
\epsfig{file=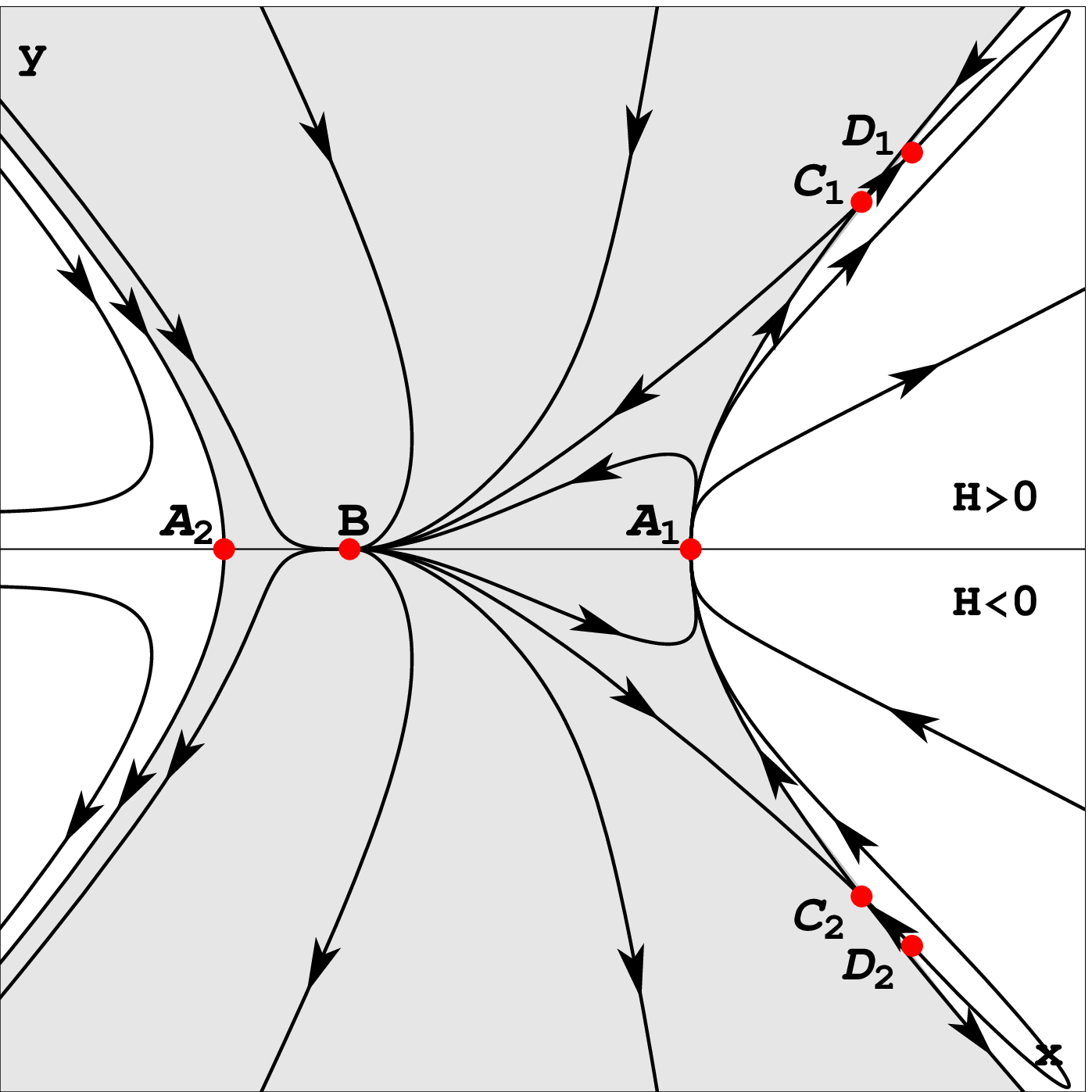,scale=0.5}
\epsfig{file=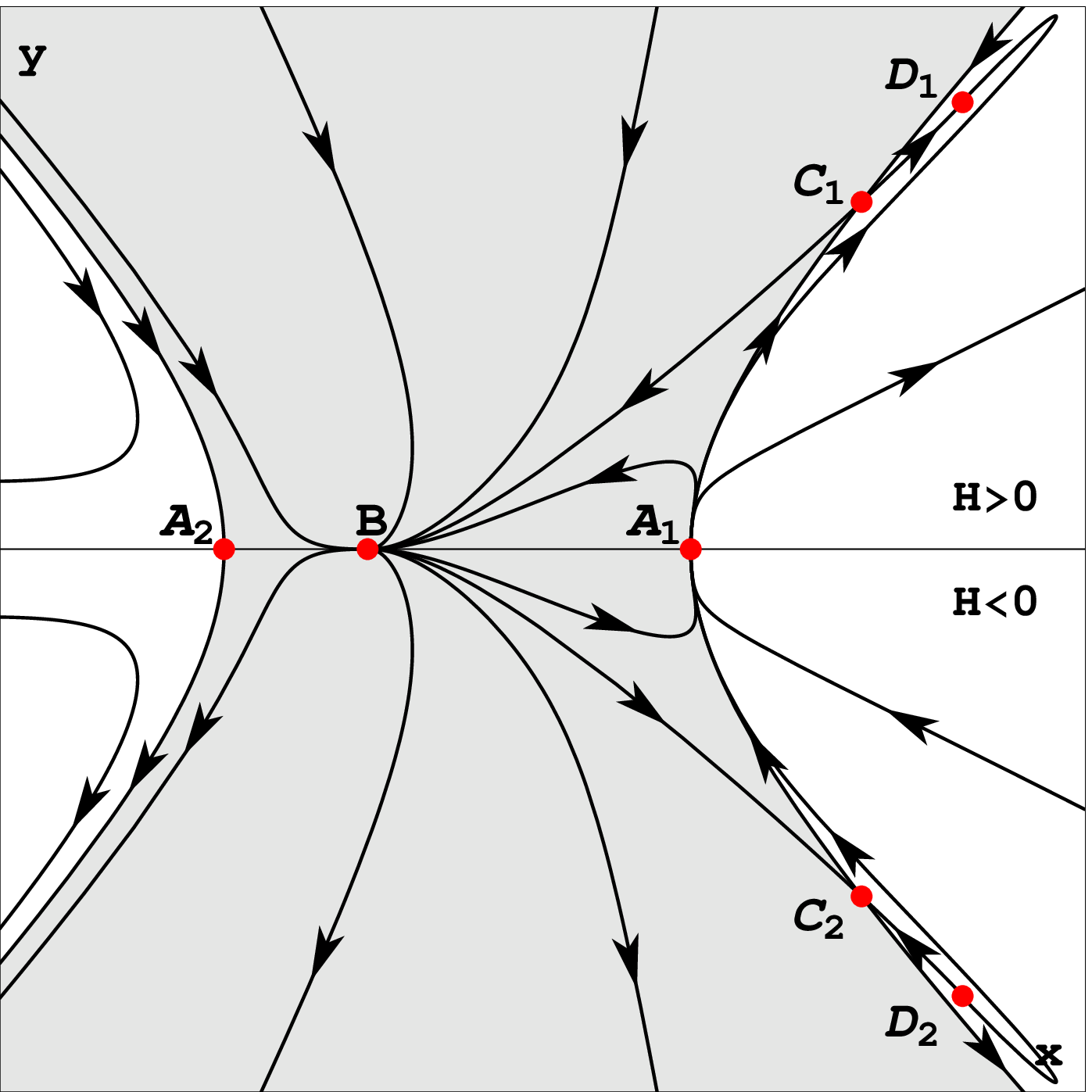,scale=0.5}
\caption{The phase plane diagrams for system \eqref{eq:sysquad} in low energy string theory limit $\om=-1$ for various matter content: $-1>w_{m}>-1.2717...$ (left panel) the critical point $D_{1}$ is a stable node while the critical point $D_{2}$ is unstable, $w_{m}<-1.2717...$ (right panel) the critical point $D_{1}$ is a stable spiral while the critical point $D_{2}$ is an unstable spiral.}
\label{fig:8}
\end{center}
\end{figure}

\begin{figure}
\begin{center}
\epsfig{file=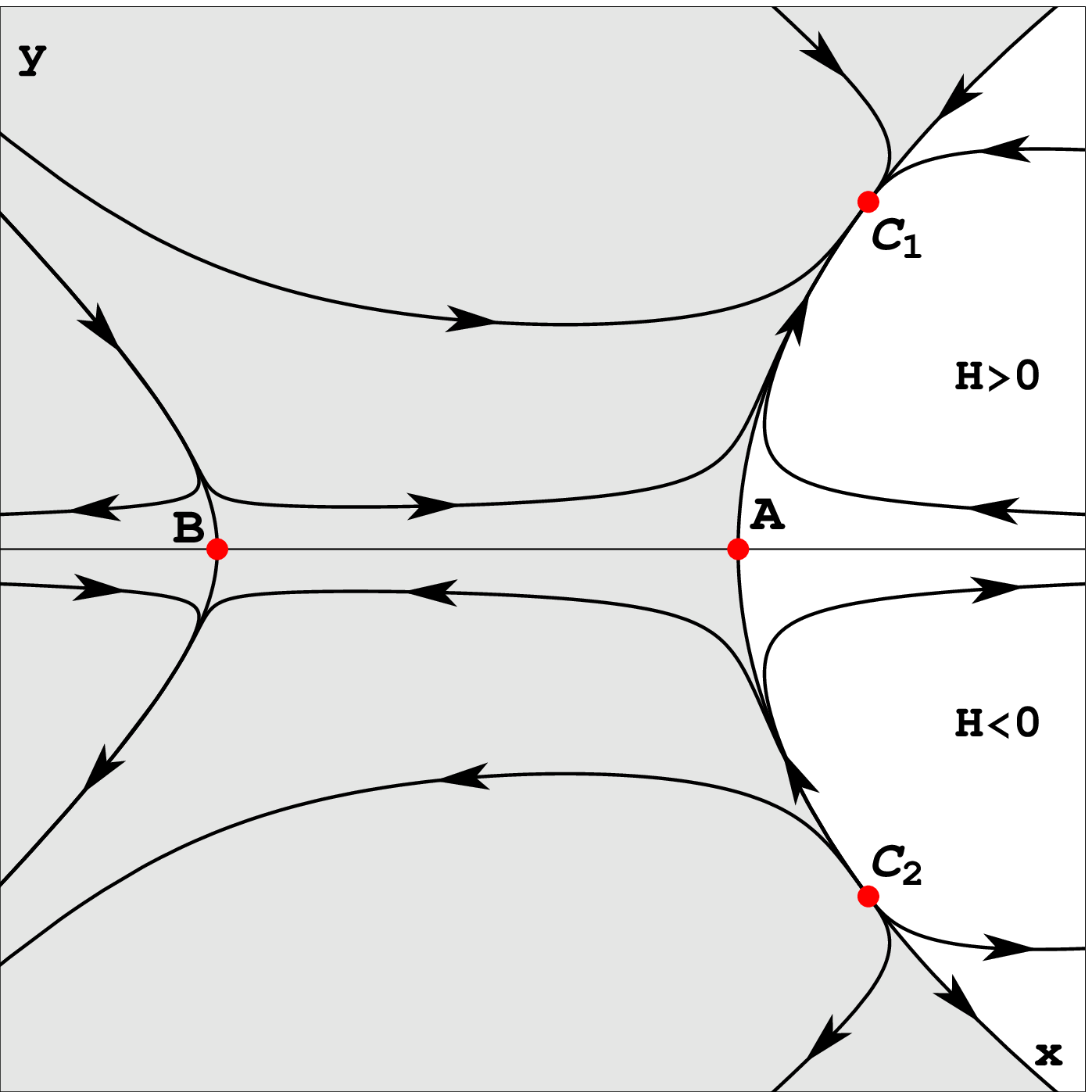,scale=0.5}
\epsfig{file=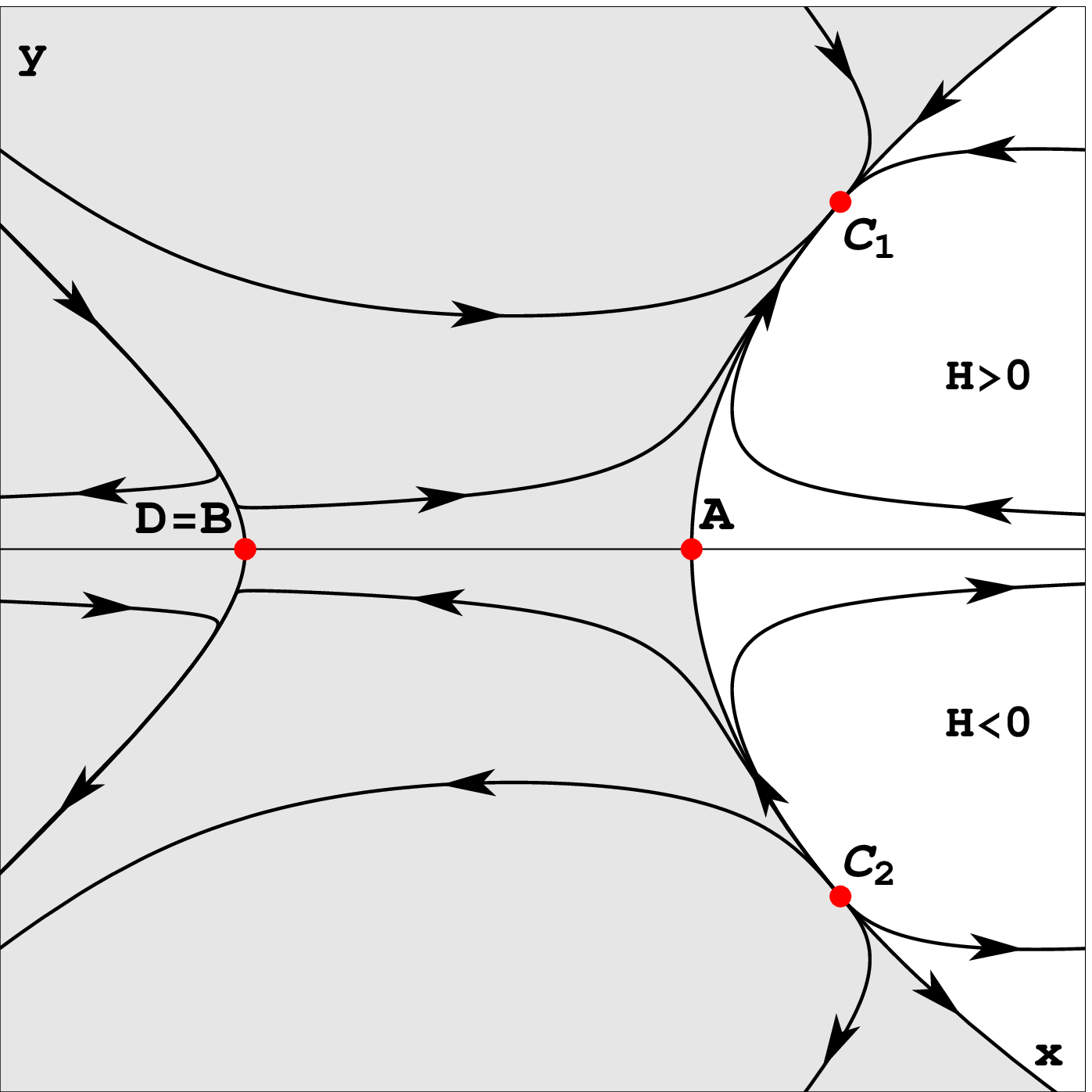,scale=0.5}
\epsfig{file=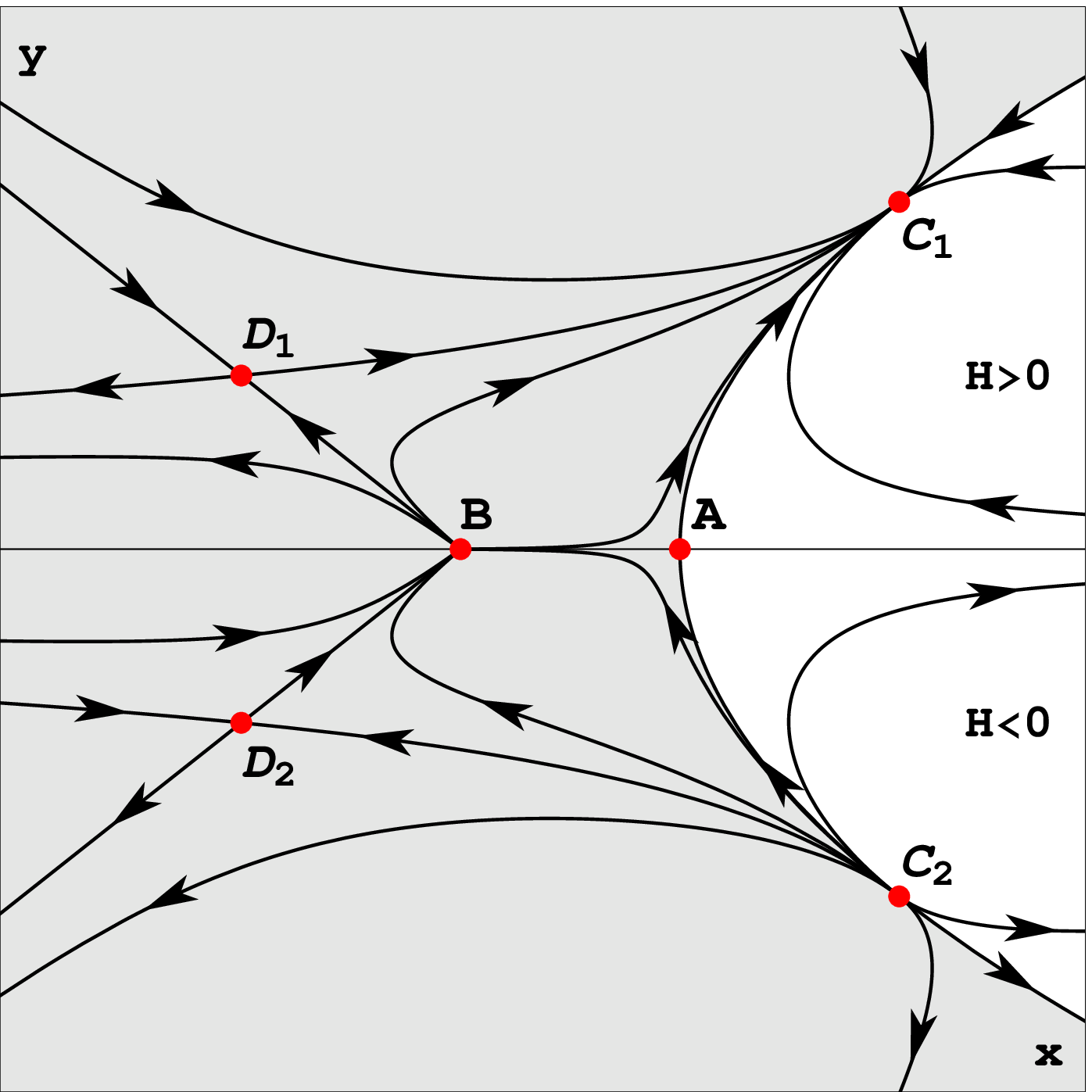,scale=0.5}
\epsfig{file=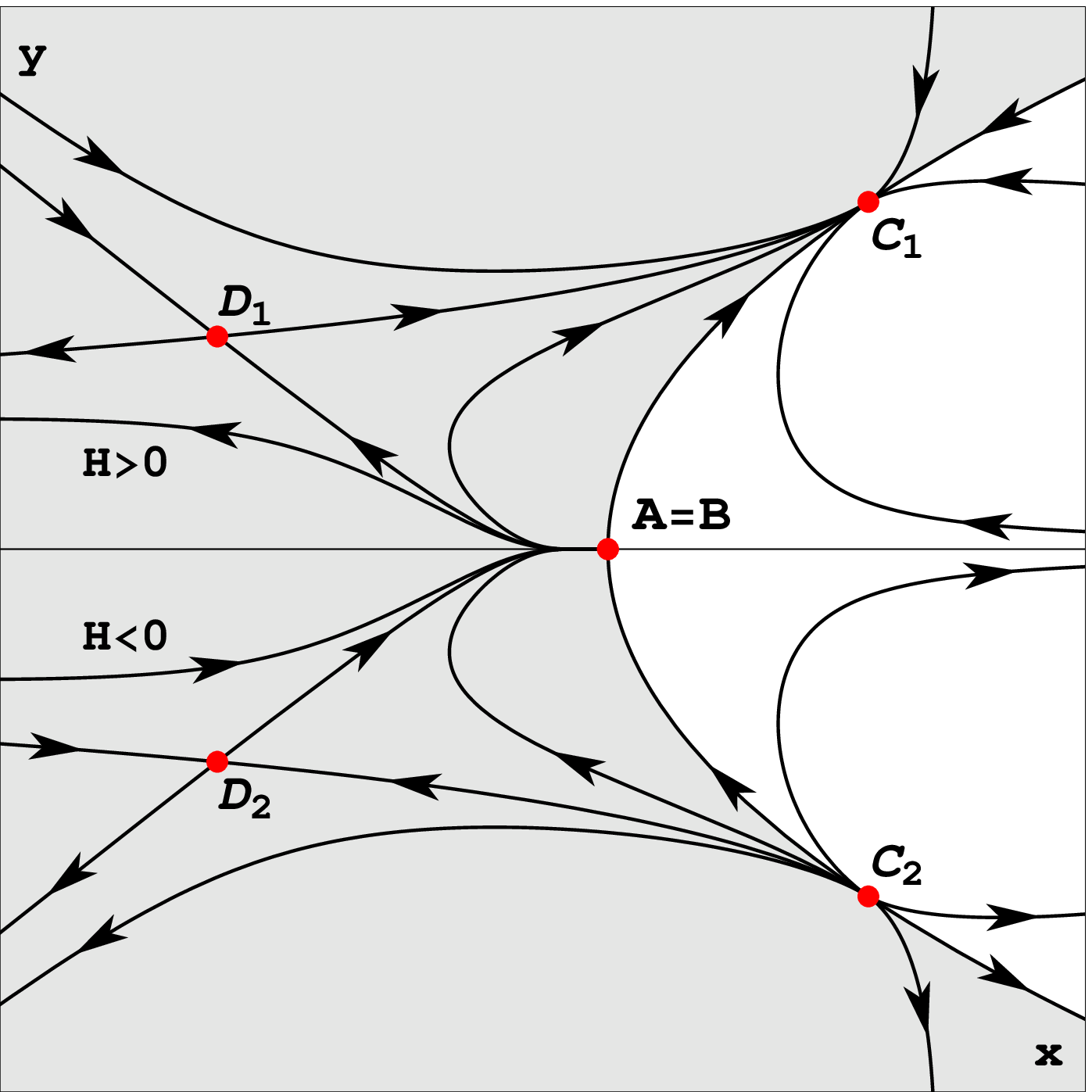,scale=0.5}
\caption{The phase plane diagrams for system \eqref{eq:sysquad} in the so-called ``O'Hanlon theory'' $\om=0$ for various matter content: $w_{m}> 5/3$ (top left), $w_{m}=5/3$ (top right), $2/3<w_{m}<5/3$ (bottom left), $w_{m}=2/3$ (bottom right).}
\label{fig:9}
\end{center}
\end{figure}

\begin{figure}
\begin{center}
\epsfig{file=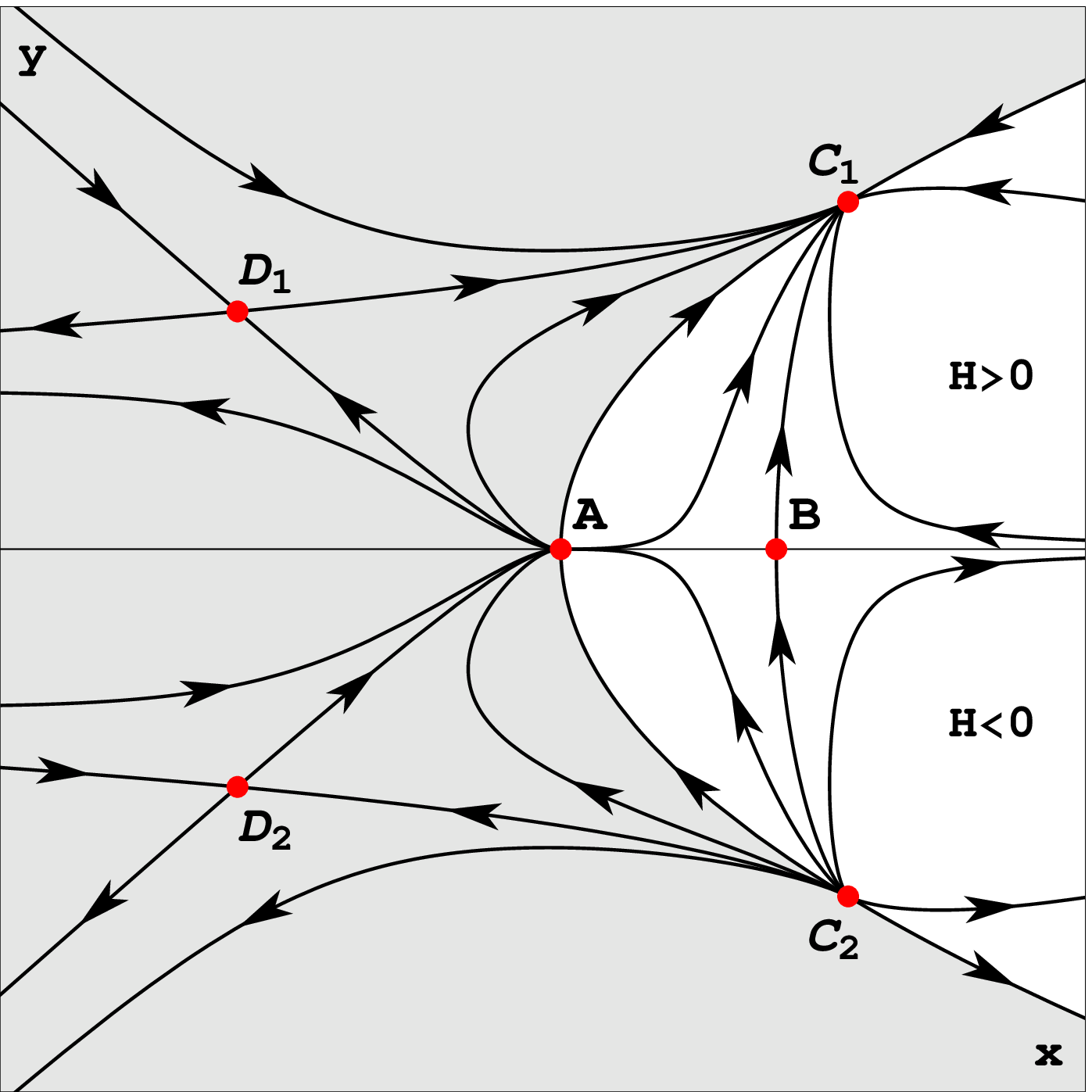,scale=0.5}
\epsfig{file=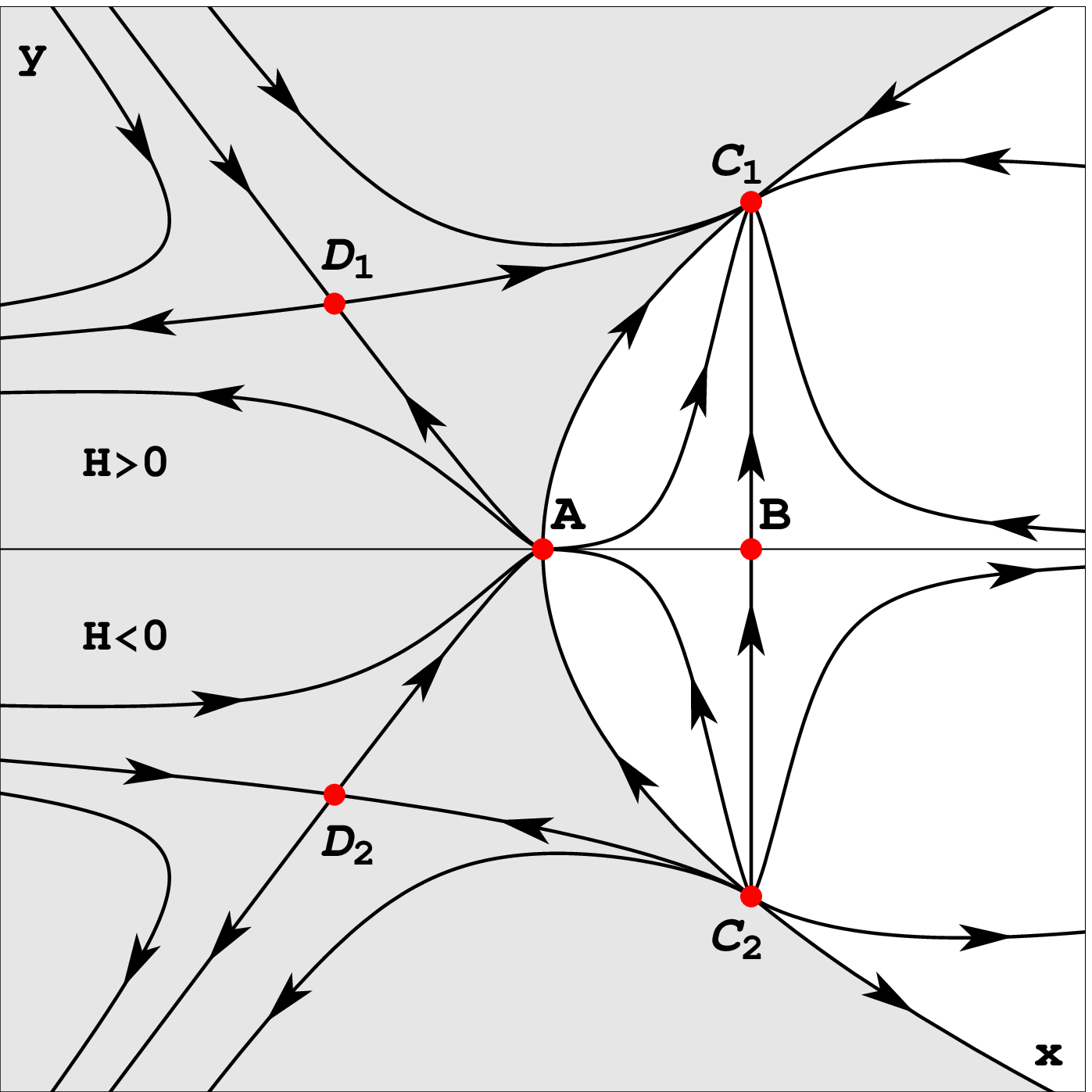,scale=0.5}
\epsfig{file=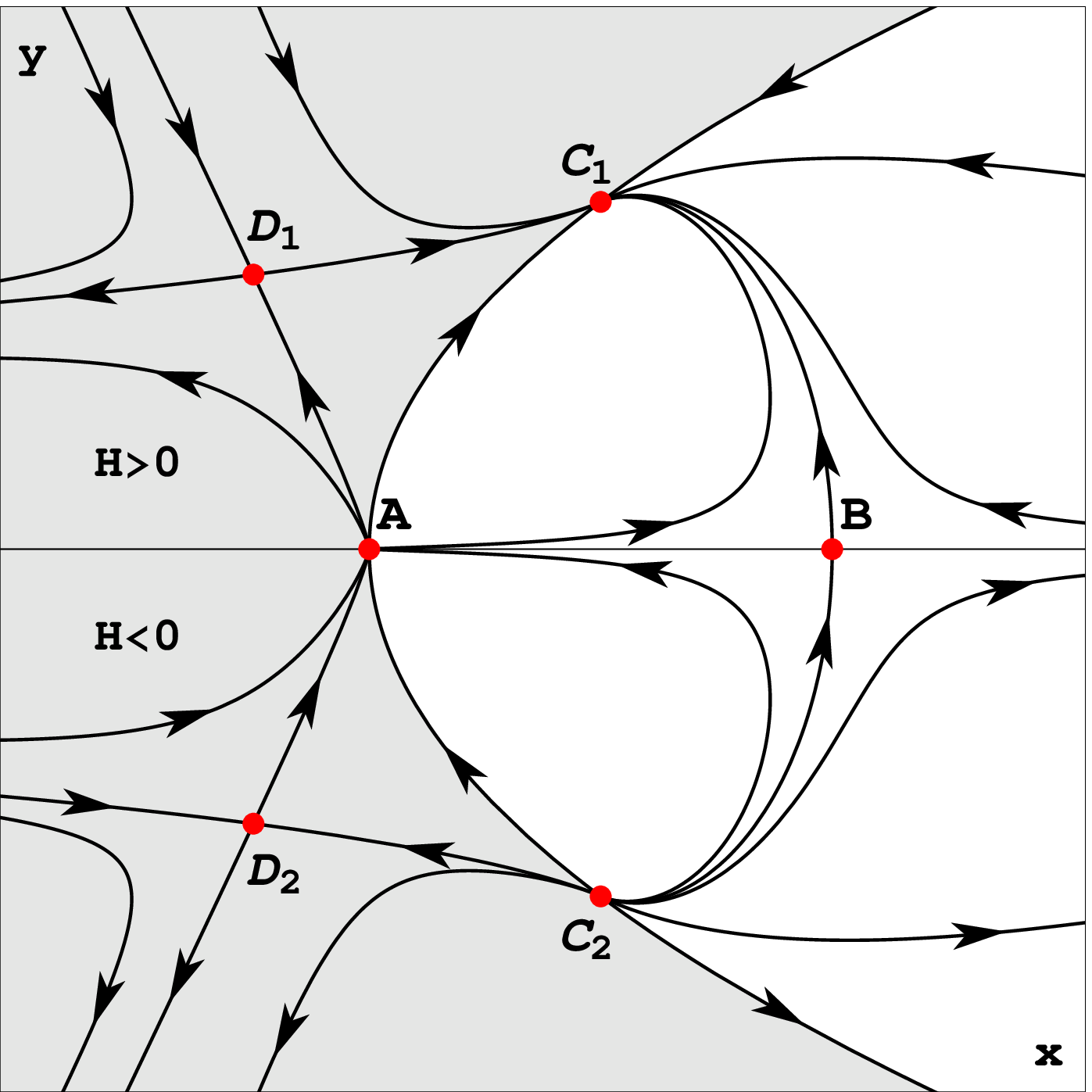,scale=0.5}
\epsfig{file=fig9.eps,scale=0.5}
\caption{The phase plane diagrams for system \eqref{eq:sysquad} in the so-called ``O'Hanlon theory'' $\om=0$ for various matter content: $1/3<w_{m}<2/3$ {top left}, $w_{m}=1/3$ (top right), $w_{m}=0$ (bottom left), $w_{m}=-1$ (bottom right).}
\label{fig:10}
\end{center}
\end{figure}

\begin{figure}
\begin{center}
\epsfig{file=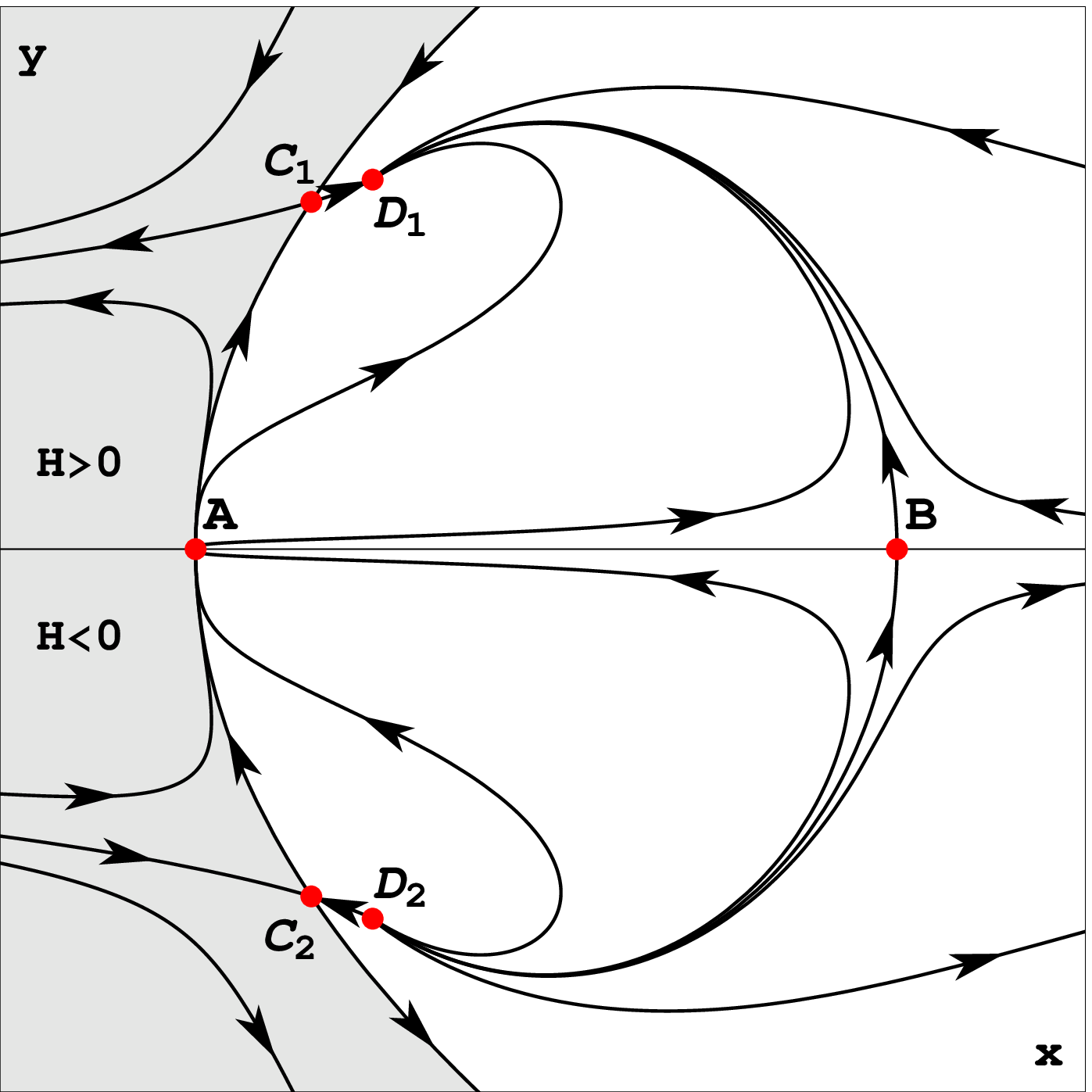,scale=0.5}
\epsfig{file=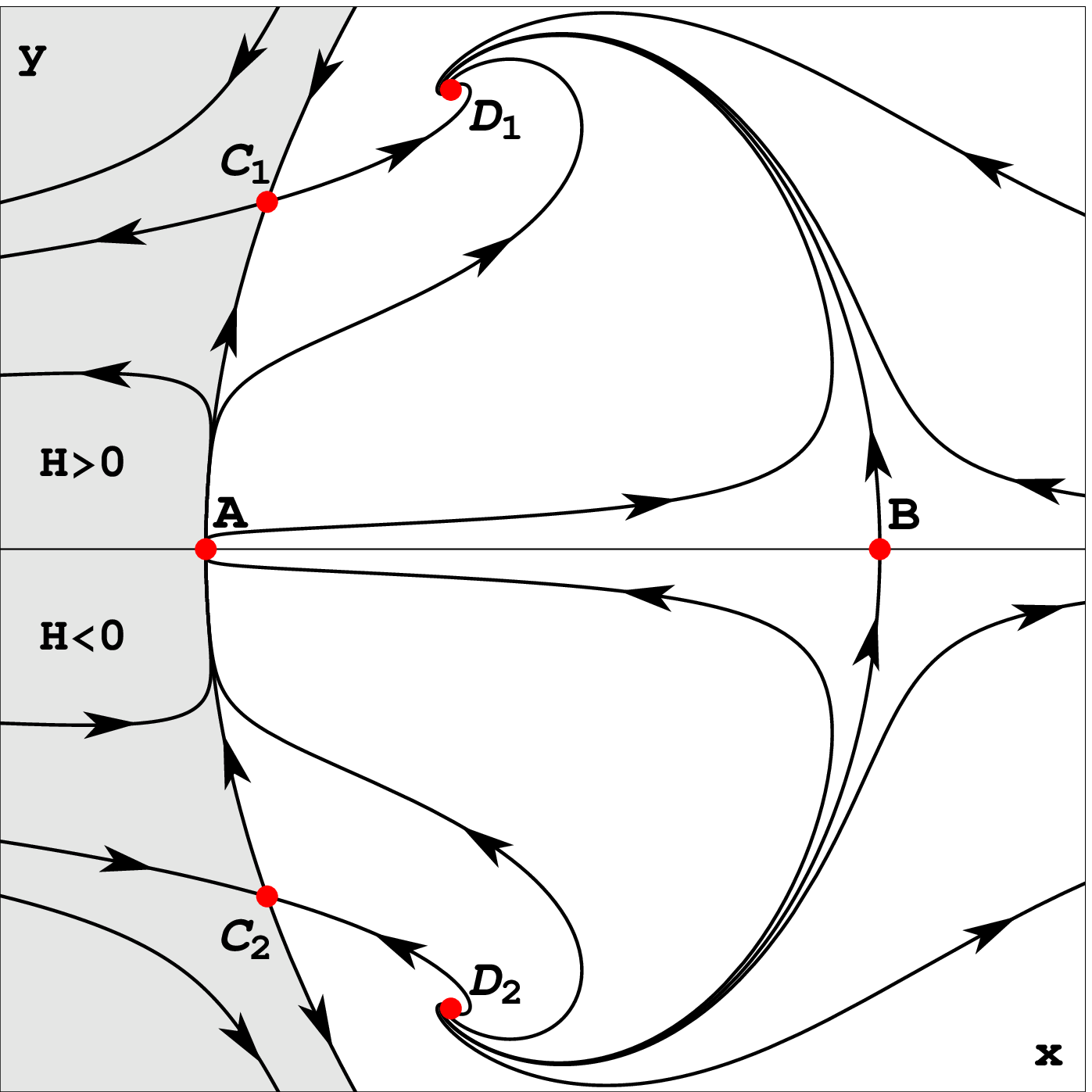,scale=0.5}
\caption{The phase plane diagrams for system \eqref{eq:sysquad} in the so-called ``O'Hanlon theory'' $\om=0$ for various matter content: $\frac{1}{15}(5-8\sqrt{10})<w_{m}<-1$ (left) , $w_{m}<\frac{1}{15}(5-8\sqrt{10})$ (right).}
\label{fig:11}
\end{center}
\end{figure}

In figures \ref{fig:1} and \ref{fig:2} we presented the phase space diagrams
for the model filled with dust matter $w_{m}=0$ and all possible values of the
BD parameter $\om$. In figure \ref{fig:3} the dynamics of the model filled with
radiation is presented. Dynamics of the model with the cosmological constant is
visualized in figures \ref{fig:4} and \ref{fig:5}. Figures \ref{fig:6},
\ref{fig:7} and \ref{fig:8} present the structure of the phase space for a
low-energy string theory limit $\om=-1$ for all possible values of the
barotropic matter equation of state parameter, while figures \ref{fig:9},
\ref{fig:10} and \ref{fig:11} give structure of the phase space for the
so-called ``O'Hanlon theory'' with $\om=0$.

All the phase space diagrams represent two types of evolution. In upper half of
the phase space the trajectories of expanding models are located, while in
lower half contracting. Direction of the arrows indicates direction of the
cosmological time $t$.

In all diagrams the phase space is divided into two regions, physical (white)
one and nonphysical one (shaded). The condition for physical region is given by
\eqref{eq:concond}
\begin{equation}
\Omega_{m} = \frac{8\pi\rho_{m}}{3\phi H^{2}} = 1+x-\frac{\om}{6}x^{2}-y^{2} > 0\,.
\end{equation}
The trajectories located at the border, i.e.~where $\Omega_{m}=0$, are the
trajectories of the vacuum models. Note that there is a special case when the
meaning of the regions can be reversed. In the case when $\phi<0$ and in scalar
field potential function $V_{0}<0$, the physical interpretation of the regions
is reversed. It is very interesting that in this case interpretation of the
scalar field as the effective gravitational coupling constant leads to
anti-gravity and de Sitter state is still a stable node type critical point. 

For example, on the first diagram in figure \ref{fig:3} we presented evolutional
paths of the model filled with radiation $w_{m}=1/3$ and for positive value of
Brans-Dicke parameter $\om>0$. The white region in the physical one where
$\phi>0$ and $\Omega_{m}>0$. The paths for contracting universes start their
evolution at the critical point $C_{2}$ corresponding to anti-de Sitter state
and end at the critical points $A_{1}$ or $A_{2}$ for vanishing scale factor.
Carefully chosen initial conditions lead to the critical point $B$. This
critical point represents a saddle type critical point and single evolutional
path lead form $C_{2}$ to it. For arbitrary initial conditions in the case of
initially expanding universes the final state is the de Sitter state at the
critical point $C_{1}$. 

Most of the phase space diagrams are captioned with some range of the model
parameters. This indicates that all the phase space diagrams in given range are
topologically equivalent. Some diagrams are plotted for specific values of the
model parameters to visualize the notion of structural instability, when two or
more critical points merge leading to non-hyperbolic critical point.

\subsection{Variability of \boldmath{$G_{\text{eff}}(\phi)$} and its asymptotic values}
\label{subsec:singularities}

Using the action integral for the Brans-Dicke theory \eqref{eq:action} the field $\phi$ can be identified with the inverse of the effective gravitational coupling
\begin{equation}
G_{\text{eff}}(\phi)=\frac{1}{\phi}
\end{equation}
which is a function of the spacetime location.

Using the linearized solutions in the vicinity of the critical points we are able to find the linearized solutions for the field $\phi$ and then the asymptotic values of the effective gravitational coupling at the critical points.

From the definition of the phase space variable $x$ we have that
\begin{equation}
x(\tau) = \frac{\ud \ln{\phi}}{\ud \tau}
\end{equation}
where $\frac{\ud}{\ud \tau}=\frac{\ud}{\ud\ln{a}}$. From dynamical system analysis we have obtained the linearized solutions of the system in the vicinity of the critical points, which can be used to construct the linearized solutions for the field $\phi$. Integrating latter equation with known linearized solution one obtains 
\begin{equation}
\label{eq:phi_tau}
\frac{\phi(\tau)}{\phi(0)} = \exp{\Big(\int_{0}^{\tau}x(\tau')\ud\tau'\Big)}
\end{equation}
behavior of the field $\phi$.

In the vicinity of the critical points: $x^{*}_{1}=\frac{3\mp\sqrt{3(3+2\om)}}{\om}$, $y^{*}_{1}=0$, denoted on the phase space diagrams as $A_{1,2}$ accordingly, from \eqref{eq:lin_1a} we have
\begin{equation}
x_{1}(\tau) = x^{*}_{1}+\Delta x \exp{\Big(\big(3(1-w_{m})+x^{*}_{1}\big)\tau\Big)}\,.
\end{equation}
Inserting this linearized solution into the equation \eqref{eq:phi_tau} one obtains 
\begin{equation}
\frac{\phi_{1}(a)}{\phi_{1}(a^{(i)}_{1})} \approx \exp{\Bigg(x^{*}_{1}\ln{\bigg(\frac{a}{a^{(i)}_{1}}\bigg)}-\frac{\Delta x}{3(1-w_{m})+x^{*}_{1}}\Bigg(1-\bigg(\frac{a}{a^{(i)}_{1}}\bigg)^{3(1-w_{m})+x^{*}_{1}}\Bigg)\Bigg)}\,,
\end{equation}
and resulting behavior of the field $\phi$ up to linear terms in $|\Delta x|\ll 1$ is
\begin{equation}
\frac{\phi_{1}(a)}{\phi_{1}(a^{(i)}_{1})} \approx \bigg(\frac{a}{a^{(i)}_{1}}\bigg)^{x^{*}_{1}}
\left(1-\frac{\Delta x}{3(1-w_{m})+x^{*}_{1}}\Bigg(1-\bigg(\frac{a}{a^{(i)}_{1}}\bigg)^{3(1-w_{m})+x^{*}_{1}}\Bigg)\right)\,.
\end{equation}
The critical point $A_{1}$ can be either an unstable node or a saddle type critical point for different values of $\om$ and $w_{m}$ parameters. The asymptotic value of the field $\phi$ can be calculated only for the unstable node critical point. Taking limit one obtains
\begin{equation}
\lim_{a\to0}\frac{\phi_{1}(a)}{\phi_{1}(a^{(i)}_{1})} = +\infty\,,
\end{equation}
which indicates that the effective gravitational coupling constant vanishes at this point
\begin{equation}
G_{\text{eff}}(\phi) \to 0\,.
\end{equation}

Different situation takes place for the second critical point $A_{2}$ which can be of any type for different values of the model parameters. For the critical point of an unstable node type (when both eigenvalues are positive) we obtain
\begin{equation}
\lim_{a\to0}\frac{\phi_{1}(a)}{\phi_{1}(a^{(i)}_{1})} = 0\,,
\end{equation}
while for a stable node (where both eigenvalues are negative) we have
\begin{equation}
\lim_{a\to\infty}\frac{\phi_{1}(a)}{\phi_{1}(a^{(i)}_{1})} = 0\,,
\end{equation}
which indicates that in both asymptotic states the effective gravitational coupling constant tends to infinity
\begin{equation}
G_{\text{eff}}(\phi) \to \infty\,.
\end{equation}

In the vicinity of the critical point $x^{*}_{2}=\frac{1-3w_{m}}{1+\om(1-w_{m})}$, $y^{*}_{2}=0$, denoted on the phase space diagrams as $B$ we have
\begin{equation}
x_{2}(\tau) = x^{*}_{2}+\Delta x \exp{\Big(\big(-\frac{3}{2}(1-w_{m})-\frac{1}{2}x^{*}\big)\tau\Big)}\,.
\end{equation}
Inserting this solution in to equation \eqref{eq:phi_tau} one obtains
\begin{equation}
\frac{\phi_{2}(a)}{\phi_{2}(a^{(i)}_{2})} \approx \exp{\Bigg(x^{*}_{2}\ln{\bigg(\frac{a}{a^{(i)}_{2}}\bigg)} - \frac{\Delta x}{-\frac{3}{2}(1-w_{m})-\frac{1}{2}x^{*}_{2}}\Bigg(1-\bigg(\frac{a}{a^{(i)}_{2}}\bigg)^{-\frac{3}{2}(1-w_{m})-\frac{1}{2}x^{*}_{2}}\Bigg)\Bigg)}
\end{equation}
and dynamics of the field $\phi$ up to linear terms in $|\Delta x|\ll1$ is described by
\begin{equation}
\frac{\phi_{2}(a)}{\phi_{2}(a^{(i)}_{2})} \approx \bigg(\frac{a}{a^{(i)}_{1}}\bigg)^{x^{*}_{2}}
\left(1-\frac{\Delta x}{-\frac{3}{2}(1-w_{m})-\frac{1}{2}x^{*}_{2}}\Bigg(1-\bigg(\frac{a}{a^{(i)}_{1}}\bigg)^{-\frac{3}{2}(1-w_{m})-\frac{1}{2}x^{*}_{2}}\Bigg)\right)\,.
\end{equation}
Inspection of dynamics in the vicinity of this critical point indicates that in the case of an unstable node type critical point (both eigenvalues are positive $\lambda_{1}>0$ and $\lambda_{2}>0$) we obtain two different types of behavior of the effective gravitational coupling constant. For model parameters leading to $x^{*}_{2}>0$ we have
\begin{equation}
\lim_{a\to0}\frac{\phi_{2}(a)}{\phi_{2}(a^{(i)}_{2})} = 0\,,
\end{equation}
indicating that
\begin{equation}
G_{\text{eff}}(\phi) \to \infty\,.
\end{equation}
While in the case when $x^{*}_{2}<0$ we have
\begin{equation}
\lim_{a\to0}\frac{\phi_{2}(a)}{\phi_{2}(a^{(i)}_{2})} = \infty\,,
\end{equation}
indicating that
\begin{equation}
G_{\text{eff}}(\phi) \to 0\,.
\end{equation}
The situation is reversed when we have a stable node critical point (both eigenvalues are negative $\lambda_{1}<0$ and $\lambda_{2}<0$). When $x^{*}_{2}<0$ we obtain
\begin{equation}
\lim_{a\to\infty}\frac{\phi_{2}(a)}{\phi_{2}(a^{(i)}_{2})} = 0\,,
\end{equation}
indicating that
\begin{equation}
G_{\text{eff}}(\phi) \to \infty\,.
\end{equation}
On the other hand, while $x^{*}_{2}>0$ but $\lambda_{1}+x^{*}_{2}<0$ we obtain
\begin{equation}
\lim_{a\to\infty}\frac{\phi_{2}(a)}{\phi_{2}(a^{(i)}_{2})} = \infty\,,
\end{equation}
indicating that
\begin{equation}
G_{\text{eff}}(\phi) \to 0\,.
\end{equation}

In the vicinity of the critical point corresponding to the de Sitter expansion (contraction):
$x^{*}_{3}=0$, $y^{*}_{3}=\pm1$, denoted on the phase space diagrams as $C_{1}$ and $C_{2}$ accordingly, we have
\begin{equation}
\begin{split}
\frac{\phi_{3}(a)}{\phi_{3}(a^{(i)}_{3})} \approx 
\exp\Bigg(&\frac{1}{3}\Delta x + \frac{1-3w_{m}}{3(1+w_{m})(3+2\om)}\Omega_{m,i} - \\&- \frac{1}{3}x_{3}(a) -\frac{1-3w_{m}}{3(1+w_{m})(3+2\om)}\Omega_{m,i}\bigg(\frac{a}{a^{(i)}_{3}}\bigg)^{-3(1+w_{m})}\Bigg)\,,
\end{split}
\end{equation}
where $x_{3}(a)$ is linearized solution \eqref{eq:dS_a} in the vicinity of critical point corresponding to de Sitter type evolution.

In the case of stable node critical point we can calculate an asymptotic value of the scalar field $\phi$ at de Sitter state
\begin{equation}
\lim_{a\to\infty}\frac{\phi_{3}(a)}{\phi_{3}(a^{(i)}_{3})} = \exp{\Bigg(\frac{1}{3}\Delta x + \frac{1-3w_{m}}{3(1+w_{m})(3+2\om)}\Omega_{m,i}\Bigg)}\,,
\end{equation}
which indicates the positive constant value of the field and hence the positive constant value of the effective gravitational coupling constant.

Initial conditions in linearized solutions are arbitrary chosen in the vicinity of critical point. One can choose as initial conditions the present values (assuming that at present the evolution of universe is close to de Sitter state). Then the field $\phi$ takes the asymptotic value
\begin{equation}
\lim_{a\to\infty}\frac{\phi_{3}(a)}{\phi_{3}(a_{0})} = \exp{\bigg(\frac{1}{3}x(a_{0})+\frac{1-3w_{m}}{3(1+w_{m})(3+2\om)}\Omega_{m,0}\bigg)}\,,
\end{equation}
with respect to the present value $\phi_{3}(a_{0})$ and $\Omega_{m,0}$ is the present value of the barotropic matter density parameter. Remembering that the phase space variable $x$ is connected with the effective gravitational coupling
\begin{equation}
x(a_{0}) = \frac{\dot{\phi}}{H\phi}\bigg|_{0} = - \frac{\dot{G}}{H G}\bigg|_{0}\,,
\end{equation}
we obtain the value at the de Sitter state 
\begin{equation}
G_{dS} = G_{0} \exp{\Bigg(\frac{1}{3}\frac{\dot{G}}{H G}\bigg|_{0}-\frac{1-3w_{m}}{3(1+w_{m})(3+2\om)}\Omega_{m,0}\Bigg)}\,,
\end{equation}
where $G_{0}$ and $\Omega_{m,0}$ are the present values of the effective gravitational coupling and the barotropic matter density parameter, accordingly.

The last two critical points under considerations are : $x^{*}_{4}=-\frac{3}{2}(1+w_{m})$ , $y^{*}_{4}=\pm\sqrt{\frac{1}{8}\big(5-3w_{m}+3\om(1-w_{m}^{2})\big)}$, denoted on the phase space diagrams as $D_{1}$ and $D_{2}$, accordingly.

Using the linearized solution in the vicinity of these points \eqref{eq:lin_4a} and from \eqref{eq:phi_tau} we obtain the linearized solution for the scalar field
\begin{equation}
\begin{split}
\frac{\phi_{4}(a)}{\phi_{4}(a^{(i)}_{4})} \approx \left(\frac{a}{a^{(i)}_{4}}\right)^{-\frac{3}
{2}(1+w_{m})} 
\exp\Bigg( & \frac{A_{1}(\Delta x -A_{2}\Delta y)}{\lambda_{1}\det{P}}
\Bigg(\bigg(\frac{a}{a^{(i)}_{4}}\bigg)^{\lambda_{1}}-1\Bigg)  -\\ 
-&\frac{A_{2}(\Delta x -A_{1}\Delta y)}{\lambda_{2}\det{P}}\Bigg(\bigg(\frac{a}{a^{(i)}_{4}}
\bigg)^{\lambda_{2}}-1\Bigg)\Bigg)\,.
\end{split}
\end{equation}
As in the previous discussion concerning asymptotic value of the Hubble function \eqref{eq:hubble_4} we have that the critical point under consideration is stable (the real parts of the eigenvalues $\lambda_{1}$ and $\lambda_{2}$ are negative) only for the phantom matter $w_{m}<-1$. Thus we obtain the asymptotic value of the scalar field at the critical point
\begin{equation}
\lim_{a\to\infty}\frac{\phi_{4}(a)}{\phi_{4}(a^{(i)}_{4})} = \infty\,,
\end{equation}
indicating that the effective gravitational coupling vanishes
\begin{equation}
G_{\text{eff}}(\phi) \to 0\,,
\end{equation}
asymptotically at the critical points under considerations.

\subsection{Analysis at infinity}

In order to obtain complete information about the dynamical behavior of the dynamical system \eqref{eq:sysquad} we need to perform the dynamical analysis at the infinity of the phase space. 

To investigate dynamics of the system \eqref{eq:sysquad} at infinity we need to introduce two charts of projective variables 
\begin{equation}
\begin{split}
& U_{(1)}\quad : \quad u = \frac{1}{x}\,,\quad v=\frac{y}{x}\,,\\
& U_{(2)}\quad : \quad u = \frac{x}{y}\,, \quad v=\frac{1}{y}\,.
\end{split}
\end{equation}
The dynamical system in the chart $U_{(2)}$ gives no critical points thus indicating no asymptotic states at infinity. We will be interested only in the dynamics in the chart $U_{(1)}$. 

The system \eqref{eq:sysquad} in the chart $U_{(1)}$ takes the following form
\begin{equation}
\label{eq:sysquadcomp}
\begin{split}
\frac{\ud u}{\ud\eta} & = 3\left( u^{2}+u-\frac{\om}{6} -\left(u^{2}+u-\frac{\om}{6}-v^{2}\right)\frac{2+\om(1+w_{m})}{3+2\om}\right) - \\ & \quad -3 u\left(u^{2}+u-\frac{\om}{6}- v^{2}\right)\frac{1-3w_{m}}{3+2\om}\,, \\
\frac{\ud v}{\ud\eta} & = 3v\left(\frac{1}{2}+u-\left(u^{2}+u-\frac{\om}{6}-v^{2}\right)\frac{1-3w_{m}}{3+2\om}\right)\,,
\end{split}
\end{equation}
where new 'time' $\eta$ is defined as
\begin{equation}
\label{eq:timecomp}
\frac{\ud}{\ud\eta}=u(\eta)\frac{\ud}{\ud\ln{a}}\,.
\end{equation}
The critical points of dynamical system are stationary states for which the right hand sides vanish. In the system under considerations we have only single critical point at infinity but for two different sets of the model parameters :
\begin{equation}
\begin{split}
I : & \quad u^{*} = 0\,, v^{*} = 0\,, \quad\text{for}\quad \om =0\,,\\
II : & \quad u^{*} = 0\,, v^{*} = 0 \,, \quad \text{for} \quad \om=-\frac{1}{1-w_{m}} \quad \text{and} \quad w_{m}\ne\frac{1}{3}\,.
\end{split}
\end{equation}
The eigenvalues of the linearization matrix calculated for both sets of parameters are: $I$ : $\lambda_{1}=1$ , $\lambda_{2}=\frac{3}{2}$  and $II$ : $\lambda_{1}=-\frac{1}{2}$, $\lambda_{2}=1$ indicating that the critical point under consideration in the first case is an unstable node in `time' $\eta$ while in the second case is a saddle type critical point.

\begin{figure}
\begin{center}
\includegraphics[scale=0.65]{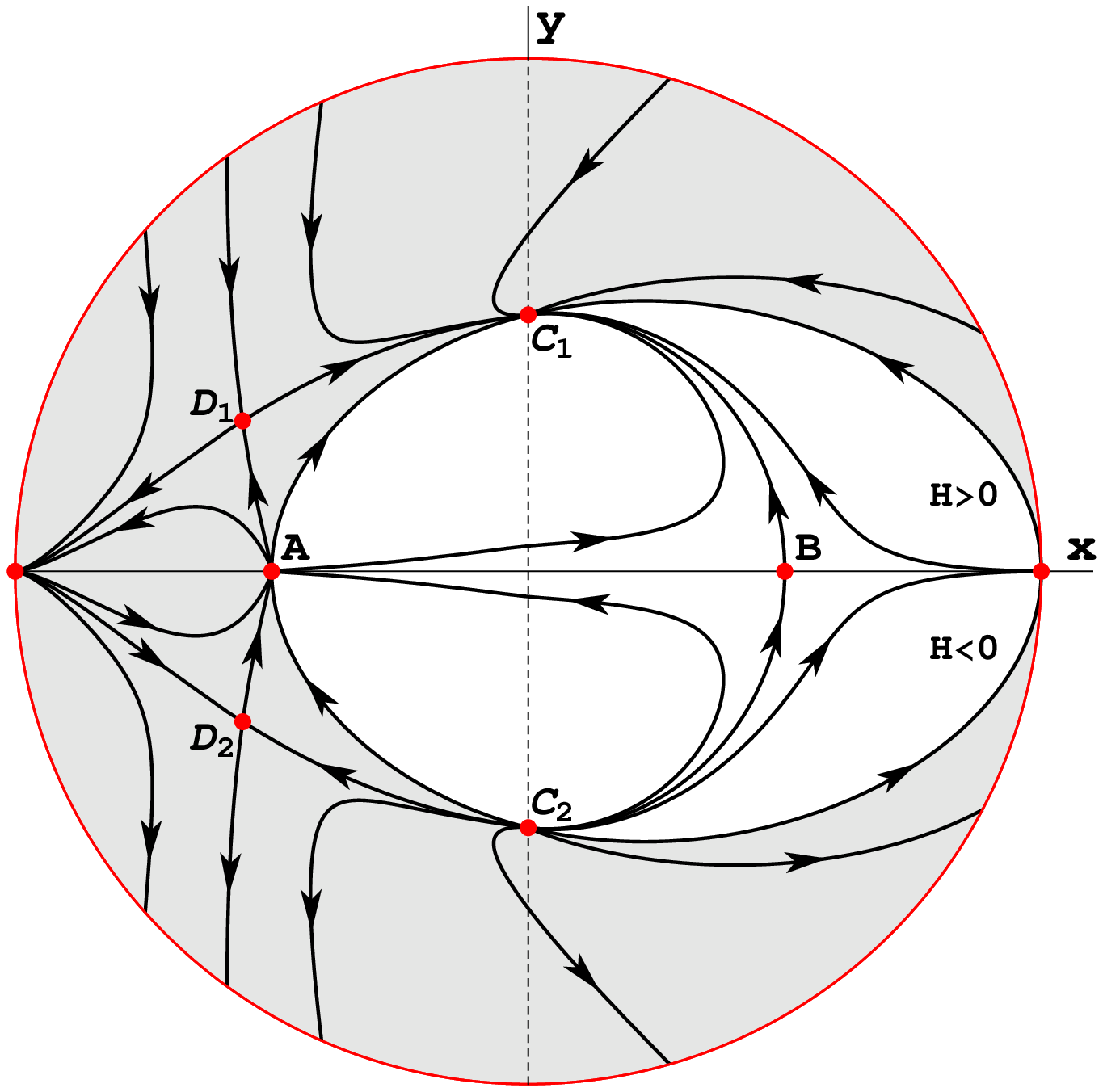}
\includegraphics[scale=0.65]{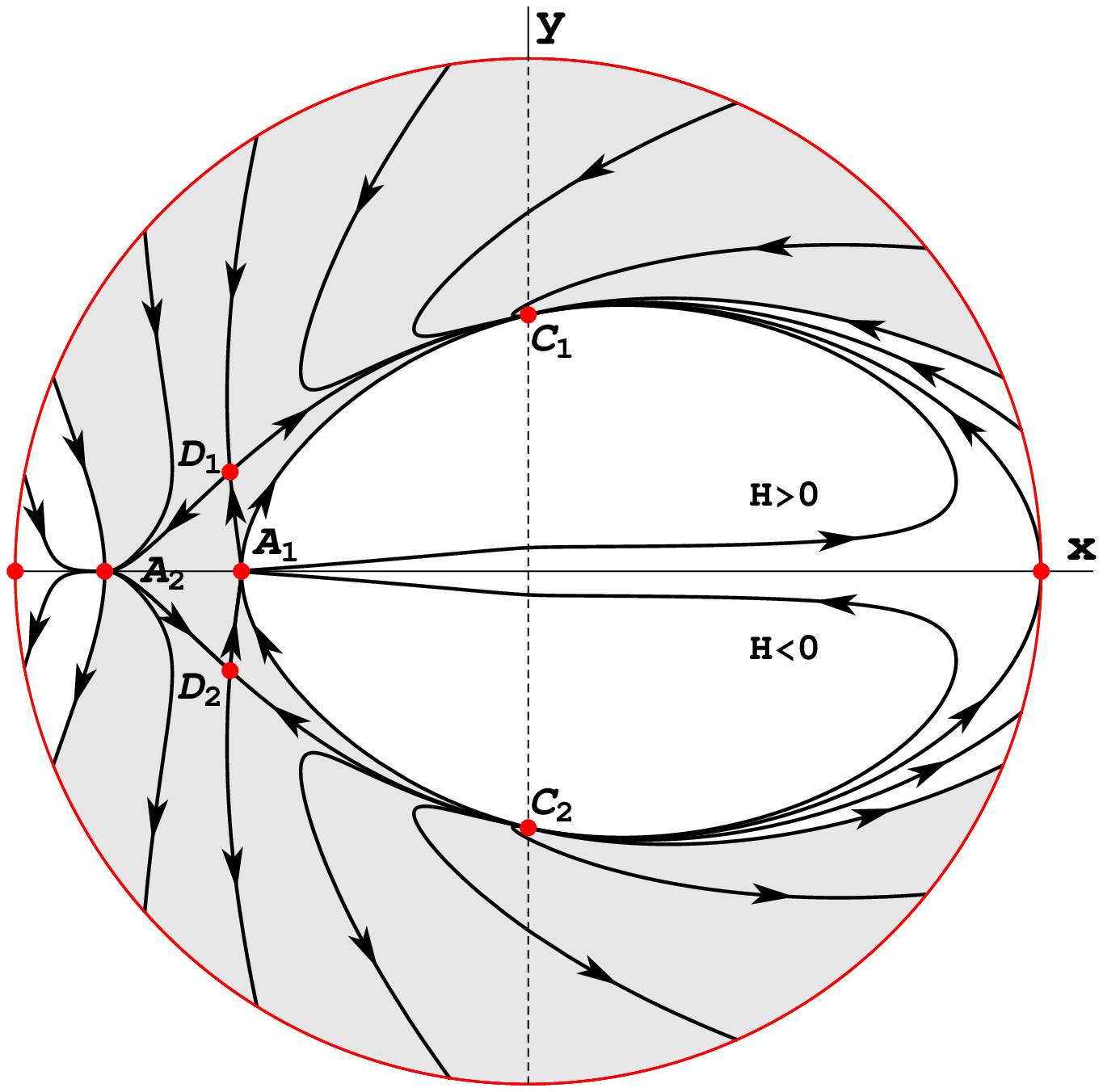}
\caption{Diagrams of the evolutional paths in the phase space compactified with circle at infinity for the model filled with dust matter $w_{m}=0$ and : $\om=0$ (upper) , $\om=-1$ (lower). In both cases we have two critical points at infinity $x\to\pm\infty$ , $y\to0$. In the first case, one is stable during expansion and the second one is unstable during expansion. In the second case the critical points are of a saddle type.}
\label{fig:12}
\end{center}
\end{figure}

\begin{figure}
\begin{center}
\includegraphics[scale=0.65]{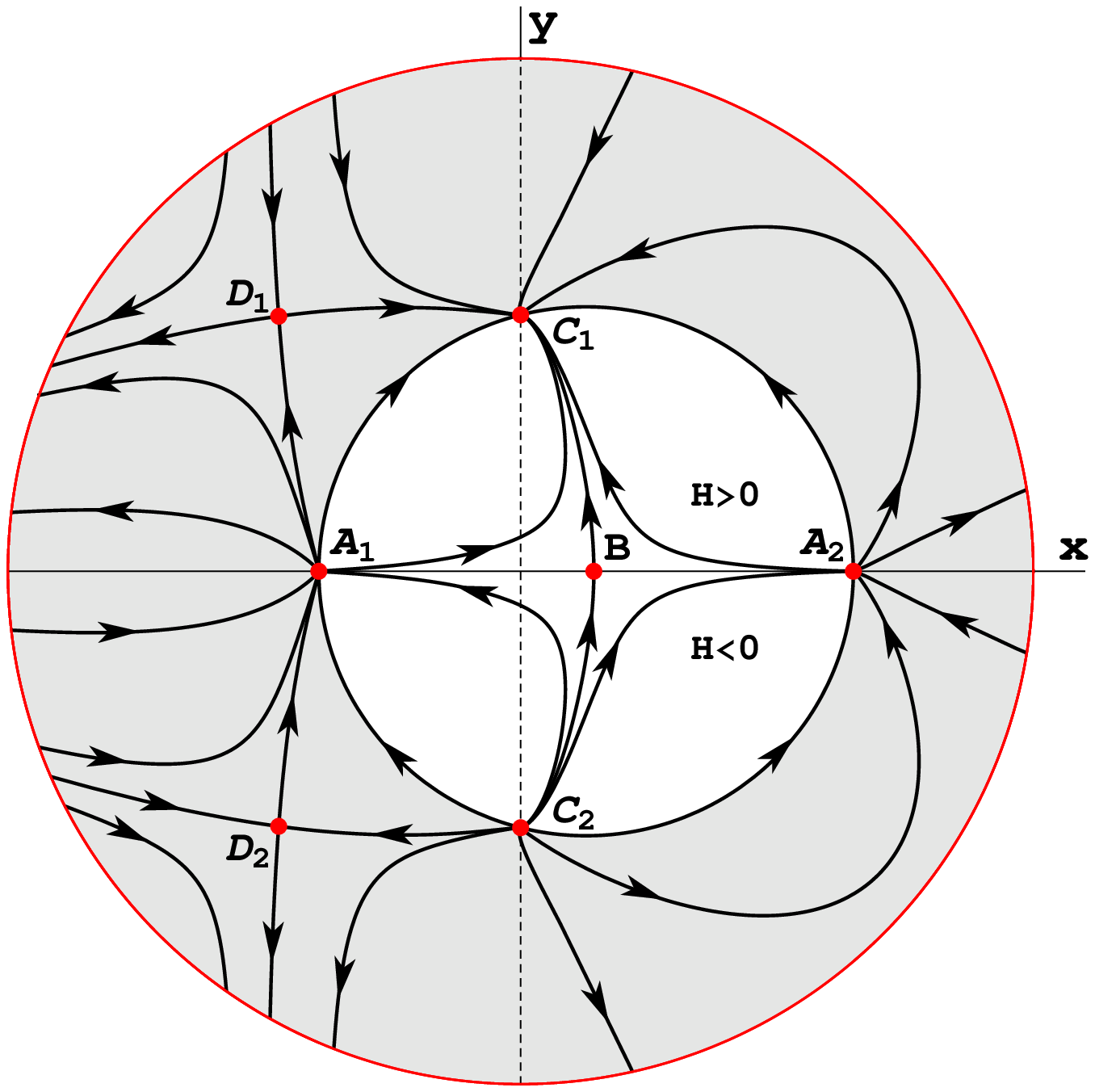}
\includegraphics[scale=0.65]{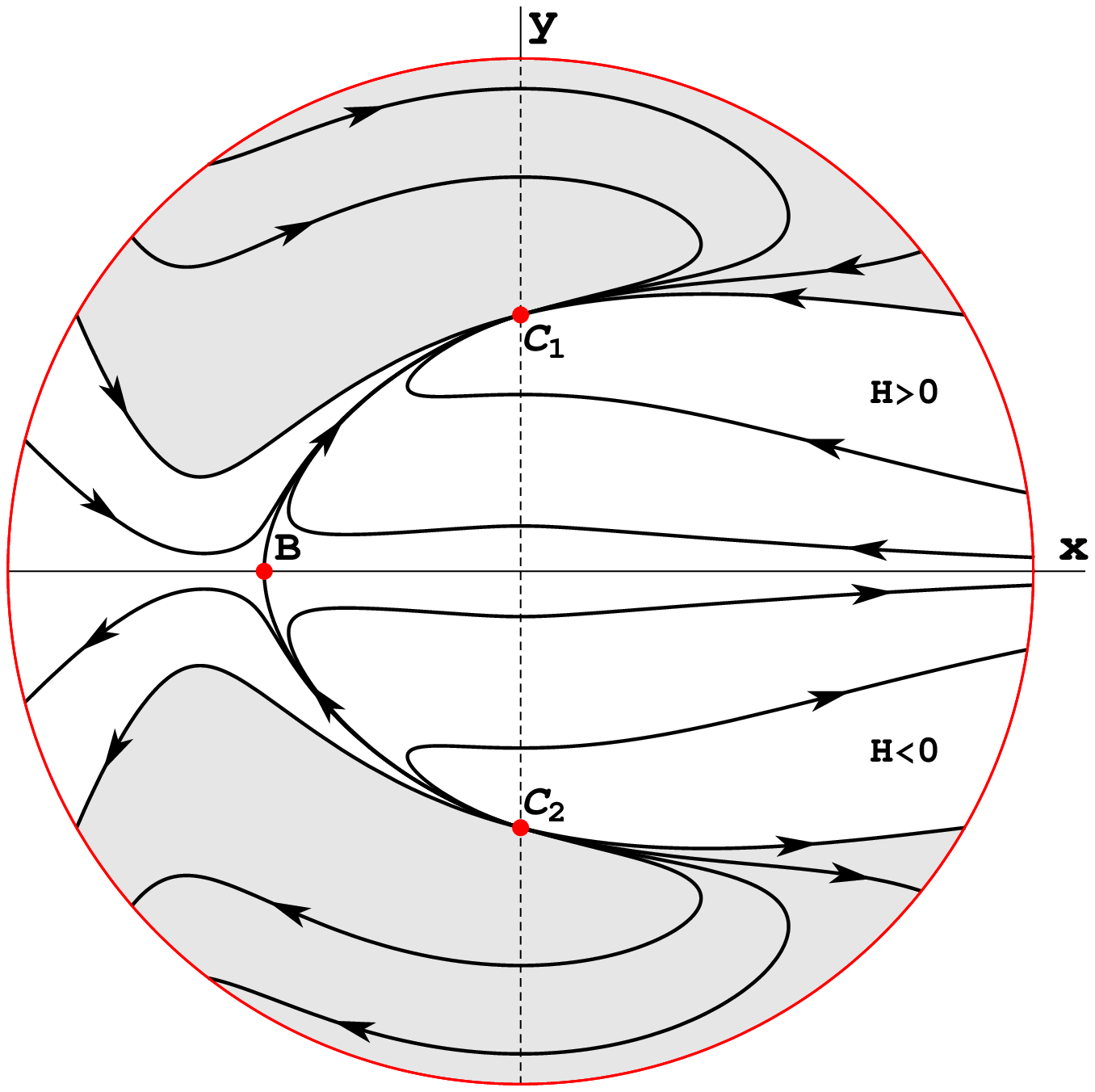}
\caption{Diagrams of the evolutional paths in phase space compactified with circle at infinity for the model filled with dust matter $w_{m}=0$ and : $\om>0$ (upper) , $\om<-5/3$ (lower). Diagrams plotted for fixed values ($\om=5$ and $\om=-2$), all phase space diagram in given range are topologically equivalent. In these cases we do not have any critical points at infinity. The circle at infinity consists of bounces during the evolution of the universe.}
\label{fig:13}
\end{center}
\end{figure}

In figure \ref{fig:12} we present the phase space diagrams equipped with a circle at infinity for the two sets of model parameters giving rise to two possible types of the critical points at infinity. The general cases are presented in figure \ref{fig:13}. Direction of the arrows indicates direction of the cosmological time.

For the model parameters giving rise to an unstable node critical point at infinity, the linearized solutions in the vicinity of the critical point are
\begin{equation}
\label{eq:solcomp}
\begin{split}
u(\eta) & = \Delta u \exp{(\lambda_{1}\eta)}\,,\\
v(\eta) & = \Delta v \exp{(\lambda_{2}\eta)}\,,
\end{split}
\end{equation}
where $\lambda_{1}=1$ and $\lambda_{2}=\frac{3}{2}$ are the eigenvalues for the linearization matrix calculated at the critical point for the conditions $I$.

Taking the time transformation \eqref{eq:timecomp} one can show that the critical point under considerations is achievable for the infinite value of `time' $\eta$ but for a finite value of the scale factor. We have
\begin{equation}
\int_{a^{(i)}}^{a^{*}} \ud\ln{a} = \Delta u \int_{0}^{-\infty}\exp{(\eta)}\ud\eta = -\Delta u < \infty\,.,
\end{equation}
and finally, value of the scale factor at the critical point is
\begin{equation}
a^{*} = a^{(i)} \exp{(-\Delta u)}\,.
\end{equation}

The initial conditions $\Delta u$ and $\Delta v$ in solutions \eqref{eq:solcomp} are of the same order and from the definition of the compactified variables we obtain
\begin{equation}
y = \frac{v}{u} = \frac{\Delta v}{\Delta u} \exp{(\frac{1}{2}\eta)}\,
\end{equation} 
and the behavior of the Hubble function
\begin{equation}
H = \frac{\Delta u}{\Delta v}\sqrt{\frac{V_{0}}{3}\phi}\exp{(-\frac{1}{2}\eta)}\,.
\end{equation}
Next, from the `time' transformation and the definition of the variable $u$ we obtain 
\begin{equation}
\phi=\phi^{(i)}\exp{(\eta)}\,.
\end{equation}
Inserting this solution to the previous one we obtain that
\begin{equation}
H = \frac{\Delta u}{\Delta v}\sqrt{\frac{V_{0}}{3}\phi^{(i)}} = \text{const}\,,
\end{equation}
the Hubble function in the vicinity of the critical point at infinity of the phase space is constant in linear approximation giving rise to the de Sitter evolution. As we have shown that the critical point is reached for a finite value of the scale factor, this process can be understood similarly to rigid ball bounce off a wall where the role of wall is played by the energy conservation condition.

\begin{figure}
\begin{center}
\includegraphics[scale=0.71]{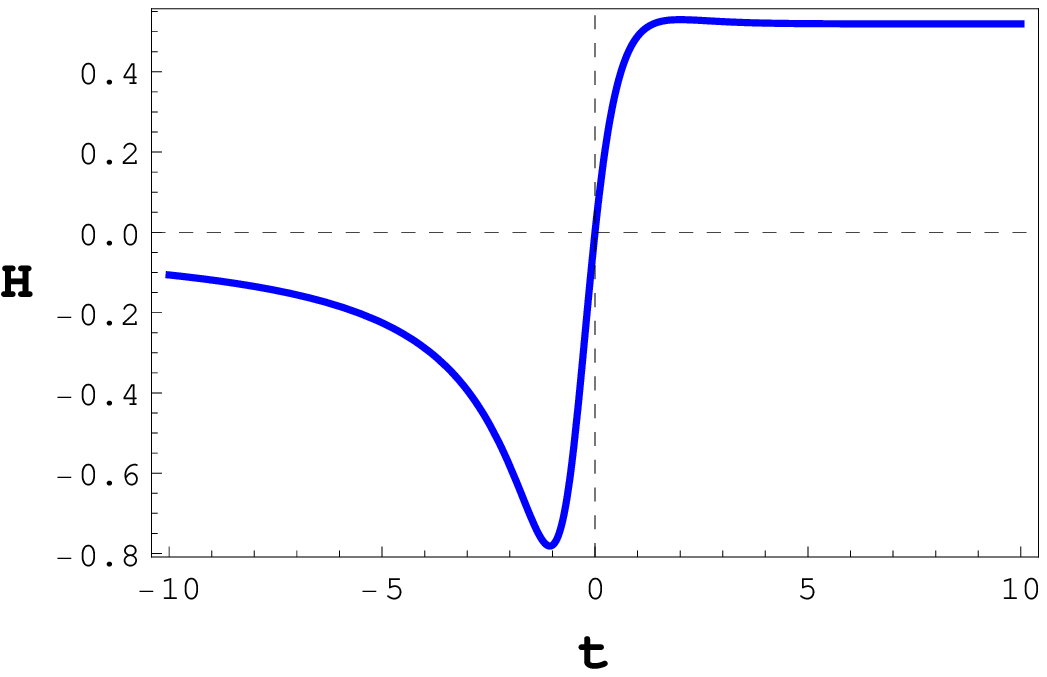}\hspace{3mm}
\includegraphics[scale=0.68]{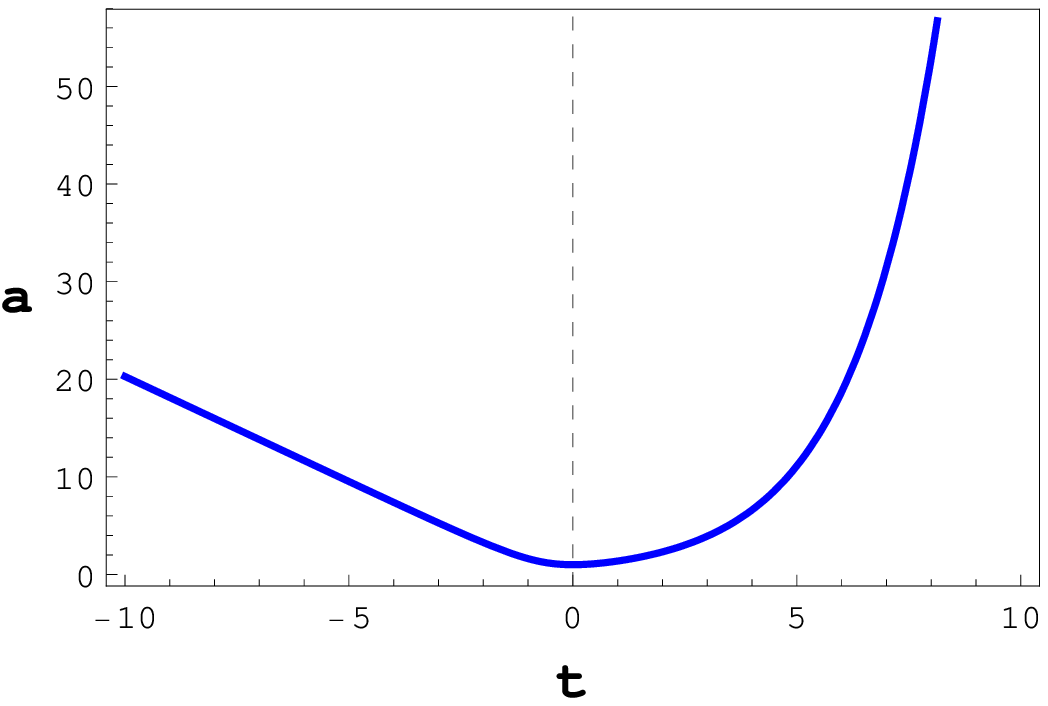}\\\hspace{-3mm}
\includegraphics[scale=0.73]{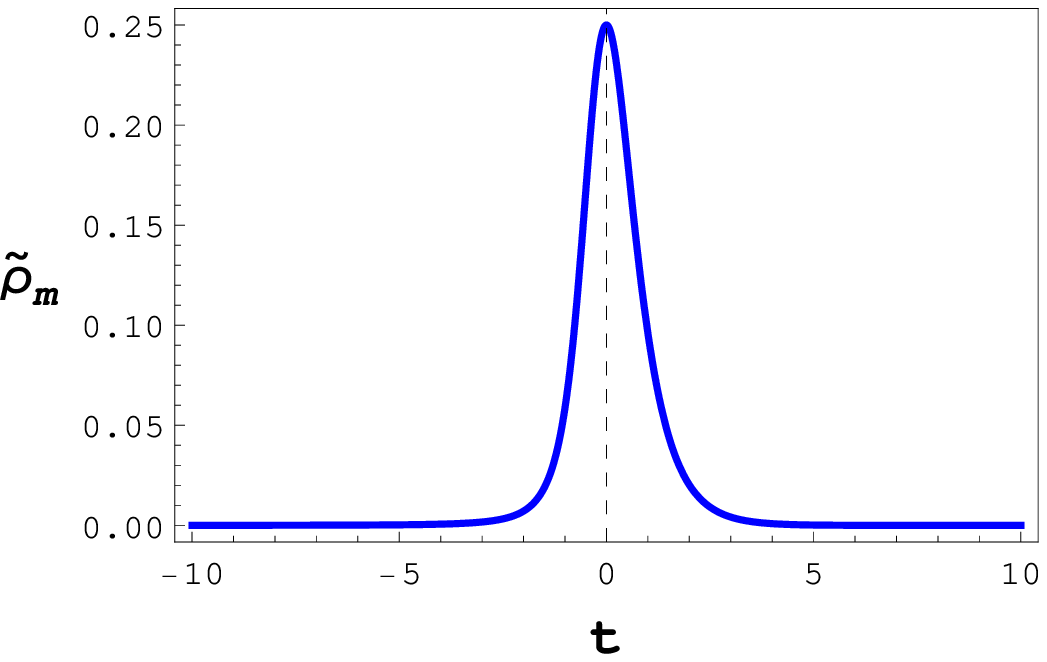}\hspace{2mm}
\includegraphics[scale=0.71]{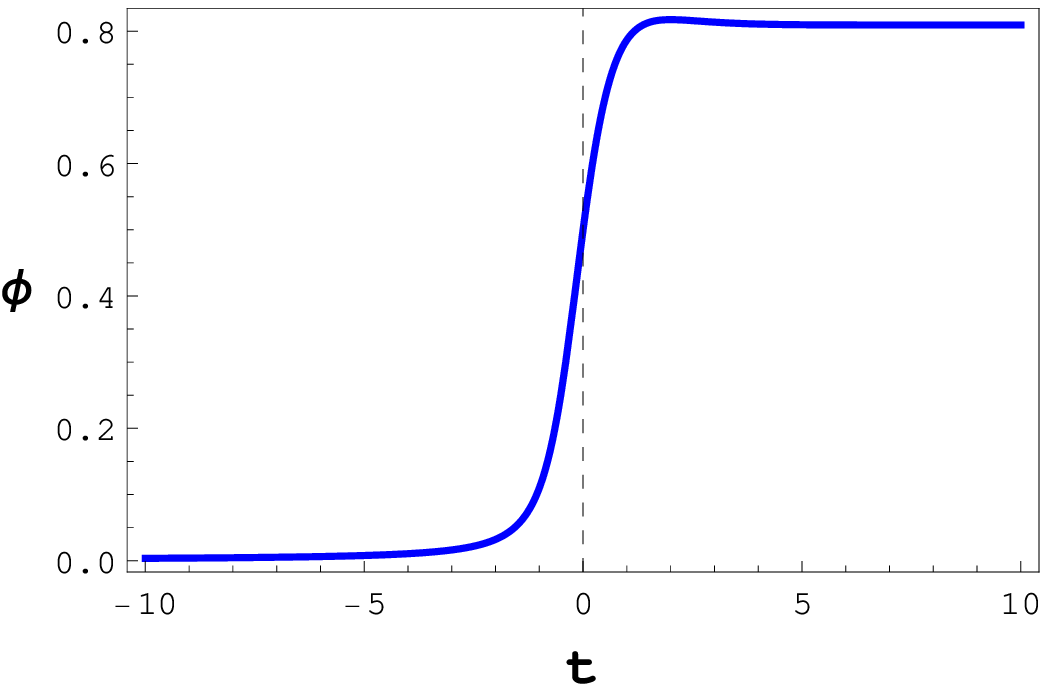}
\caption{Cosmological time evolution of the Hubble function $H$, the scale factor $a$, the barotropic matter density $\widetilde{\rho}_{m}=8\pi\rho_{m}$ and the scalar field $\phi$ for sample evolutional trajectory with a bounce and the model parameters $\om=-2$  and $w_{m}=0$. The initial conditions taken at the bounce are : $\phi_{(i)}=1/2$, $a_{(i)}=1$, $H_{(i)}=0$ and $\dot{\phi}_{(i)}=1/2$. The behavior of the scale factor is non-symmetric in time evolution. The Hubble function and the scalar field approach non zero constant value both in past and future. Note that in the past at anti-de Sitter state the value of scalar field is much smaller than in the future indicating that at anti-de Sitter state the effective gravitational coupling constant was much larger.}
\label{fig:14}
\end{center}
\end{figure}

Using compactified polar coordinates $(r,\theta)$ defined as
\begin{equation}
x=\frac{r}{1-r}\cos{\theta}\,,\quad y=\frac{r}{1-r}\sin{\theta}\,,
\end{equation}
one obtains that for an arbitrary set of model parameters the circle at infinity consists of critical points, i.e.~the stationary solutions are in the form $r=1$ and an arbitrary $\theta$. The only possibility for the phase space variables $(x,y)$ to acquire the infinite value while their ratio is still finite is through the vanishing Hubble function $H=0$. Thus the circle at infinity in the phase space $(x,y)$ consists of bounces during the evolution of universe. To demonstrate our claim we have plotted in figure \ref{fig:14} the cosmological time evolution of the Hubble function $H(t)$, the scale factor $a(t)$, the scalar field $\phi(t)$ and the barotropic matter density for the sample trajectory with the initial conditions taken at a bounce $H_{(i)}=0$.

\begin{figure}
\begin{center}
\includegraphics[scale=0.65]{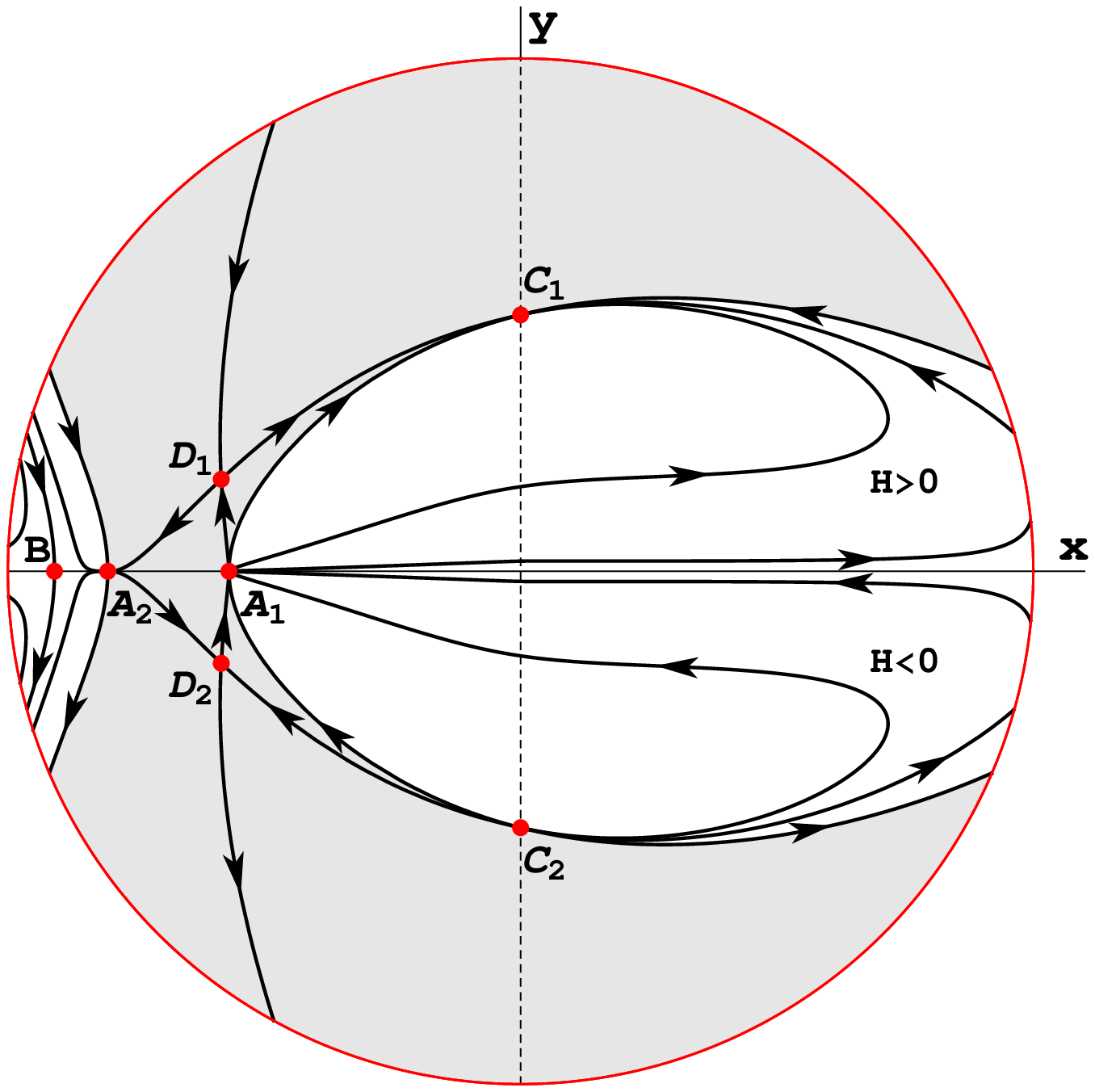}
\includegraphics[scale=0.65]{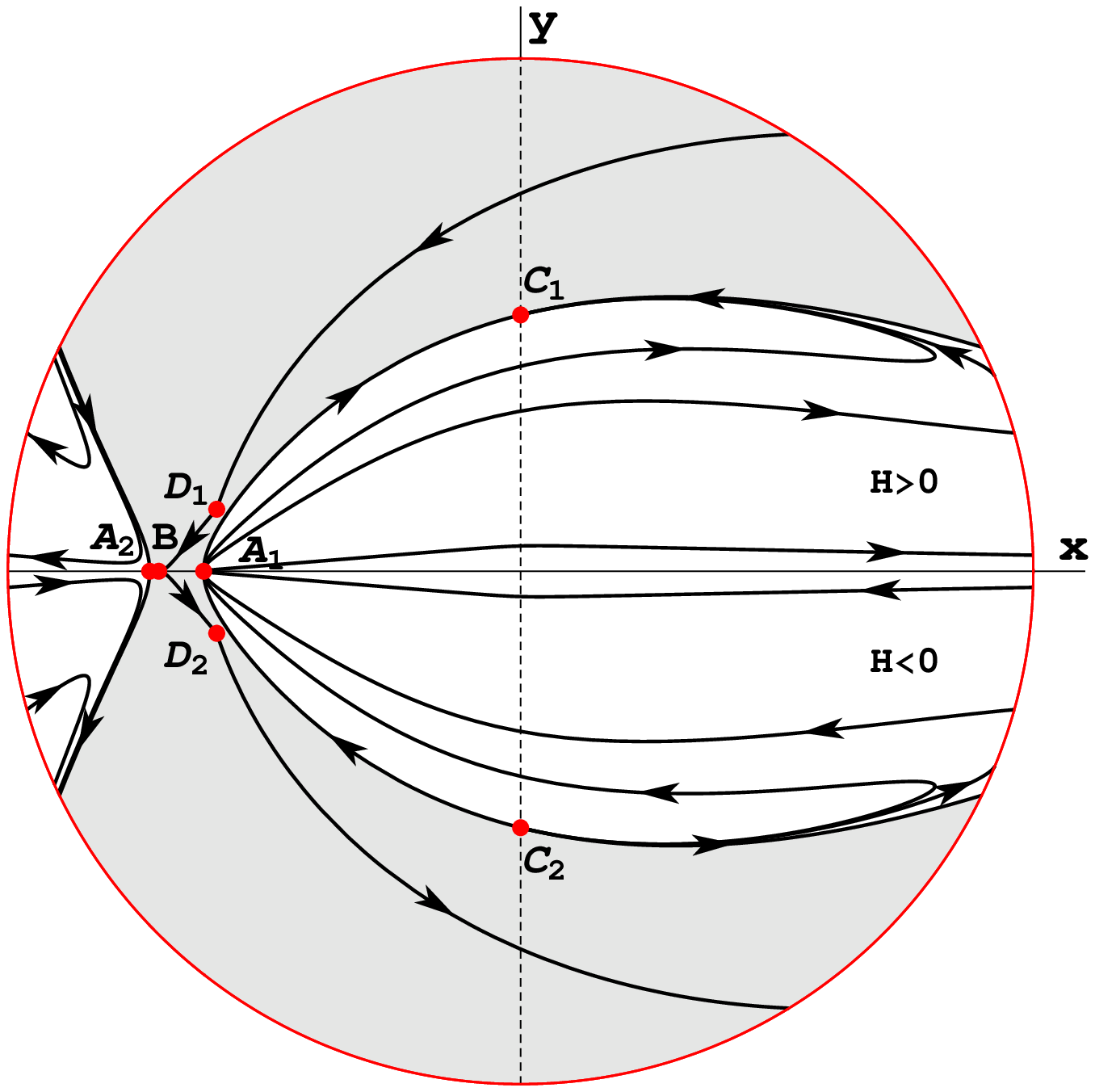}
\caption{Diagrams of the evolutional paths in phase space compactified with circle at infinity for the model filled with dust matter $w_{m}=0$ and : $-4/3<\om<-1$ (upper) , $-3/2<\om<-4/3$ (lower). The circle at infinity consists of bounces during the evolution of the universe. For clarity of the presentation we omitted trajectories in the nonphysical region (see the bottom left diagram on figure~\ref{fig:1} and the top left diagram on figure~\ref{fig:2}).}
\label{fig:15}
\end{center}
\end{figure}

\begin{figure}
\begin{center}
\includegraphics[scale=0.4]{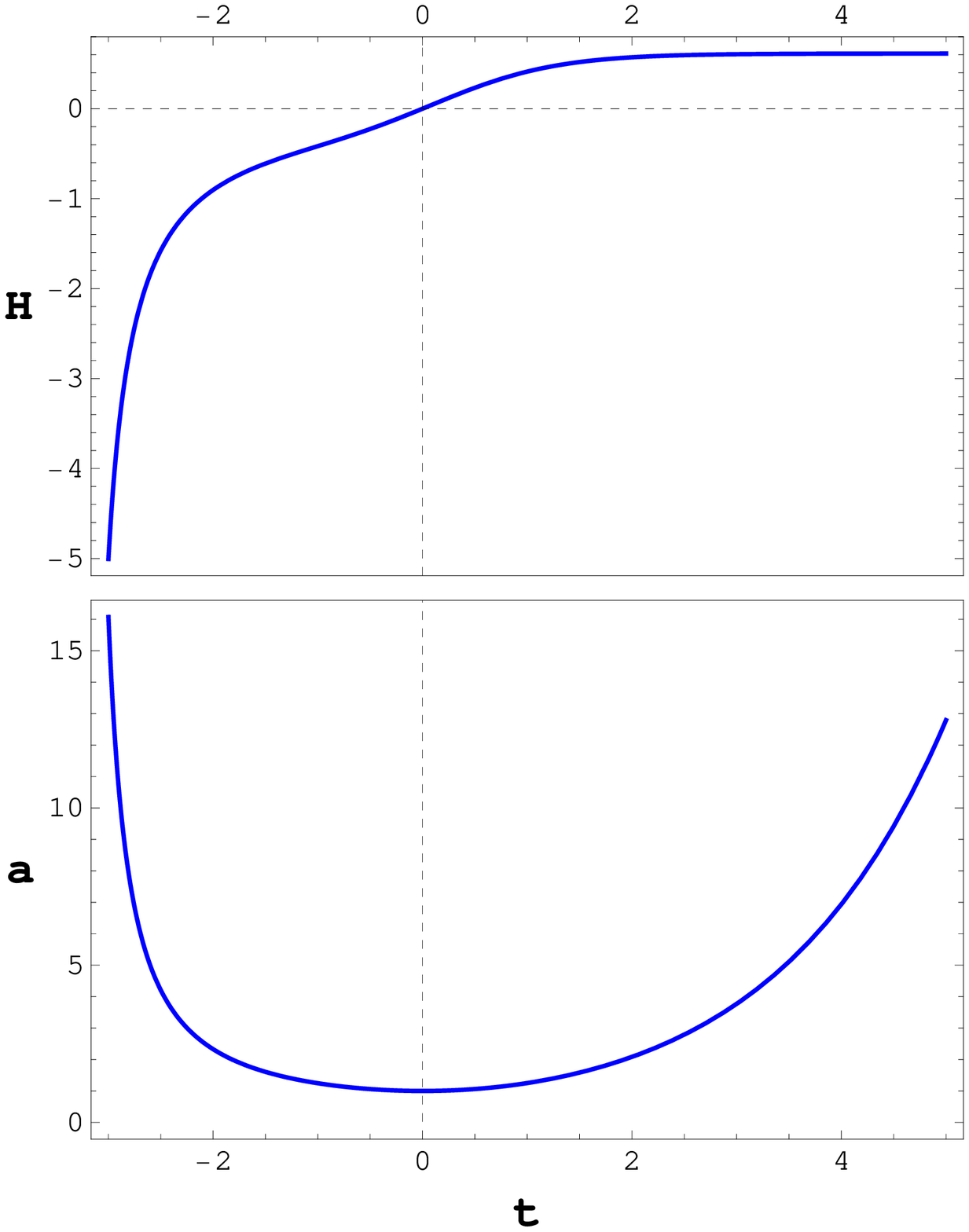}
\includegraphics[scale=0.4]{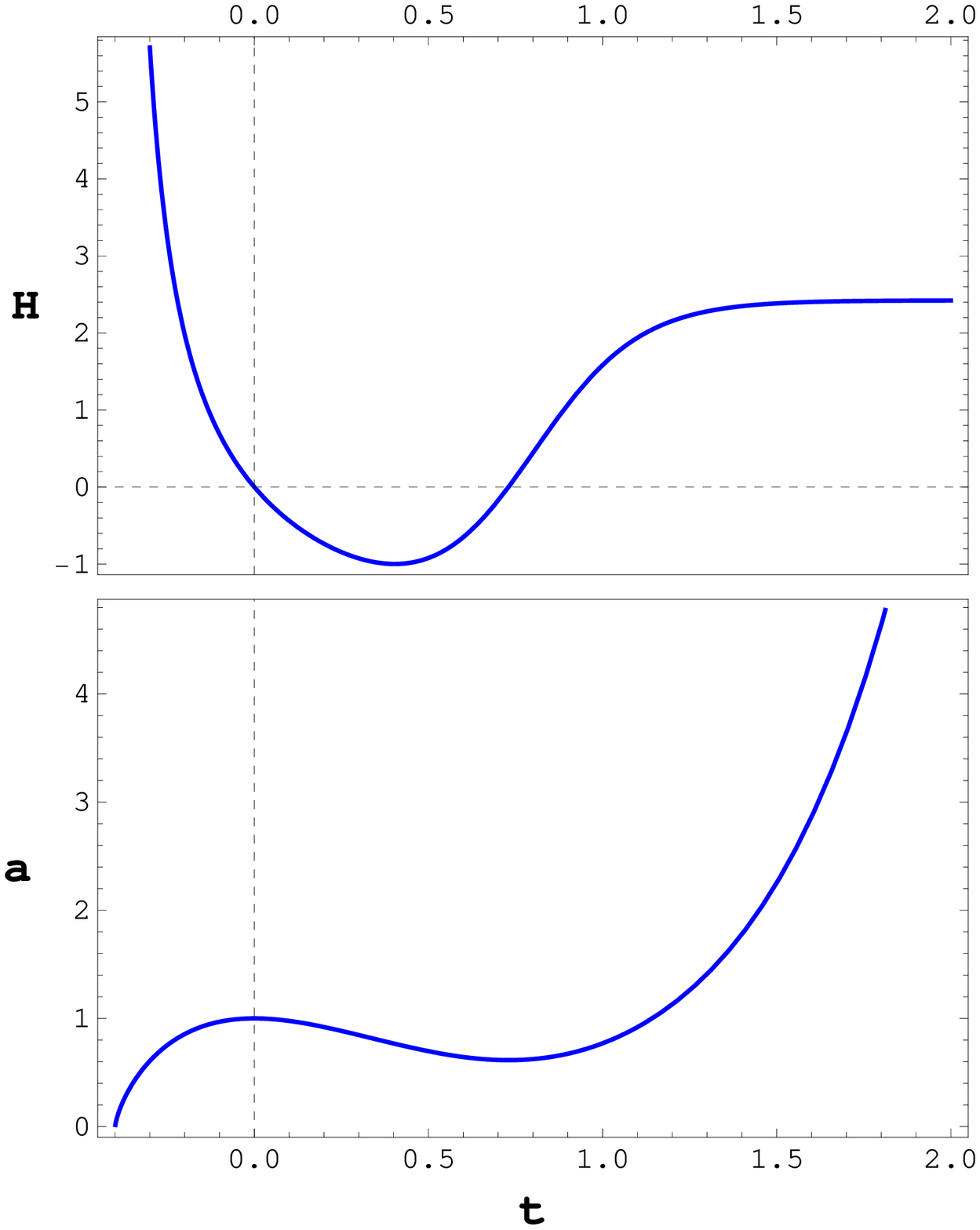}
\caption{Cosmological time evolution of the Hubble function $H$, the scale factor $a$ for sample evolutional trajectories with one bounce (left panel $-4/3<\om<-1$) and two bounces (right panel $-3/2<\om<-4/3$) in the model with dust matter $w_{m}=0$. The initial conditions taken at the bounce are : $H_{(i)}=0$, $\phi_{(i)}=1/2$ and $\dot{\phi}_{(i)}=1/2$ (left panel) and $\dot{\phi}_{(i)}=1$ (right panel). Note the coexistence of the bounces and the singularity in the second case.}
\label{fig:16}
\end{center}
\end{figure}

The bouncing solutions are admissible for $\om<0$ (see the lower diagram on figure \ref{fig:12}, figure \ref{fig:15} and \ref{fig:16}), while for $\om<-3/2$ there is possibility to obtain bouncing solutions connecting the anti-de Sitter state with the de Sitter state (see the lower diagram on figure \ref{fig:13}). However the Brans-Dicke theory contains a ghost for $\om<-3/2$ and, thus, it is nonphysical in this region of the parameter of the theory.

\section{Conclusions}
\label{sec:conclusions}

There are, in principle, two possible ways to explain the current accelerated expansion 
of the Universe. In the first approach, some substantial form of matter is
postulated and standard FRW equations are anticipated. In the second approach
some modifications of Einstein equations are postulated instead of substantial matter of
unknown form. One can call the first approach physical and the second one
geometrical. In this scheme the cosmological constant $\Lambda$ proposition is
located in the middle of both approaches. In our paper we study dynamics of the
Brans-Dicke cosmological models in the framework of dynamical systems theory. Following this theory,
we are looking for the global dynamics visualized in the phase space. In
our case we reduce the dynamics to a $2$--dimensional dynamical system. Then we find
the critical points and linearize the system around these singular solutions.
We obtain the phase diagrams on the plane. The advantage of such a visualization
of dynamics is that we obtain all evolutionary paths for all admissible initial
conditions. In the system under consideration, $\om$ and the coefficient of equation
of state for the barotropic matter play the role of bifurcation parameters. We
study these bifurcations in detail to show how the structure of the phase space
changes under the influence of these parameters.

Following the Hartman-Grobman theorem \cite{Perko:2001} a linear part of the
system is a good approximation in the vicinity of the critical point. From this
fact one can derive the $H^{2}(a)$ relation important in cosmography which
identifies the model with different form of dark energy. In this way on can
translate the geometrical approach to dark energy into the substantial
approach. In our case we have obtained, starting from the geometrical approach
to the dark energy problem, the corresponding substantial form which
can consequently be used in testing and selection of cosmological models by
astronomical data.

Inclusion of the substantial form of cosmological constant $w_{m}=-1$ leads to
structurally unstable dynamical system, because of the presence of the
degenerated critical point, which we treat as a exceptional one in the ensemble
of all. Therefore such models form a bifurcation set in the ensemble.

The main aim of the paper was to explore the phase plane of the Brans-Dicke
cosmology with a scalar field potential and barotropic matter. Investigation
shows that the structure of the phase plane changes under changing model
parameters (the parameter $\om$ and the equation of state parameter $w_{m}$).
We distinguish some representative cases of the phase diagrams. For the general
form of the scalar field potential function dynamics of the model, due to a suitable
choice of the state variables, can be reduced to a $3$--dimensional dynamical system which in
the special case of the quadratic potential function reduces to a $2$-dimensional dynamical system.
For this system, the phase space structure is studied in details. While the
structure of the phase plane depends on the intervals for the model parameters,
the generic feature of this phase plane is the presence of de Sitter
attractor. From linearization of the system around this state we obtain the
$H^{2}(a)$ relation in linear approximation, which almost corresponds to the
$\Lambda$CDM model. In this correspondence the equation of state coefficient
should be assumed arbitrary small but positive (dust matter is excluded).

Our consideration illustrates that nontrivial topology of the phase space can play important role in dynamical investigation of cosmology. The boundary of physical region of the phase space is formed by the trajectories representing vacuum solutions one can glue (with corresponding twist) opposite sides of this boundary. As a result we obtain a Moebius strip as a model of a regularized phase space.

Dark matter has an emergent character, i.e.~the corresponding term containing $\om$ in the
$H^2(a)$ relation is mimicking a dark matter. Acceleration of the Universe also
occurs in Brans-Dicke theory. The detailed analysis of the structure of phase plane
shows the appearance of three other critical points. Among them there is a
scaling solution important in the quintessence cosmology in the context of the
coincidence problem.

Our investigation of dynamics of the simplest Brans-Dicke cosmological models shows in general how rich evolutional scenarios can be. The phase space structure admits a scaling solution important in the context of the acceleration conundrum as well as new scenarios of very early evolution are predicted by the model itself. Especially we have shown that scaling solution are represented by a global attractor in the phase space. At this state both energy densities of the scalar field and matter becomes proportional which can explain why today they are comparable if they are considered in terms on density parameters.

We have shown that predicted by the model classical bounce is generic property and is stable under small changes of initial conditions in the phase space both in the theory with a ghost ($\om<-3/2$) and without a ghost ($-3/2<\om<0$). This type of evolution appears if the loop quantum effects are included \cite{Bojowald:2001xe}. In the Brans-Dicke cosmology it is purely classical effect. Moreover this bounce is asymmetric in cosmic time which can be used in description of direction of the time itself \cite{Zeh:book}. We discovered bouncing trajectories and their genericity after introducing of nontrivial topology of phase space (see figures \ref{fig:12}, \ref{fig:13} and \ref{fig:15}) as a consequence of removing degeneration at infinity. This indicates that a topological structure is important in exploration of dynamics of the cosmological models.
On the other hand when the physical trajectories of the Brans-Dicke models are in a bounded region of the phase space then the structural stability can be discussed in strict way as the property of the system itself.

\acknowledgments{
We are very grateful to the organizers of the 49th Winter School of Theoretical Physics
``Cosmology and non-equilibrium statistical mechanics'', L{\c a}dek-Zdr{\'o}j, Poland, February
10-16, 2013, especially to prof.~Andrzej Borowiec for invitation and opportunity to present 
part of this work. 
We thank Konrad Marosek for discussion and comments.

The research of O.H. was funded by the
National Science Centre through the postdoctoral internship award (Decision
No.~DEC-2012/04/S/ST9/00020).
}

\bibliographystyle{JHEP}
\bibliography{../bd_theory,../darkenergy,../quintessence,../quartessence,../astro,../dynamics,../standard,../inflation,../sm_nmc,../singularities,../lqc}

\end{document}